\documentclass[10pt,prd,preprintnumbers,floatfix,aps,nofootinbib,showpacs,twocolumn,superscriptaddress,noshowpacs]{revtex4} 

\usepackage{epsfig}
\usepackage{url}
\usepackage{hyperref}
\usepackage[normalem]{ulem} 
\usepackage{latexsym}
\usepackage{epsfig}
\usepackage{amsmath}
\usepackage{amssymb}
\usepackage{wasysym}
\usepackage{graphicx}
\usepackage{verbatim}
\usepackage{enumerate,mdwlist}%lists
\usepackage[titletoc]{appendix}
\usepackage{amsfonts}
\usepackage{tikz} %diagrams
\usetikzlibrary{calc}
\usepackage{pgfplots}
\usepackage[export]{adjustbox}
\usepackage{bbold}

\usepackage[normalem]{ulem}%strikethrough text

\newcommand{\lp}{\left(}
\newcommand{\rp}{\right)}
\newcommand{\lb}{\left[}
\newcommand{\rb}{\right]}

\newcommand{\pa}{\parallel}
\newcommand{\pe}{\perp}

\newcommand{\ba}{\begin{eqnarray}}
\newcommand{\ea}{\end{eqnarray}}
\newcommand{\be}{\begin{equation}}
\newcommand{\ee}{\end{equation}}
\newcommand{\bpm}{\begin{pmatrix}}
\newcommand{\epm}{\end{pmatrix}}

\newcommand{\Msun}{M_\odot}
\newcommand{\M}{\mathcal{M}}
\newcommand{\R}{\mathcal{R}}
\newcommand{\Lag}{\mathcal{L}}

\newcommand{\D}{{\mathcal{D}}}

\newcommand{\mcA}{{\mathcal{A}}}
\newcommand{\mcB}{{\mathcal{B}}}
\newcommand{\mcO}{{\mathcal{O}}}
\newcommand{\Od}{{\mathcal{O}}}
\newcommand{\mcG}{{\mathcal{G}}}

\newcommand{\gw}{\mathrm{gw}}

\newcommand{\rS}{r_s}%{r_\mathrm{s}}
\newcommand{\rV}{r_V}%{r_\mathrm{V}}
\newcommand{\rF}{r_4}
\newcommand{\rT}{r_2}
\newcommand{\LF}{\Lambda_4}
\newcommand{\LT}{\Lambda_2}
\newcommand{\Ln}{\Lambda_n}

\newcommand{\pFphi}{p_{4\phi}}

\newcommand{\pFX}{p_{4X}}
\newcommand{\pFXX}{p_{4XX}}
\newcommand{\pFXn}{p_{4X^n}}
\newcommand{\pT}{p_{2}}
\newcommand{\pTX}{p_{2X}}
\newcommand{\pTXX}{p_{2XX}}

\newcommand{\G}{{\mathcal{G}}}%effective metric
\newcommand{\vp}{\varphi}%Scalar Pert
\newcommand{\hb}{\bar{h}}%trace reversed
\newcommand{\bD}{\bar{\mathcal{D}}}%differential op
\newcommand{\hT}{\tilde{h}}%tilde pert
\newcommand{\gmn}{g_{\mu\nu}}%g_munu
\newcommand{\gab}{g_{\alpha\beta}}%g_\alpha\beta
\newcommand{\Gmn}{G_{\mu\nu}}%G_munu
\newcommand{\hmn}{h_{\mu\nu}}%h_munu
\newcommand{\hbmn}{\hb_{\mu\nu}}%hb_munu
\newcommand{\hTmn}{\hT_{\mu\nu}}%ht_munu
\newcommand{\dpp}{\lp\D_{\vp\vp}\rp}%D_vpvp
\newcommand{\para}{\parallel}

\newcommand{\Mpl}{M_{\text{Pl}}}

\newcommand{\ud}[2]{^{#1}_{\phantom{#1} #2}}
\newcommand{\du}[2]{_{#1}^{\phantom{#1} #2}}

\definecolor{grey}{rgb}{0.4,0.4,0.4}
\definecolor{dullmagenta}{rgb}{0.4,0,0.4}
\definecolor{darkblue}{rgb}{0,0,0.4}
\definecolor{midblue}{rgb}{0,0,0.5}
\definecolor{midred}{rgb}{0.5,0,0}
\definecolor{orange}{rgb}{1,0.5,0}
\definecolor{lightbrown}{rgb}{0.75,0.5,0.25}
\definecolor{tan}{cmyk}{0.14,0.42,0.56,0}
\definecolor{djunglegreen}{cmyk}{0.99,0,0.52,0}
\definecolor{lightgreen}{rgb}{0,1,0}
\definecolor{olivegreen}{cmyk}{0.64,0,0.95,0.40}
\definecolor{midgreen}{rgb}{0.0,0.675,0.0}
\definecolor{darkgreen}{rgb}{0,0.5,0}

\newcommand{\pP}{\langle\Phi\rangle}
\newcommand{\pPP}{\langle\Phi^{2}\rangle}

\usepackage[all]{xy} %array diagrams
\usepackage{amsfonts}

\begin{document}

\title{Gravitational wave lensing beyond general relativity:\\ birefringence, echoes and shadows}

\author{Jose Mar\'ia Ezquiaga}
\email{ezquiaga@uchicago.edu; \\ NASA Einstein Fellow}
\affiliation{Kavli Institute for Cosmological Physics and Enrico Fermi Institute, The University of Chicago, Chicago, IL 60637, USA}

\author{Miguel Zumalac\'arregui}
\email{miguel.zumalacarregui@aei.mpg.de}
\affiliation{Max Planck Institute for Gravitational Physics (Albert Einstein Institute) \\
Am Mühlenberg 1, D-14476 Potsdam-Golm, Germany}
\affiliation{Berkeley Center for Cosmological Physics, LBNL and University of California at Berkeley, \\
Berkeley, California 94720, USA}

\begin{abstract}
Gravitational waves (GW), as light, are gravitationally lensed by intervening matter, deflecting their trajectories, delaying their arrival and occasionally producing multiple images. 
In theories beyond general relativity (GR), new gravitational degrees of freedom add an extra layer of complexity and richness to GW lensing. We develop a formalism to compute GW propagation beyond GR over general space-times, including kinetic interactions with new fields. Our framework relies on identifying the dynamical \textit{propagation eigenstates} (linear combinations of the metric and additional fields) at leading order in a short-wave expansion. We determine these eigenstates and the conditions under which they acquire a different propagation speed around a lens. Differences in speed between eigenstates cause \textit{birefringence phenomena}, including time delays between the metric polarizations (orthogonal superpositions of $h_+,h_\times$) observable without an electromagnetic counterpart. In particular, \textit{GW echoes} are produced when the accumulated delay is larger than the signal's duration, while shorter time delays produce a \textit{scrambling} of the wave-form. We also describe the formation of \textit{GW shadows} as non-propagating metric components are sourced by the background of the additional fields around the lens.  
As an example, we apply our methodology to quartic Horndeski theories with Vainshtein screening and show that birefringence effects probe a region of the parameter space complementary to the constraints from the multi-messenger event GW170817. In the future, identified strongly lensed GWs and binary black holes merging near dense environments, such as active galactic nuclei, will fulfill the potential of these novel tests of gravity.
\end{abstract}

\date{\today}

% \pacs{
% %  04.30.-w %Gravitational waves
%  04.30.Nk %GW propagation and interactions
%  04.50.Kd, %modified gravity
%  95.36.+x, %Dark Energy
%  98.80.-k %cosmology
%  }

% \keywords{gravitational waves propagation, modified gravity}

\maketitle

 \tableofcontents
 
%---------
%SECTION: INTRODUCTION
%--------
\section{Introduction}

The detection of gravitational wave (GW) signals from black-hole and neutron-star mergers provides a direct probe of Einstein's general relativity (GR) and fundamental properties of gravity. 
These tests have far reaching implications for cosmology, probing the accelerated expansion of the universe and dark energy models in a manner complementary to ``traditional''  observations based on electromagnetic (EM) radiation  \cite{Ezquiaga:2018btd}.
Observations are sensitive to how GWs are emitted and detected, as well as their propagation through the universe.
GW emission and detection occurs in small scales by cosmological standards, in dense regions and near massive objects. In contrast, propagation can occur over vastly different regimes, and allows small effects to compound over very large distances.

GW propagation beyond GR is fairly well understood in the averaged cosmological space-time, described by the Friedmann-Robertson-Walker (FRW) metric. GWs are well described by linear perturbations due to the small amplitude of GWs away from the source. 
The high degree of symmetry of FRW solutions ensures the decoupling of scalar, vector and tensor perturbations, automatically isolating the propagating degrees of freedom, with deviations from GR represented by a handful of terms in the propagation equation.
These facts greatly simplify the study of GWs, making it tractable even for highly complex theories beyond GR.

%-FIGURE  SUMMARY DIAGRAM-
\begin{figure*}[t!]
\centering 
\includegraphics[width = 0.95\textwidth,valign=t]{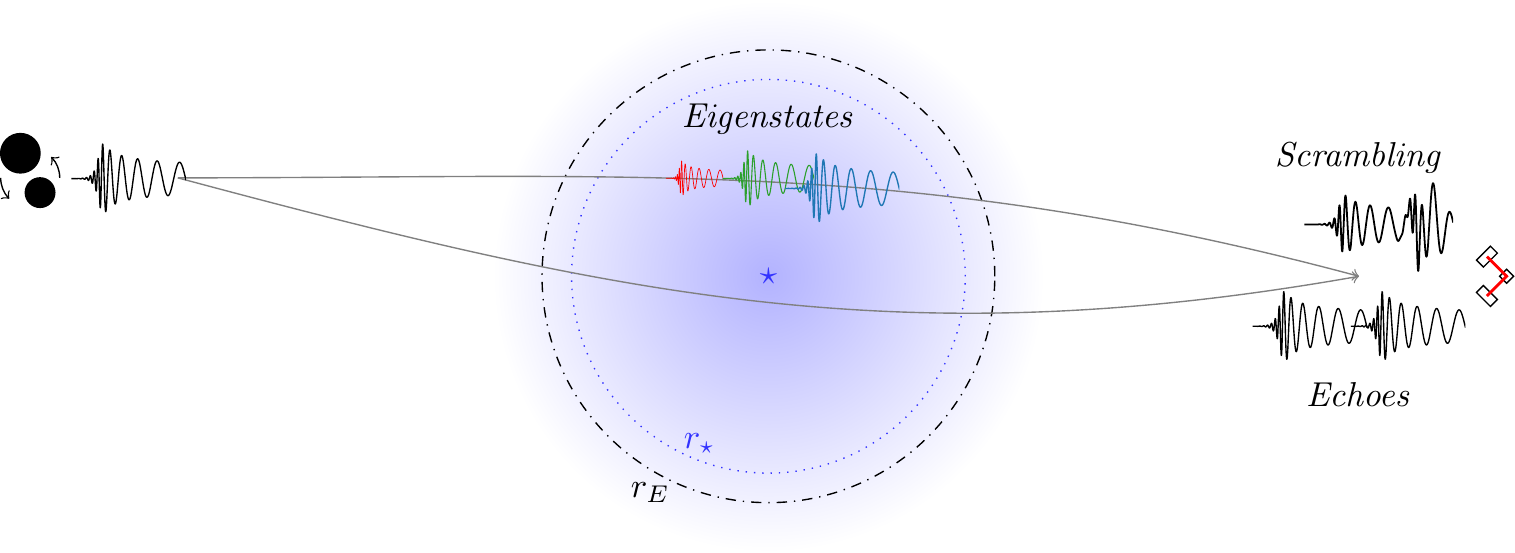}
 \caption{Schematic diagram of the gravitational wave lensing beyond general relativity. A GW emitted by a binary black hole splits into its propagation eigenstates (wave-forms in color) when it enters the region enclosed by $r_\star$ where modify gravity backgrounds are relevant (note that in general this scale can be different from the scale of strong lensing, i.e. the Einstein radius $r_E$). Depending of the time delays between the propagation eigenstates the signal detected could be scrambled or echoed. If the GW travels closer than the Einstein radius, multiple images could be formed as indicated by the gray solid trajectories. }
 \label{fig:diagram_summary}
\end{figure*}
%---

Corrections to FRW GW propagation have been well studied and have provided some of the most powerful tests of gravitational theories. Such is the case of the anomalous GW speed, measured to a precision $\vert c_g- c \vert\lesssim \mathcal{O}(10^{-15})$ \cite{2041-8205-848-2-L13} with the binary neutron star merger GW170817. This measurement poses a phenomenal challenge to a broad class of dark energy theories \cite{Ezquiaga:2017ekz,Creminelli:2017sry,Baker:2017hug,Sakstein:2017xjx}, well beyond next-generation cosmological observations \cite{Alonso:2016suf}. 
Other tests such as GW damping \cite{Deffayet:2007kf} are limited by precision in the luminosity distance measurement and the population of standard sirens, with the weak constraints from GW170817 \cite{Arai:2017hxj} expected to improve in the future \cite{Lagos:2019kds,Belgacem:2017ihm,Belgacem:2019pkk,Belgacem:2018lbp,Baker:2020apq}. 
In addition, FRW GW propagation can be used to constrain interactions with additional cosmological fields \cite{Jimenez:2019lrk} such as tensor  \cite{Max:2017flc} or multiple vector fields \cite{Caldwell:2016sut}, but only when the additional fields have a tensor structure. 
Despite of these achievements and prospects, tests of the propagation of GWs over FRW are intrinsically limited in probing gravity theories by the same simplifications that made them tractable in the first place.

Lensing of GWs offers important opportunities to test GR in at least three distinct ways.  
1) In minimally coupled theories, lensing of electromagnetic radiation only probes the solution of the metric. In contrast, lensing of GWs tests the gravitational sector directly, including the fundamental degrees of freedom, their properties and interactions.
2) New propagating degrees of freedom are in most cases isolated by the FRW symmetries: even the simplest gravitational lenses break these symmetries and introduce new interactions with new gravitational fields (e.g. scalars).  
3) Finally, beyond FRW effects can introduce new scales and affect the gravitational polarizations ($+,\times$) differently, providing signatures that do not require an electromagnetic (EM) counterpart. This enables tests from black-hole (BH) binaries, applicable to more events and at higher redshift. 
Specific examples of these features are explored in this work.

The well studied and rich phenomenology of gravitational lensing highlights the importance of understanding GW propagation beyond FRW in testing GR. 
Phenomena ranging from galaxy shape distortions, to multiple imaged sources, to the integrated Sachs-Wolfe effect are nowadays routinely used to probe dark energy and gravity. 
As detections of lensed GWs will become increasingly likely \cite{10.1093/mnras/sty2145,PhysRevD.97.023012,10.1093/mnras/sty411}, modelling GW propagation beyond the FRW approximation will become critical to fully use rapidly growing catalogues of GW events to explore the intervening matter and its gravitational effects. 
As we will discuss here, theories beyond GR extend the range of gravitational lensing phenomena even further.

\subsection{Summary for the busy reader}

In this work we study the lensing of GWs beyond GR. We develop a general framework to study the GW propagation over general space-times, identify novel effects and forecast constraints on specific gravity theories. Our main results can be summarized as follow:
\begin{itemize}
\item \emph{Core concept:} over general space-times different gravitational degrees of freedom \emph{mix} while they propagate. Each \emph{propagation eigenstate} is a linear superposition of different polarizations that evolves independently.  
Eigenstates with different speeds cause GW \emph{birefringence}.  
Non-propagating modes can also be sourced inducing \emph{GW shadows}. We present our formalism in sections \ref{sec:formalism} and \ref{sec:lensing_beyond_gr}.
\item \emph{Novel phenomena:} at leading order, the main observables are time delays between propagation eigenstates and with respect to light. Delays larger than the GW signal produce \emph{echoes}. Time delays shorter than the signal cause interference patterns, \emph{scrambling} the wave-form. We investigate these phenomena in section \ref{sec:observational_effects}, where we also discuss the observational prospects. Particularly interesting events for these tests correspond to identified strongly-lensed multiple images and binary black holes merging close to a super-massive black hole.
\item \emph{An example, screening in Horndeski:} a natural arena for these lensing modifications are gravity theories with screened environments. We obtain the propagation eigenstates of Horndeski gravity over general space-times in section \ref{sec:horndeski}. We then study the lensing time delays induced by Vainshtein screening in section \ref{sec:screening}. 
\item \emph{Detection prospects:} These novel lensing effects could be critical to test gravity theories beyond GR. For our simple quartic Horndeski example theory we already find large sectors of the parameter that could be constrained beyond GW170817. Dedicated analyses could be applied to past and future LIGO-Virgo data. These birefringence tests do not require electromagnetic counterparts.
\end{itemize}

A schematic diagram of the effects of lensing beyond GR is presented in Fig. \ref{fig:diagram_summary}. A GW traveling near the lens splits in the different propagation eigenstates. If the modified gravity theory and background configuration around the lens is such that the eigenstates have different speeds, the overall GW signal could split into sub-packets after crossing the lens potential leading to echoes in the detector. If the time delay between the eigenstates is shorter than the duration of the signal, there will be interference effects producing a scrambling of the detected signal.

%--------
%SECTION: GENERAL CONSIDERATIONS
%--------
\section{The problem: a general theory for gravitational radiation}
\label{sec:formalism}

For any given gravity theory, the propagation of GWs can be determined from the equations of motion (EoM) for the linearized perturbations, which are obtained expanding around the background metric
\begin{equation}
 g_{\mu\nu}^{\rm tot} =  g_{\mu\nu} + h_{\mu\nu}\,.
\end{equation}
For concreteness, we will focus our discussion to metric theories of gravity with an additional scalar field, although our arguments could be easily extended to other types and number of fields. 
Expanding similarly the scalar field around the background solution
\begin{equation}
\phi^{\rm tot} =  \phi + \vp\,,
\end{equation}
the evolution of GWs $h_{\mu\nu}$ and the additional gravitational degree of freedom $\vp$ will follow a set of coupled equations
\begin{align}
&\lp \D_{hh}\rp_{\mu\nu}^{~~\alpha\beta}h_{\alpha\beta} + \lp\D_{h\vp}\rp_{\mu\nu}\varphi = 0\,, \label{eq:schematic_tensor}\\
&\lp \D_{\vp\vp}\rp\vp + \lp\D_{\vp h}\rp^{\alpha\beta}h_{\alpha\beta}=0\,, \label{eq:schematic_scalar}
\end{align}
where each of the differential operators depend on background quantities and in order to distinguish among them we have introduced a sub-index to indicate which perturbations the operator is acting on in the action. Therefore, the propagation could be modified with respect to GR by \emph{(i)} new interaction terms leading to $\D_{hh}\neq\D_{\rm GR}$, \emph{(ii)} the mixing with $\vp$ and \emph{(iii)} the modification of the effective background in which GWs propagate. 

Any of these modifications makes solving the propagation of GWs significantly more complicated than in GR.  
The essence of the problem will be identifying the \emph{propagation eigenstates} which diagonalize the EoM. In general, we will encounter two main obstacles with respect to the standard approach: fixing the gauge (section \ref{sec:gauge_fix}) and identifying the radiative degrees of freedom (DoF) (section \ref{sec:radiative_dof}).
We will also introduce the short-wave expansion (section \ref{sec:wkb_expansion}). 

%GAUGE FIXING
\subsection{Gauge fixing}\label{sec:gauge_fix}

The richer structure of the propagation equations beyond GR affects the gauge fixing procedure. 
In synthesis, one can always fix the transverse gauge
\begin{equation}\label{eq:gauge_transverse}
\nabla^\mu  h_{\mu\nu} =0\,,
\end{equation}
but not simultaneously set the traceless condition 
\begin{equation}
h = g^{\mu\nu}h_{\mu\nu}=0 \,.
\end{equation}
Imposing the traceless condition throughout relies on $h$ obeying the same equation as the residual gauge, which is not true in general beyond GR.

A gauge transformation is a diffeomorphism $x^\mu\to x^\mu + \xi^\mu$ that preserves the form of the background metric $g_{\mu\nu}$. It acts on the metric perturbation as
\begin{equation}\label{eq:gauge_metric_transform}
 h_{\mu\nu} \to h_{\mu\nu} + 2\nabla_{(\mu}\xi_{\nu)}\,,
\end{equation}
where derivatives and contractions involve the background metric $g_{\mu\nu}$.
We will start with the transverse condition (\ref{eq:gauge_transverse}), defined relative to $g_{\mu\nu}$.\footnote{We will discuss the generalization to a transverse condition with respect to a different metric in appendix \ref{app:gauge_alternative}.} 
The transverse condition transforms as
\begin{equation}
 \delta\left(\nabla^\mu h_{\mu\nu}\right) = \Box \xi_\nu +  R\du{\nu}{\mu}\xi_\mu\,,
\end{equation}
where the Ricci tensor of the background metric stems from a re-arrangement of covariant derivatives.
The transverse condition is imposed by $\xi^\mu(x)$ satisfying
\begin{equation}\label{eq:gauge_transverse_fix}
\Box \xi_\nu + R\du{\nu}{\mu}\xi_\mu = -\nabla^\mu h_{\mu\nu}\,.
\end{equation}
The above choice does not completely fix the gauge, as any additional transformation $x^\mu\to x^\mu+\zeta^\mu(x)$ will preserve the transverse condition if
\begin{equation}\label{eq:gauge_residual}
\Box \zeta_\nu + R\du{\nu}{\mu}\zeta_\mu = 0\,.
\end{equation}
This equation fixes the time evolution of the residual gauge.

Let us now investigate whether we can eliminate the trace of the metric $h$ using the residual gauge. Using the trace of Eq. (\ref{eq:gauge_metric_transform}), eliminating the trace requires
\begin{equation}\label{eq:gauge_trace_fixing}
 \nabla_\mu\zeta^\mu = \frac{1}{2}h\,.
\end{equation}
Although at some initial time we can always fix the amplitude of $\zeta_\mu$ to satisfy this condition, Eq. (\ref{eq:gauge_trace_fixing}) will only be preserved if the trace has the same causal structure that $\zeta_\mu$. This problem occurs in GR in the presence of sources ($R_{\mu\nu}\neq0$) and the trace cannot be eliminated \emph{globally}. However, the difference beyond GR is that one cannot even fix the trace \emph{locally}, because $h$ will be subject to a different differential operator. 
A similar conclusion was obatined in \cite{Rizwana:2016qdq} in the context of $f(R)$ gravity.

%RADIATIVE DoF
\subsection{Identifying the radiative degrees of freedom}
\label{sec:radiative_dof}

The presence of additional fields complicates the identification of the propagating degrees of freedom. On the one hand, the background field mixes the metric perturbations in new ways. In the case of a scalar field this is achieved with their derivatives, for example $\nabla^\mu\phi\nabla^\nu\phi\cdot h_{\mu\nu}$ or $\nabla^\mu\nabla^\nu\phi\cdot h_{\mu\nu}$. On the other hand, the extra perturbations have their evolution coupled with $h_{\mu\nu}$. This means that the decomposition in radiative and non-radiative DoF will be background dependent and in general not possible in a covariant language. Moreover, the new interaction terms could source the non-radiative modes even in vacuum. Thus, we have to keep track of all the constraints and propagation equations. 

In a local region of space-time, we are in the limit of linearized gravity and can decompose the 10 metric perturbations around flat space as\footnote{This procedure can also be applied around a curved background provided that $g_{0i}\ll g_{00},g_{ij}$.}
\begin{equation} \label{eq:3+1_dof}
\begin{split}
ds^2=-(1+2\Phi)&dt^2+w_i(dtdx^i+dx^idt) \\
&+((1-2\Psi)\delta_{ij}+2s_{ij})dx^idx^j\,,
\end{split}
\end{equation}
where $\Phi$ is a scalar (1DoF), $w_i$ is a vector (3DoF), $s_{ij}$ is a traceless tensor (5DoF) and $\Psi$ is a scalar (1DoF). 
As discussed before, some of these perturbations are non-physical and can be removed fixing the gauge. Under a gauge transformation, the above perturbations change as
\begin{align}
\Phi&\rightarrow\Phi+\partial_0\xi^0\,, \\
w_i&\rightarrow w_i+\partial_0\xi_i-\partial_i\xi^0\,, \\
\Psi&\rightarrow\Psi-\frac{1}{3}\partial_i\xi^i\,, \\
s_{ij}&\rightarrow s_{ij}+\partial_{(i}\xi_{j)}-\frac{1}{3}\partial_k\xi^k\delta_{ij}\,.
\end{align}
We can always set the spatial transverse gauge $\partial^is_{ij}=0$ as 
\begin{equation}
\nabla^2\xi_j+\frac{1}{3}\partial_j\partial_i\xi^i=-2\partial^is_{ij}\,.
\end{equation}
We can also use $\xi^0$ to set $\Phi=0$ or the vector components to be transverse $\partial^iw_i=0$. These choices do not exploit the residual gauge freedom, but will be enough for our purposes.

In the spatial transverse gauge $s_{ij}$ contains the two transverse-traceless polarizations $h_+$ and $h_\times$. In this language, the fact that the background scalar mixes the tensor modes translates into $\Phi$, $\Psi$ and $w_i$ not being set to zero by the constraint equations. In general, the non-radiative DoF will be sourced by both $s_{ij}$ and $\vp$, which themselves mix during the propagation.

%PROPAGATION EIGENSTATES
\subsection{Short-wave approximation}\label{sec:wkb_expansion}

As a working hypothesis we will consider that the wave-length of the GWs is small compared to the typical spatial variation of the background fields.  
That is, we will make a \emph{short-wave} or \emph{WKB} approximation \cite{misner1973gravitation}, expanding the metric perturbation as  
\be \label{eq:wkb_metric}
\hmn= \lp\mcA^{(0)}_{\mu\nu}+\epsilon\mcA^{(1)}_{\mu\nu}+\mcO\lp\epsilon^2\rp\rp e^{i\frac{\theta}{\epsilon}}\,,
\ee
and the scalar wave
\be \label{eq:wkb_scalar}
\vp= \lp\mcA^{(0)}_s+\epsilon\mcA^{(1)}_s+\mcO\lp\epsilon^2\rp\rp e^{i\frac{\theta}{\epsilon}}\,,
\ee
where we have introduced a set of amplitudes $\mcA^{(n)}$, a phase $\theta$ and a small dimensionless parameter $\epsilon$.%
\footnote{$\epsilon$ is used for book-keeping only and can be set to one when the different orders in the calculation have been collected.}

The short-wave expansion leads naturally to the wave-vector definition
\be
k_\mu=\frac{\partial \theta}{\partial x^\mu}\,,
\ee
from the gradient of the phase. 
The leading order observables will be the phase evolution and propagation eigenstates, which are determined by the second derivative operators. In other words, we will be solving the mixing in the kinetic terms. Next to leading order contributions will introduce corrections to the amplitude and further mixings. We leave their analysis for future work.

At leading order in derivatives, solving the propagation entails diagonalizing an $11\times11$ matrix
\begin{equation}
\D_{ab}V_{b}=0\,,
\end{equation}
where $D_{ab}$ is a matrix of second order differential operators and $V_b$ is a vector containing the 10 metric components $\hmn$ plus the scalar degrees of freedom $\vp$. Fortunately, as we discussed in section \ref{sec:radiative_dof}, locally we can reduce this to a $3\times3$ problem. We will generically refer to the propagation eigenstates as $H_J$ with $J=1,2,3$. Moreover, we define $\hat{\M}$, the \emph{mixing matrix} changing from the basis of interaction eigenstates $(h_\times,h_+,\vp)$ to the basis of propagation eigenstates $(H_1,H_2,H_3)$:
\be
\bpm H_1 \\ H_2 \\ H_3 \epm = \hat{\M}\bpm h_+ \\ h_\times \\ \vp\epm \,.
\ee

In addition, we will focus in the regime where the stationary phase approximation holds, that is, when the time delay between the lensed images is larger than the duration of the signal.  
A hard limit on the stationary phase approximation is the onset of diffraction and wave effects \cite{Takahashi:2003ix}, which occurs when the multiple images interfere or the wavelength of the GW $\lambda_\gw$ is of the order of the Schwarzschild radius of the lens $\rS = 2GM_L/c^2$. For a compact binary this can be translated into
\begin{equation}\label{eq:diffraction_limit}
 \frac{M_L}{M_\odot}\lesssim 10^5 \left(\frac{f_\gw}{\rm Hz}\right)^{-1}\,,
\end{equation}
where $f_\gw$ is the frequency of the GW. In the band of ground-based detectors, wave optics is only relevant for lenses $M_L\lesssim 100-1000\Msun$. At lower frequencies (e.g. LISA and other space-borne GW detectors) diffraction effects are produced by heavier lenses.

%--------
%SECTION: PROPAGATION EIGENSTATES IN HORNDESKI THEORIES
%--------
\section{GW lensing beyond General Relativity}
\label{sec:lensing_beyond_gr}

From the previous section we learned that over general backgrounds GW degrees of freedom mix during the propagation. Therefore, the first step to study lensing beyond GR is to identify the propagation eigenstates. In section \ref{sec:lensing_eigenstates} we will use an example theory to identify propagation eigenstates as a combination of different polarizations, travelling at different speeds. 
This speed difference leads to \emph{birefringence} (polarization-dependent deflection and \emph
{time delays}), which are discussed in section \ref{sec:lensing_pheno}. 
The observational consequences will be discussed later, in section \ref{sec:observational_effects}.

%SECTION: PROPAGATION EIGENSTATES IN HORNDESKI THEORIES
\subsection{Propagation Eigenstates}
\label{sec:lensing_eigenstates}

In order to build intuition about kinetic mixing, let us consider a particular example. We will keep the discussion general for the moment and later show how this example materializes in a concrete class of scalar-tensor theories (see section \ref{sec:horndeski}). 
Let us further assume that we have already solved the constraint equations and we are left with $h_+$, $h_\times$ and $\vp$. At leading order, the equations for the propagating modes can then be written schematically as%
\footnote{This is not the most general situation since there could also be an induced mixing between $h_+$ and $h_\times$ (we will discuss some examples in section \ref{sec:horndeski_non-luminal}). However this example contains the relevant phenomenology while allowing for analytic diagonalization.}
\begin{equation}\label{eq:eq_LO}
\left(
 \begin{array}{ccc}
  G_{hh} & 0 & G_{+s} \\
  0 & G_{hh} & G_{\times s} \\
  G_{+s} & G_{\times s} & G_{ss}
 \end{array}
\right)
\left(
\begin{array}{c}
 h_+ \\
 h_\times \\
 \varphi
\end{array}
\right)
\equiv
\hat{\mathcal{D}}
\left(
\begin{array}{c}
 h_+ \\
 h_\times \\
 \varphi
\end{array}
\right)
=0\,,
\end{equation}
where the coefficients of the kinetic matrix $\hat{\mathcal{D}}$ can be read off by, in general, comparing with the covariant equations. In Fourier space and normalizing the fields canonically, we have
\begin{eqnarray} 
 G_{hh} = \omega^2 - c^2_{ij} k^i k^j\,
 &,&\; 
 G_{ss} = \omega^2 - c^{s2}_{ij}k^ik^j \label{eq:standard_dispersion}\\
 G_{+s} = k^2 M_\phi \cos(2\phi)\,
 &,&\; 
 G_{\times s} = k^2 M_\phi\sin(2\phi) \label{eq:mixing_coefficients}
\end{eqnarray}
where $k^2 = \omega^2-c_m^2 \vec k^2$ (the factor $k^2$ indicates the mixing vanishes, on shell, for modes propagating at the speed of light)
and $M_\phi$ controls the mixing between the tensor and scalar modes. 
For solutions to exist the determinant of the kinetic matrix  $\det(\hat{\mathcal{D}})=G_{hh}(G_{hh}G_{ss}-M_\varphi^2 k^2)$ needs to be non-zero.%

The propagation eigenfrequencies of the system are given by the characteristic equation $ \det(\hat{\mathcal{D}} - \lambda_i \mathbb{1}) = 0$ 
and choosing $\omega$ so that $\lambda_i(\omega_i)=0$, or equivalently 
% (substituting $\lambda_i=0$)
\begin{equation}\label{eq:eigenvalues}
 G_{hh}\big(G_{hh}G_{ss}-M_\phi^2 k^4\big) =0\,.
\end{equation}
In the absence of mixing ($M_\phi=0$), the propagation of each mode is determined by the standard dispersion relations (\ref{eq:standard_dispersion}), which allows a non-luminal speed for scalars and tensors. 

The propagation eigenmodes can be obtained by solving
\begin{equation}
 (\hat{\mathcal{D}}-\lambda\mathbb{1})\vec v_i = \hat{\mathcal{D}}(\omega_i)\vec v_i = 0\,.
\end{equation}
(the second equality enforces the on-shell relation $\lambda_i(\omega_i)=0$). 
In other words, the propagation eigenstates can be defined through the mixing matrix $\hat{\M}$ that relates them to the interaction eigenstates,
\be \label{eq:change_basis}
\bpm H_1 \\ H_2 \\ H_3 \epm= \bpm v_{1+} & v_{1\times} & v_{1\vp} \\ v_{2+} & v_{2\times} & v_{2\vp} \\ v_{3+} & v_{3\times} & v_{3\vp} \epm \bpm h_+\\ h_\times \\ \vp\epm\,,
\ee
where the rows are precisely the eigenvectors $\vec v_i$. Note that because the equations of motion (\ref{eq:eq_LO}) define a symmetric matrix, the matrix of eigenvectors is orthogonal and we can simply invert this mapping by $\vec h = \hat \M ^{T} \vec H$. 
It is useful to define the phase speeds as
\begin{equation}
 c^2_h = \frac{1}{\vec k^2}c^2_{ij}k^i k^k\,,\; c^2_s = \frac{1}{\vec k^2}c^{s2}_{ij}k^i k^k\,,
\end{equation}
where the directional dependence on $\hat k$ has been omitted.
We will study the case in which the GW speed is not modified before presenting the general calculation.

\subsubsection{Equal speed case $c_h=c_m$} 
\label{sec:ch_eq_cm}

In the case in which the GW speed $c_h$ equals the mixing speed $c_m$ the eigenvalue equation simplifies considerably:
\begin{equation}
\left(\omega^2-c_m^2 \vec k^2\right)^2 \left((1-M_\phi^2)\omega^2-\vec k^2(c_s^2-c_m^2 M_\phi^2)\right) =0\,,
\end{equation}
One can then check that the eigenmodes propagating with speed $c$ correspond to the two metric polarizations. 

The third eigenmode is a combination of the scalar and metric perturbation
\begin{equation}
 \vec v_3 =(-M_\phi \cos (2 \phi ),-M_\phi \sin (2 \phi ),1)\to (0,0,1)\,,
\end{equation}
propagating with speed
\begin{equation}
 c_3^2 = \frac{c_s^2-c_m^2 M_\phi^2}{1-M_\phi^2}\to c_s^2\,,
\end{equation}
where the arrow represents the limit of small mixing $M_\phi^2/(c_h^2-c_s^2)\ll1$. Note that the mixing can turn the scalar speed imaginary, triggering a gradient instability.

Similarly when $c_s=c_m$ the diagonalization simplifies. In this case, we obtain $c_1=c_h$, $c_3=c_m$ and 
\be
c_2=\frac{c_m^2M_\phi^2-c_h^2}{M_\phi^2-1}\,.
\ee
The second eigenmode is then 
\begin{equation}
 \vec v_2 =(\cos (2 \phi ),\sin (2 \phi ),-M_\phi)\,.
\end{equation}
Thus, $M_\phi$ controls the amplitude of the induced scalar perturbation.

\subsubsection{General case $c_h\neq c_m$}

The situation is more involved in the general case when the tensor and mixing speed are not the same. The characteristic equation is 
\begin{equation}
  (\omega^2-c_h^2\vec k^2) \Big((\omega^2-c_h^2\vec k^2)(\omega^2-c_s^2\vec k^2)-M_\phi^2(\omega^2-c_m^2\vec k^2)^2
  \Big) = 0\,,
\end{equation}
(if either $c_s,c_h$ are equal to $c_m$ then one of the terms factorizes and we're back to the previous case).
The first parenthesis indicates that one eigenstate will propagate with speed $c_1=c_h$. The two remaining modes are mixed, and their speeds, $c_2,c_3$ are determined by equating the second parenthesis to zero.
It is useful to define the sum and difference of the square of the mixed modes velocities
\begin{eqnarray}
 \Sigma &\equiv& c_2^2+c_3^2 = \frac{c_h^2+c_s^2-2c_m^2M_\phi^2}{1-M_\phi^2}\,, \\
 \Delta &\equiv& c_2^2-c_3^2 = \frac{\sqrt{(\Delta c_{hs}^2)^2+4 M_\phi^2 \Delta c_{hm}^2\Delta c_{sm}^2}}{1-M_\phi^2}\,,
\end{eqnarray}
where we define the difference in the speeds $\Delta c_{ij}^2=c_i^2-c_j^2$ and one should recall that $c_i = \omega_i/|\vec k|$. Then the eigenstates and their velocities are given by
\begin{enumerate}
 \item \uline{Pure metric polarization:} 
 \begin{equation}
 \vec v_1 = 
 \left(
 \begin{array}{c}
 -\sin(2\phi) \\ \cos(2\phi) \\0 
 \end{array}
\right)\,
\,,\qquad 
 c_1^2 = c_h^2\,.
 \end{equation}
 $\vec v_1$ is the combination of $h_+,h_\times$ orthogonal to the scalar field shear and its propagation speed corresponds to the tensor speed without mixing.
  
 \item \uline{Mostly-metric polarization:}
  \begin{eqnarray}
  \vec v_2 &=& \left(
  \begin{array}{c}
  \cos(2\phi) \\ \sin(2\phi) \\ 
  M_\phi\frac{2c_h^2-\Delta-\Sigma}{\Sigma+M_\phi^2\Delta-c_h^2-c_s^2}
  \end{array}
  \right)
  \,,
  \  
 c_2^2 = \frac{1}{2}\left(\Sigma+\Delta \right).
 \end{eqnarray} 
 $\vec v_2$ is thus a combination of tensorial and scalar polarizations with a propagation speed different from $c_h^2$. In the limit of small mixing $M_\phi^2\ll1$ one obtains
  \begin{eqnarray}
  \vec v_2 &\to&  %low mixing limit
 \left(
  \begin{array}{c}
  \cos(2\phi) \\ \sin(2\phi) \\ M_\phi\frac{c^2-c_h^2}{c_h^2-c_s^2}
  \end{array}
  \right) + \cdots
  \,,
  \\
 c_2^2 &\to& c_h^2 + M_\phi^2\frac{(\Delta c_{hm}^2)^2}{\Delta c_{hs}^2} + \cdots\,,
 \end{eqnarray} 
 where it is then clear that $\vec v_2$ reduces to the combination of $h_+,h_\times$ orthogonal to $\vec v_1$ when $M_\phi/\Delta c_{hs}^2\to 0$. 
    
\item \uline{Mostly-scalar polarization:}
  \begin{eqnarray}
   \vec v_3 &=& \left(
   \begin{array}{c}
   M_\phi\cos(2\phi) \\ M_\phi\sin(2\phi) \\ -M_\phi^2\frac{2c_h^2+\Delta-\Sigma}{c_h^2+c_s^2+M_\phi^2\Delta-\Sigma} 
   \end{array}
   \right)
  \,,\ 
  c_3^2 = \frac{1}{2}\left(\Sigma-\Delta \right)\,.
 \end{eqnarray}
 $\vec v_3$ is also a combination of tensorial and scalar polarizations with a propagation speed different from $c_s^2$. When the mixing is small one finds
  \begin{eqnarray}
   \vec v_3 &\to&  
   \left(
   \begin{array}{c}
   0 \\ 0 \\ \frac{c_s^2-c_h^2}{c^2-c_s^2}
   \end{array}
   \right) +\cdots
  \,,
  \\
  c_3^2 &\to& c_s^2 - M_\phi^2\frac{(\Delta c_{sm}^2)^2}{\Delta c_{hs}^2}  + \cdots\,.
 \end{eqnarray}
  $\vec v_3$ it reduces to the scalar polarization when $M_\phi/\Delta c_{hs}^2\to 0$. One should note that in this definition it has been assumed $c_h^2>c_s^2$, otherwise $\vec v_2$, $\vec v_3$ are swapped.
\end{enumerate}

Two quantities will be specially relevant in the following discussion: $\Delta c^2_{10}\equiv c_1^2 - c^2$, the speed difference between the pure-metric eigenstates and electromagnetic signals; and $\Delta c^2_{21}\equiv c_2^2 - c_1^2$, the difference between the mostly-metric and pure-metric eigenstates. 
In the limit of small mixing the second one can be expressed as
\be
\Delta c_{21}^2= M_\phi^2\frac{(\Delta c_{hm}^2)^2}{\Delta c_{hs}^2} + \mcO\lp M_\phi^3\rp\,.
\ee
A difference in the propagation speed between the first two propagation eigenstates leads to a polarization dependent propagation in the interaction basis. In other words, there could be birefringence in the detected GW signals. 
Therefore, we will generically refer to differences in the propagation with respect to light as \emph{multi-messenger}, while the differences among propagation eigenstates will be referred as \emph{birefringent}.

%-FIGURE DIAGRAM-
\begin{figure*}[t!]
\centering 
\includegraphics[width = 0.98\textwidth,valign=t]{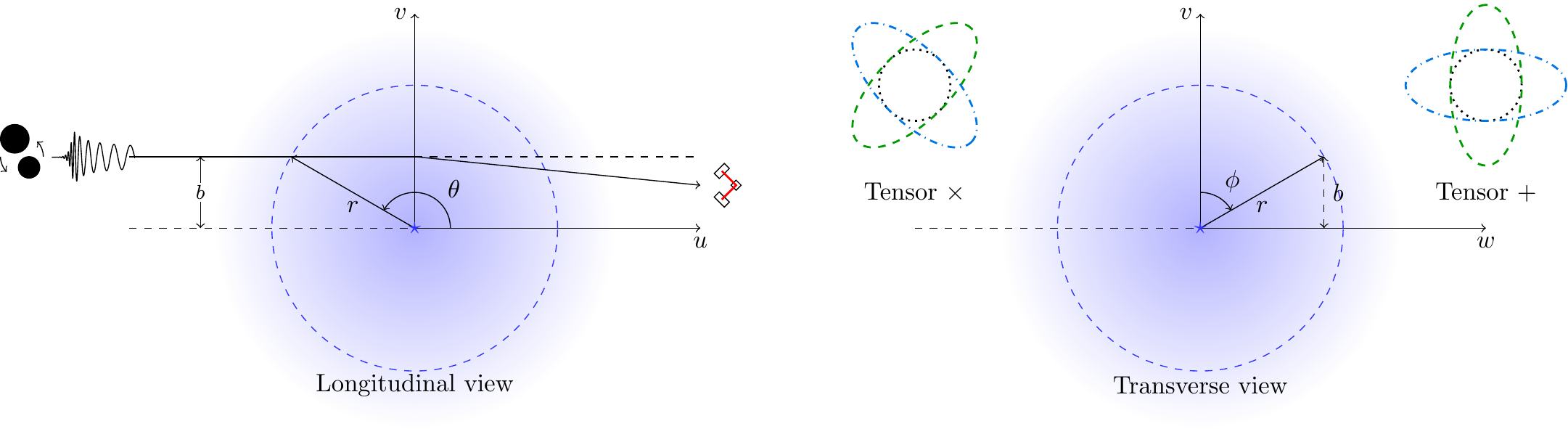}
 \caption{Schematic view of the gravitational wave propagation. The undeflected GW trajectory corresponds to the solid black line for an impact parameter $b$, as plotted on the left. The transverse view is presented on the right together with a representation of the effect of a tensor mode crossing a circle of test particles. At any given point the GW is located at a radius $r$ and angular positions $\theta$ and $\phi$.} 
 \label{fig:diagram_prop}
\end{figure*}
%---

\subsection{Birefringence, GW deflection and time delays}
\label{sec:lensing_pheno}

%Observables
There are 4 signals whose propagation can be studied at leading order in GW lensing beyond GR: electromagnetic radiation (or standard model particles) traveling at speed $c_0\equiv c$ and 3 propagation eigenstates traveling at speeds $c_1,c_2,c_3$, which depend on the interaction basis speeds $c_h,c_s,c_m$ and the mixing $M_\phi$. 
A gravitational lens will imprint a deflection and time delay, which might differ between each signal. 
In addition lensing will (de-)magnify the images and introduce a characteristic phase shift for images that cross caustics \cite{Schneider_book,Ezquiaga:2020gdt}. 
Here we will discuss deflection angles briefly, before focusing on the implications of time delays. In the following we will assume sources and lenses in the geometric optics limit, where the wavelength of the GW is much smaller than the Schwarzschild radius of the source $\lambda_\text{gw}\ll \rS=2GM_L/c^2$.

One should note that in general there will be two types of effects in modified gravity:  an \emph{anomalous speed effect} due to the modified effective metrics in which each eigenstate propagates and a \emph{universal effect} due to the modified Newtonian potentials stemming from $\Phi,\,\Psi$, whose relationship with the matter distribution might differ via modified Poisson equations. The anomalous speed effect will affect the deflection angle and time delays of each propagation eigenstate differently (e.g. birefringence). The universal effect is the same for all polarizations and ultra-relativistic matter signal due to the equivalence principle. 
Traditional lensing analyses in modified gravity have focused on the universal effect, searching for deviations in the gravitational potentials $\Phi \neq \Psi$ (see e.g. \cite{Mukherjee:2019wfw}). Here we focus on the novel effects due to the anomalous speed of the propagation eigenstates.

\subsubsection{Deflection angle}

Let us consider the deflection of a ray/signal propagating in the $u$ direction. 
The eikonal equation for the phase of the propagation eigenmode $I$, cf Ref. \cite[Eq. 3.15]{Schneider_book}, reads
\begin{equation}\label{eq:general_k_equation}
 \dot k_{\alpha} = -\frac{1}{2} (\partial_\alpha g_I^{\mu\nu})k_\mu k_\nu
 = -\frac{1}{2} \frac{\partial c_I^2(\vec x,\hat k)}{\partial x^\alpha} |\vec k|^2\,,
\end{equation}
where $\dot k_{\alpha}$ is a derivative w.r.t. the affine parameter and the second equality assumes a static metric and canonical normalization (i.e. $g_I^{\mu\nu}k_\mu k_\nu = -\omega^2 + c_I^2(\vec x,\hat k)|\vec k|^2$ and using the fact that $k,x$ are independent variables).

Expanding on small deviations around the unperturbed trajectory $k_\alpha = k_\alpha^{(0)} + k_\alpha^{(1)}+\cdots$ the (small) deflection angle is  
\begin{equation} \label{eq:deflec_angle}
 \vec{\hat{\alpha}}_I \approx \frac{\vec k^{(1)}}{|\vec k^{(0)}|} 
\approx -\frac{1}{2}\int du \, \vec\nabla_\perp c_I^2(\vec x,\hat k)\big|_{r(u),\hat u}\,,
\end{equation}
where the integral is obtained in the \textit{Born approximation} by evaluating Eq. (\ref{eq:general_k_equation}) on the unperturbed trajectory $x^\alpha \approx x^\alpha_{(0)}$ and specializing to a spherical lens. We have defined the propagation direction $\hat k \propto \hat u$, the radial distance $r^2(u) = u^2+b^2$ and the gradient perpendicular to the propagation direction $\vec\nabla_\perp$ (note that we can always define $\vec k_{(0)}\cdot\vec k_{(1)}=0$). The geometry of the problem is summarized in Fig. \ref{fig:diagram_prop}.    

Equation (\ref{eq:deflec_angle}) can be used to compute the deflection angle for light and ultrarelativistic particles with minimal coupling to the metric. In that case, the effective velocity induced by the perturbed potential $\Phi$ and $\Psi$ (using the previously mention canonical normalization)
\be
\frac{c_{0, \rm eff}^2(x)}{c^2}=\frac{1-2\Psi(x)}{1+2\Phi(x)}\,,
\ee
leads to the standard expression in terms of the metric potential
\be \label{eq:deflec_angle_light}
\vec{\hat{\alpha}}_0\approx \int\vec\nabla_\perp(\Phi+\Psi) du\,.
\ee
In the case of GR sourced by non-relativistic matter $\Phi=\Psi$ and one recovers the standard result $\vec{\hat{\alpha}}_\text{GR}\approx 2\int\vec\nabla_\perp\Phi du$.
In theories without GW birefringence all eigenstates are deflected by $\alpha_0$.

Birefringence will cause the deflection angle between two eigenstates $I,J$ to differ by
\begin{equation} \label{eq:deflection_angle_diff}
 \Delta \hat{\alpha}_{IJ} 
\approx -\frac{1}{2}\int du \, \nabla_\perp \Delta c_{IJ}^2(\vec x,\hat k)\big|_{r(u),\hat u}\,,
\end{equation}
and vanishes in the limit of equal speed as expected.
Typical GR deflection angles are small, on the scale of $\theta_E \sim \text{arcsec}\sqrt{M/10^{12}M_\odot}$ for strongly lensed cosmological sources. %missing 1.5 \sqrt{D_ls/(D_l D_s H_0)}
These deflections are hard to resolve even for the most precise optical telescopes.
GW detectors have rather low angular resolution that is many orders of magnitude lower than what it would be required to detect a different incoming direction for different polarizations (although there are ambitious projects for high resolution GW astronomy in the next decades \cite{Baker:2019ync}). On the other hand, GW detectors have excellent time resolution, making time delays between gravitational polarizations a much more robust observable. 

\subsubsection{Time delays}

There are three independent time delays that a given lens can imprint on the observables, $\Delta t_{01}, \Delta t_{12}, \Delta t_{23}$. Each time delay will be the sum of a Shapiro term (difference in speeds locally) and a geometric contribution (difference in travel distance): 
\begin{equation}\label{eq:time_delay_general}
 \Delta t_{IJ} \equiv \int d u\left(\frac{1}{c_I}-\frac{1}{c_J}\right) + \Delta t_{IJ}^{\rm geo}\,,
\end{equation}
where we used the Born approximation discussed above (recall that the propagation speed will in general depend on the position as well as the propagation direction of the signal). Let us now discuss how the deflection angle (\ref{eq:deflection_angle_diff}) leads to the geometric time delay.

\begin{figure}[t!]
\centering 
\includegraphics[width = 0.95\columnwidth,valign=t]{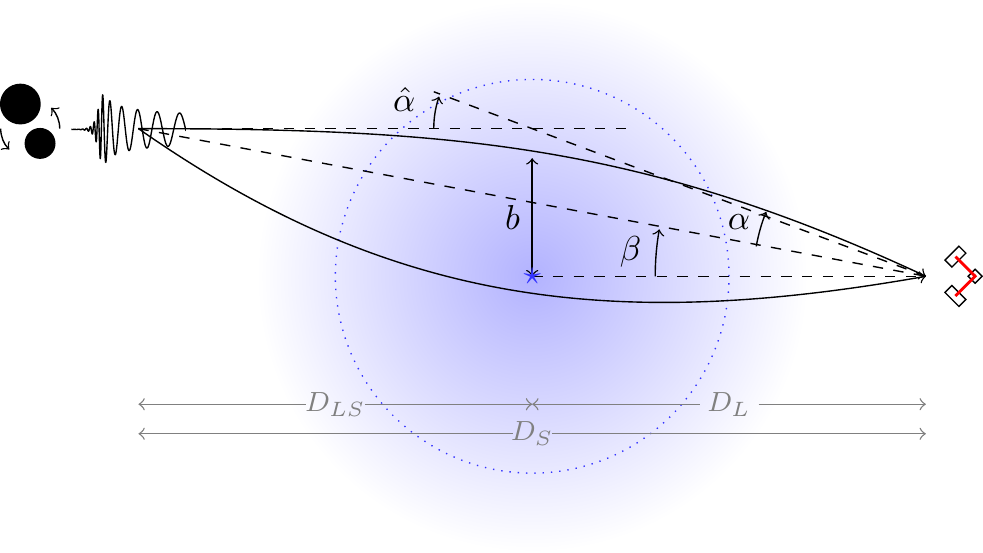}
 \caption{Diagram of the source-lens geometry under consideration. The trajectory of the GW (solid black line) is curved due to the lens with a deflection angle $\hat{\alpha}$. The true angular position of the source is $\beta$, while the observer sees the lensed image at $\alpha+\beta$. $D_{L}, D_{S}, D_{LS}$ are, respectively, the angular diameter distances to the lens, source, and between the lens and the source $D_S$. $b$ is the closest distance of the GW to the lens. }
 \label{fig:diagram_angles}
\end{figure}

%lens equation and geometric time delay
Assuming a single lens and spherical symmetry, each propagation eigenstate obeys its own lens equation
\begin{equation}
 \beta = \theta_I - \alpha_I\,,
\end{equation}
where $\beta$ is the angular position of the source (equal for all polarizations) and $ \theta_I$ are the apparent position of the source for each polarization $I$ (source and lens plane, respectively) cf. Fig. \ref{fig:diagram_angles}.
We have defined also $\alpha_I={\hat{\alpha}}_ID_{LS}/D_S$.
Here $D_{L}, D_{S}, D_{LS}$ are, respectively, the angular diameter distances to the lens, source, and between the lens and the source. In the case of multiple lenses one should substitute the source with the previous lens.
The geometric time delay due to the different angles (assuming $c_I \approx c$ over the trajectory) between two propagation eigenstates can be computed following the standard approach and a bit of trigonometry (see e.g. Ref. \cite[section 4.3]{Schneider_book}). We obtain
\begin{equation} \label{eq:geom_delay}
\Delta t_{IJ}^{\rm geo}=\frac{(1+z_L)}{2c}\frac{D_LD_{LS}}{D_S}\lp\vert\vec{\hat{\alpha}}_{I}\vert^2- \vert\vec{\hat{\alpha}}_{J}\vert^2\rp\,,
\end{equation}
where $z_L$ is the redshift of the lens. 
The order of magnitude of the delay will be determined by the dilated Schwarzschild diameter crossing time
\be
t_M=4GM_L(1+z_L)/c^3\,.
\ee
As a rule of thumb, one can use that $t_{M}\simeq10 (M_{Lz}/1\Msun)\,\mu\text{s}$, i.e. the delay is  $\sim$months, days and minutes for lenses with $M_{Lz}=10^{12}\Msun\,,\  10^{10}\Msun\,,\  10^{7}\Msun$, respectively. 
In these units the geometrical time delay can be written as
\begin{equation}
\Delta t_{IJ}^{\rm geo}=\frac{t_M}{2}\lp\vert\vec{\tilde{\alpha}}_{I}\vert^2- \vert\vec{\tilde{\alpha}}_{J}\vert^2\rp\,,
\end{equation}
where the angles $\vec{\tilde{\alpha}}_I=\vec\alpha_I/\theta_E$ are now normalized in units of the Einstein ring of a point lens
\begin{equation} \label{eq:einstein_angle}
\theta_E = \sqrt{ \frac{4 G M}{c^2 } \frac{D_{LS}}{D_L D_S}}.
\end{equation}

Assuming that the difference between the deflection angles of the different eigenstates $\Delta\alpha_{IJ}$ is small compared to the light deflection angle $\alpha_0$, Eq. (\ref{eq:deflec_angle_light}), $\Delta\alpha_{IJ}\ll\alpha_0$, we find
\be \label{eq:geom_delay_small_deflection}
\Delta t_{IJ}^\text{geo}\approx 2 \Delta t^\text{geo}_0 \frac{\Delta\hat\alpha_{IJ}}{\alpha_0}\,,
\ee
where $\Delta t_0^\text{geo}$ is the geometrical time delay induced by the gravitational potential on a wave propagating at the speed of light. This quantity depends on the distance to the source, to the lens and the mass of the lens. For a point lens it is given by
\be
\Delta t^\text{geo}_0=t_M\lp\frac{D_LD_{LS}}{b\cdot D_S}\rp\lp\frac{\rS}{b}\rp\,.%|\vec b|
\ee
From this expression it is explicit that the time delay is subject to the geometry of the lens-source. The time delay will be maximal when the lens is at intermediate distances between the source and observer.

The multi-messenger and polarization time delays, Eqs. (\ref{eq:time_delay_general}, \ref{eq:geom_delay}) constitute the most promising observables of birefringence. Their exact values depend on the effective background metric for the GWs, through the theory parameters, lens properties and the configuration of additional fields around the lens. We will now turn to the general phenomenological consequences of birefringence and its observability (section \ref{sec:observational_effects}). In section \ref{sec:screening} we will study a specific example of a theory with Vainshtein screening, with a detailed modelling of gravitational lenses.

%-FIGURE DIAGRAM-
%---

\section{Phenomenology and observational prospects}
\label{sec:observational_effects}

Let us now analyze the broad phenomenological consequences of birefringence. We will start in section \ref{sec:observational_regimes} by describing the observational regimes for different values of the time delay, with a discussion of the single and multi-lens case. 
In section \ref{sec:observational_near_lens} we will then discuss special lensing configurations, focusing on a source near a super-massive black hole.
We will address the interplay between birefringence and multiple images due to strong lensing in section \ref{sec:observational_multiple_images}.
Finally, section \ref{sec:observational_probabilities} addresses the probability of detecting GW birefringence, along with current and forecasted constraints.

\subsection{Observational Regimes: Scrambling \& Echoes}\label{sec:observational_regimes}

There are three important scales when discussing tests of GW lensing birefringence for a given event and detector network: the time resolution, the duration of the GW signal and the timescale of the observing run. Three distinct observational regimes can be established, depending of how the time delay between the propagation eigenstates $\Delta t_{IJ}$ relates to these scales.

%sensitivity and type of signals
The sensitivity to $\Delta t_{IJ}$ will be determined by modelling as well as experimental uncertainties. For the delay between EM and gravitational signals, the error $\Delta t_{0I}$ is likely dominated by assumptions about the EM counterpart. For example, when the gamma ray is emitted after a binary neutron star merger. 
In contrast, the GW emission can be modeled accurately, e.g. using a post-Newtonian expansion or numerical relativity.
Thus, delays between gravitational polarizations are mostly limited by the time resolution of the instrument, which will be of order%
\footnote{One can sharpen this estimate easily using a noise curve with $h_i\to h_i e^{f\Delta t_{12}}$ applied to each polarization, see below.}
\begin{equation}
 \sigma_{t_g} \sim f_{\rm peak}^{-1} \,,
\end{equation}
or $\sim {\rm ms}$ for current ground detectors (LIGO/Virgo). 
Finally, we note that the emission of scalar polarizations is suppressed in many theories (due to screening mechanisms), which might make the scalar (or mostly-scalar) polarization very hard to detect, precluding a measurement of $\Delta t_{3I}$.
In the following we will focus mostly on the time-delay between the pure metric and mostly-metric polarizations $\Delta t_{12}$.

The duration of the signal $T_g$ reflects how long a detector is sensitive to a given event. Depending on the mass, compact binary coalescence observed by ground detectors can last from less than a second (black hole binaries) to over a minute (neutron star binaries). Continuous signals such as rotating neutron stars (ground detectors) or mHz compact binaries (LISA) can in principle be detected as well. In those cases $T_g$ is limited by the duration of the observational campaign $T_{\rm obs}$. Here we will assume continuous observation up to $T_{\rm obs}$: a more realistic analysis should account for the detector's duty cycle (the fact that detection are regularly interrupted for several reasons) when $\Delta t_{ij}\in(T_g,T_{\rm obs})$.

\begin{figure*}
\centering 
\includegraphics[width=0.49\textwidth]{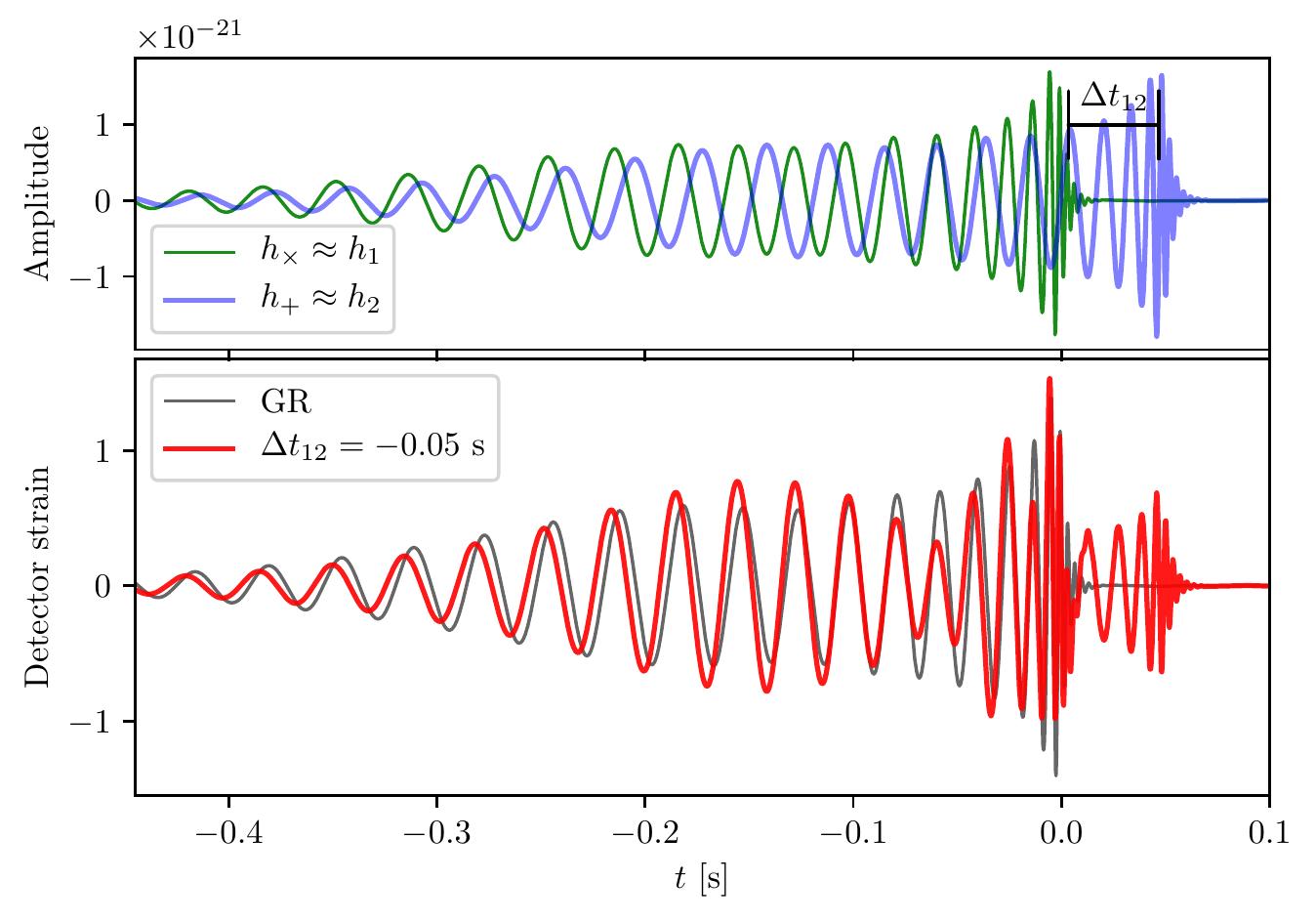}
\includegraphics[width=0.48\textwidth]{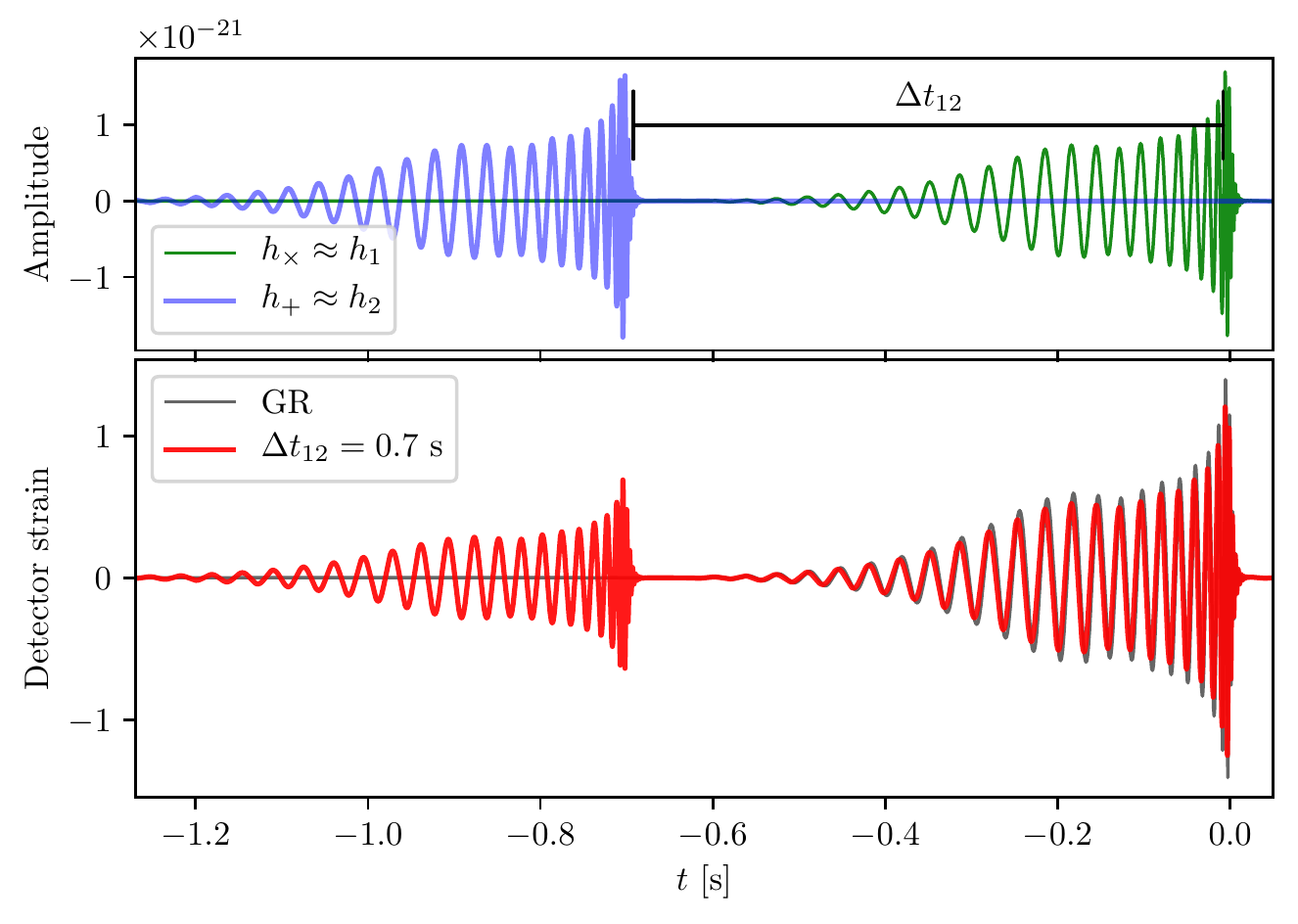}
\caption{Signatures of birefringence in the scrambling ($|\Delta t_{12}|<T_g$, left) and the splitting ($|\Delta t_{12}|>T_g$, right) regimes. The upper panels show the amplitude of the two gravitational polarizations with $\Delta t_{12} = -0.05,0.7$s, respectively. The lower panel show the strain observed on a LIGO-H1 detector with $\mathcal{A}^+=-0.38$, $\mathcal{A}^\times=0.71$ (additional detectors in the network will have different responses). The signal corresponds to two $30 M_\odot$ black holes, head-on ($\cos\iota=1$) at $500$Mpc.}
\label{fig:detector_echoes}
\end{figure*}

The following situations are possible:
\begin{itemize}
 \item \textit{Signal scrambling:} if $\sigma_{t_g}\lesssim |\Delta t_{12}|\lesssim T_g$ the signal is observed as a single event and time delay(s) between different eigenstates distort the waveform.
 \item \textit{Signal splitting/GW echoes:} if $ T_g \lesssim |\Delta t_{12}|\lesssim T_{\rm obs}$ the signal is split and each eigenstate will be observed as a separate event. The orbital parameters of different events will be related (e.g. orbital inclination/orientation), and it may be possible to associate different echoes from the same underlying event.
\item \textit{Single polarized signals:} if $ |\Delta t_{12}|\gtrsim T_{\rm obs}$, only one instance of each event can be observed. This leads to an excess of edge-on signals, relative to the expectation of random orientations.%
\footnote{Due to duty cycle/interruptions of the detector, a fraction of echoes are missed even if $ T_g \lesssim |\Delta t_{12}|\lesssim T_{\rm obs}$, leading to an excess of edge-on events. Given that source binaries are randomly inclined, knowing the antenna patter of the detector and having a large statistical sample may allow to discriminate this effect.}
\end{itemize}
One should note that the first two effects are analogous to strong lensing where multiple images can be produced and might interfere if their time delay is of the order of the signal duration (sometimes called microlensing regime). However, we stress that these are completely different effects in origin and are also governed by different physical quantities (we will comment more on those differences below). The scrambling and echoes are thus independent of strong lensing and would apply to each multiple image if present. 
Moreover, with a large network of detectors one could distinguish the different polarizations further distinguishing the two effects.

In addition, multiple lenses along the line of sight will contribute a separate time delay. Misalignment between lenses causes a difference in propagation eigenstates for each subsequent lens (e.g. different angle $\phi$). In this situation, each lens causes a separate scrambling or splitting of the signal. Let us first discuss the single lens case and then comment on the effect of multiple lenses.

\subsubsection{Single lens}

To better understand the effects of birefringence, let us consider the effect of a single lens on a head-on GW event, i.e. $\hat L \cdot \hat n \equiv \cos\iota = 1$ (this will be generalized later). In this case the $\times,+$ polarization are emitted with equal amplitude, and one can define the basis so that they are proportional to metric components of the $1,2$ propagation eigenstates (i.e. rotating the coordinates so the azimuthal angle is  $\phi=0$). 
In this case the signal after crossing the region where modify gravity effects are relevant is approximately given by 
\begin{equation}
 h_{ij} \approx h_\times(t) e^{\times}_{ij} + h_+(t-\Delta t_{12})e^{+}_{ij} + \cdots\,,
\end{equation}
where the ellipsis represent GW shadows, including those of additional polarizations.
This relationship assumes that the amplitudes are approximately equal in the interaction and propagation basis, and that the mixing with the scalar mode is subdominant. While the exact relationship requires solving the GW propagation at sub-leading order, $\omega^{-1}$, one can assume that the corrections are small, given the large frequency of GWs. This implies that the relative amplitude is unchanged in the propagation so that $h_+\sim h_2$ and $h_\times \sim h_1$. We are also not taking into account standard lensing effects (e.g. magnifications and phase shifts). All these assumptions could be generalized, but for pedagogical purposes we restrict the derivation to the simplest example. One should note too that these assumptions hold for GWs on FRW, where effects on the amplitude ($\alpha_M$) are much harder to detect than effects on the phase ($\alpha_T,m_g^2$).

The strain on a given detector is then
\begin{equation}
 h \approx \mathcal{A}^\times h_\times + \mathcal{A}^+h_+ + \cdots
\end{equation}
where $\mathcal{A}^{I}$ is the detector's response for a given polarization, given the source's position in the sky.
Figure \ref{fig:detector_echoes} shows the effect of the time delay for a binary black hole signal, both on each polarization and as seen in one detector. 
The scrambling regime $|\Delta t_{12}|<T_g$ is characterized by a time modulation of the amplitude, caused by the interference between the signals, as well as two distinct imprints from the merger, separated by $\Delta t_{12}$.
In the splitting regime two copies of the signal are detected with a delay $\Delta t_{12}$ and amplitudes given by the detector's response to each polarization.
Multiple detectors provides further means to characterize the signal via different response functions, time delays, etc... 

For this example we have considered an unlensed, non-spinning, equal-mass binary. However, some of these effects could be degenerate with binary parameters in more general systems. For example spinning, asymmetric binaries are known to introduce modulations in the wave-form. Similarly, strongly lensed multiple GWs produce multiple images that might have short time delays for certain lenses. Nonetheless, with a network of detectors one could use the polarization information to break this degeneracies. For instance, if one expects the amplitude difference between the echoes be produced by the projection on the detector's antenna pattern of each eigenstate, one could use the information on the sky localization to constrain this possibility. If both polarizations can be detected independently, the degeneracy can be completely broken.    

\subsubsection{Multiple lenses}

Multiple lenses can cause further scrambling and splitting of a GW source. Considering spherical lenses and treating their effects as independent, the relationship between the signal at the source and the detector can be approximated as 
\begin{equation} \label{eq:prop_multiple_lenses}
\vec h_{\rm d} \approx \prod_L \left[e^{i\omega t_{1}}\hat{\mathcal{M}}^{-1} 
\exp(\hat{\mathcal{T}})
\hat{\mathcal{M}}\right]_L 
\vec h_{\rm s}\,.
\end{equation}
Here $\vec h_{\rm d,s}$ is the vector of amplitudes in Fourier space in the interaction eigenstates at the detector/source. 
$\hat{\mathcal{M}}$ is the mixing matrix introduced in (\ref{eq:change_basis}), which relates the interaction $h_I,\, I\in (\times,+,\vp)$ and propagation $H_J,\, J\in (1,2,3)$ eigenstates. 
Here we have also introduced the \textit{delay matrix} which encompasses the phase evolution of the propagation eigenstates
\begin{equation}
 \exp(\hat{\mathcal{T}}) =
 \left(
 \begin{array}{ccc}
  1 & 0 & 0 \\
  0 & e^{-i\omega \Delta t_{12}} & 0 
  \\   0 & 0 & e^{-i\omega \Delta t_{13}}
 \end{array}
 \right)\,,
\end{equation}
(note that an overall factor $e^{i\omega t_1}$ has been factored out to express the results in terms of time delays). 
The subscript $L$ denotes that the quantities depend on the lens properties (mass, mass distribution) and its configuration relative to the line of sight (impact parameter $b$, azimuthal angle $\phi$).%

Schematically, equation (\ref{eq:prop_multiple_lenses}) is telling us that if a GW crosses a region near a lens, the GW propagation will be determined by the propagation eigenstates, possibly leading to time delays among them. Therefore, after crossing the first lens the initial GW wavepacket could be split in separate packets for each $H_I$. Then, if another lens is on the line of sight, each GW packet will subdivide again since the eignestates of the second lens will be in general different from the first one. In principle this process can be iterated for as many lenses are in the GW trajectory. A possible observational signature of these multiple splittings would be a significant reduction in the GW amplitude since for random orientations of the lenses the projection into the eigenstates at each lens will reduce the overall amplitude of the detected signals. 
Of course, the key question is how probable is to have this multiple encounters. We touch on the lens probabilities in section \ref{sec:observational_probabilities}.

Before moving on, we remind the reader that equation (\ref{eq:prop_multiple_lenses}) is only valid at leading order and does not take into account the modifications of the amplitudes of the propagation eigenstates. In general both the mixing matrix and eigenfrequencies depend on the spatial coordinates. This means that there would be spatially dependent corrections to the amplitudes of $\vec H$. This next to leading order corrections can be computed solving at higher order in the short-wave expansion. As previously alluded, we leave this analysis for future work.

\subsection{Source near the lens}
\label{sec:observational_near_lens}

A particular interesting source-lens configuration happens when the GW source is very close to the lens. In that case, the GW will inevitably travel in a region where the background fields are relevant and more likely to enhance birefringence effects. Due to this particular geometry, the total time delay will be dominated by the Shapiro part, since the the geometrical time delay scales with the source-lens distance $D_{LS}$.

\begin{figure}[t!]
\centering 
\includegraphics[width = 0.75\columnwidth,valign=t]{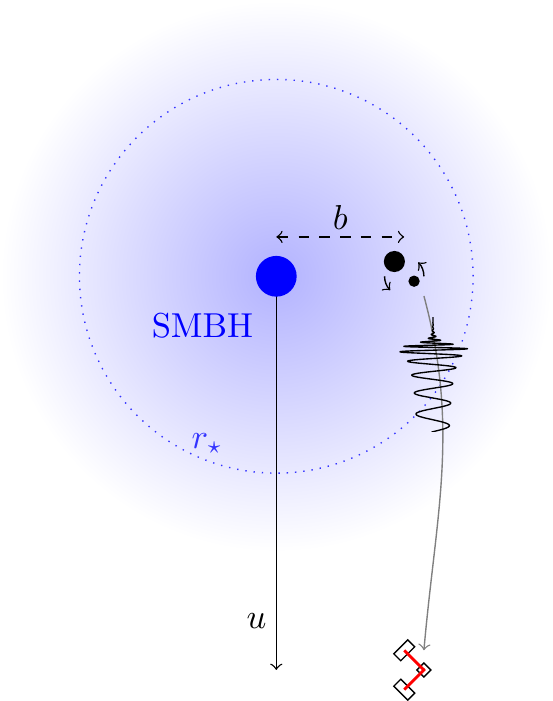}
 \caption{Diagram of a binary black hole coalescence near a super-massive black hole (SMBH). In this situation the Shapiro time delay is the dominant effect. The binary and the lens are separated by an impact parameter $b$ and the GW propagates in the $\hat{u}$ direction.}
 \label{fig:diagram_smbh_bbh}
\end{figure}

A realization of this setup will occur if a binary black hole (BBH) merge near the disk of an active galactic nucleus (AGN) (see e.g. \cite{PhysRevLett.123.181101}). 
There, compact objects are expected to accumulate in specific regions of the accretion disk, the so-called migration traps, at around $20-300\,r_s$ \cite{Bellovary:2015ifg}. 
A schematic representation of this type of systems is given in Fig. \ref{fig:diagram_smbh_bbh}, where the impact parameter of the binary $b$ is smaller than the typical scale $r_\star$ where modified gravity backgrounds become relevant. We remind the reader that this scale does not have to be related with the scale of strong lensing.

%-FIGURE DIAGRAM-
%---

Recently, a possible EM counterpart to the the heaviest BBH detected so far, GW190521 \cite{Abbott:2020tfl}, was announced in \cite{PhysRevLett.124.251102}. The interpretation of this coincident EM-GW event was that the BBH mergered within the disk of an AGN: the large kick after the merger would have produced the flare. The mass of the SMBH was estimated to be $\sim1-10\times10^{8}\Msun$, meaning that the binary might have merger at only $0.0002-0.03$pc of the SMBH. Such short distance to the lens would make this event a great candidate to test modifications of gravity. 
It is to be noted, however, GW190521 is also the furthest event so far with the largest localization volume, making the clear association of a counterpart more difficult. 
In any case, if this BBH formation channel constitutes a significant fraction of the observed events, one could use this population to very efficiently constrain the GW lensing effects beyond GR discussed here. Moreover, LISA could also see the inspiral of $\sim5-10$ events of this class during a 4 year mission (see e.g. Fig. 2 of \cite{Ezquiaga:2020tns}), in which case the dopler modulation and repeated lensing could confirm the origin \cite{DOrazio:2019fbq,Toubiana:2020drf}. A multi-band observation together with an identification of the flare after merger would make this type of BBH system a truly unique laboratory of the theory of gravity.

%turn it around
The opposite scenario of a lens near the observer is also promising to probe birefringence. One possibility is to correlate the maps of nearby gravitational lenses with sky localizations of GW events: for instance, events located behind galactic plane could be used to test theories predicting a sizeable time delay by Milky Way galaxies.
These are examples of unusual lensing setups leading to observable consequences in theories with GW birefringence. In contrast, for standard lensing configurations observable effects are predominantly caused by intervening lenses. In the remainder of the section we will focus on intervening lenses. 

{An examples of a test of gravity using lensing maps of known galaxies was used after GW170817 \cite{Boran:2017rdn}. Alternatively, one could also look at the statistical effect of large-scale inhomogeneities \cite{Garoffolo:2020vtd}.}

%SECTION: STRONG LENSING
\subsection{Strong vs. weak lensing \& multiple images}
\label{sec:observational_multiple_images}

Lensing effects depend on the source-lens geometry and can be classified into strong and weak lensing depending on whether multiple images form or not. These standard multiple images are in addition to possible echoes/splitting caused by birefringence. 
In particular, a point lens is characterized by an Einstein ring radius
\begin{equation}
r_E\approx\theta_E\cdot D_L\,,
\end{equation}
where the Einstein angle $\theta_E$ was given in (\ref{eq:einstein_angle}). 
Whenever the impact parameter of the source is of the order or smaller than the Einstein radius, $b\lesssim r_E$, we are in the regime of \emph{strong lensing} and multiple images of the same GW could be produced by the lens. In the case of having different propagation eigenstates, multiple images of each $H_I$ will be produced. In the opposite limit $b\gtrsim r_E$ we are in the regime of \emph{weak lensing} where only one image can be detected. 
{Weak lensing modify gravity effects could be constrained cross-correlating with galaxy surveys \cite{Mukherjee:2019wcg}.} 
Note that the modify gravity lensing effects are a priori independent of the "standard" lensing regimes. Depending of the theory there could be large modifications even in weak lensing. 
This was schematically depicted in figure \ref{fig:diagram_summary}, where the scale of modify gravity $r_\star$ does not correspond to $r_E$.  

For example, a GW travelling near a point mass lens will form two images with positive (+) and negative (-) parity for each propagation eigenstate. For angular positions $\beta\lesssim1$, we can quantify the dimensionless time delay $T_\pm=t_\pm/t_M$ between the two images analytically \cite{Ezquiaga:2020spg}
\begin{equation}
    T_- - T_+ =\frac{1}{2}y\sqrt{y^2+4}+\ln\lp\frac{x_+}{x_-}\rp \,,
\end{equation}
where we have defined the source angle in units of the Einstein radius $y=\beta/\theta_E$ and the images positions
\begin{equation}
x_{\pm}= \frac{1}{2}\left| y \pm \sqrt{y^2+4}\right|.
\end{equation}
This tells us that for source angles of the order of the Einstein radius, $y\sim1$, the delay between the images will be of the order of the characteristic lensing time scale $t_M$, which for lenses of $\sim10^{10}\Msun$ corresponds to a delay $\sim1$ day. If the impact parameter is much smaller than $r_E$, $y\ll1$, the delay simplifies to $T_--T_+\sim y$, which implies that it will be parametrically smaller than $t_M$. This means that for certain theories and lens-source geometry it is possible that there is a degeneracy between the delay of multiple images and the delay between the echos of different eigenstates.

The interplay between strong lensing and the anomalous speed lensing effects beyond GR will depend on the relation between the Einstein radius and the typical scale where modify gravity effect are relevant. For example, for modified gravity theories with an screening mechanism that we will study in section \ref{sec:screening}, the relevant scale to compare will be the Vainsthein radius $r_V$. 
In the regime of weak lensing, when $b\gg\theta_E$, only one image is detectable with a negligible magnification $\vert \mu_I\vert^{1/2}\simeq1$. This was our assumption for Fig. \ref{fig:detector_echoes}, where we computed the echoes and scrambling assuming only one image.

Strong lensing probabilities have been discussed in the context of advanced LIGO-Virgo extensively \cite{10.1093/mnras/sty2145,PhysRevD.97.023012,10.1093/mnras/sty411} with rates ranging between 1 every 100 or 1000 events depending on the source population and lens assumptions. For LISA, it has been shown that a few strongly lensed GW from SMBH binaries could be observed \cite{PhysRevLett.105.251101}, although the result is highly dependent on the modeling of the population of SMBHs.

%SECTION: PROBABILITIES
\subsection{Lensing probabilities}
\label{sec:observational_probabilities}

Let us now estimate the probabilities of observing GW birefringence by randomly distributed lenses. We will consider two generic dependences with the lens mass, proportional to 1) the Einstein radius and 2) a physical radius with a power-law dependence on the lens mass. We will use these simple models to compare with current GW data (assuming non-detection) and estimate the sensitivity of future observations. 

The probability of observing an event with a given property $X$ (e.g. a time delay) is \cite{Cusin:2019eyv}
\begin{equation}
 P_X = 1-e^{-\tau_X}\,,
\end{equation}
where the optical depth is
\begin{equation}\label{eq:optical_depth_def}
 \tau_X = \frac{1}{\delta\Omega}\int_0^{z_s} dV_c n(z')\sigma_{X}\,.
\end{equation}
Here $\delta\Omega$ is an element of solid angle, $dV = \delta \Omega D_L^2 \frac{dz'}{(1+z')H(z')}$ is the physical volume element given a solid angle $\delta \Omega$, $n(z')$ is the physical density of lenses and $\sigma_X$ is the angular cross section. 
We will assume all lenses have equal mass and dilute as matter, with physical number density
\begin{equation}
 n(z') = \Omega_L \frac{3 H_0^2}{8\pi G M_L}(1+z')^3\,.
\end{equation}
The lens mass distribution and other properties can be included straightforwardly in Eq. (\ref{eq:optical_depth_def}). Note that the prefactor can be written as $\frac{3H_0^2}{8\pi G M_L} = \left(\frac{4\pi}{3}\bar r ^3\right)^{-1}$ in terms of a characteristic scale
\begin{equation}
 \bar r \equiv \left(\frac{2GM_L}{H_0^2}\right)^{1/3} \approx 1.2  {\rm Mpc} \left(\frac{M_L}{10^{12}M_\odot}\right)^{1/3}\,.
\end{equation}
Here $\bar r$ is the mean separation between lenses if the universe's critical density was distributed in objects of mass $M_L$. Incidentally, $\bar r$ coincides with the Vainshtein radius for the theory studied in section \ref{sec:screening} for parameters $\LF = H_0$, $\pFphi=1$.

The angular cross section $\sigma_X$ represents the area around a lens for which a propagation effect $X$ is observable, where we take that
\be \sigma_X = \pi \theta_X^2\,,\ee
i.e. effects are detectable for angular deviations $\leq \sigma_X$ away from a lens.
This form assumes spherical symmetry and that the effects are easier to detect closer to the lens, as it is expected for example from modify gravity screening backgrounds. If the effect $X$ becomes undetectable for a smaller angle $\theta_{0}$ (e.g. transitioning from the scrambling to the echoes regime) then the cross section would be $\sigma_X=\pi(\theta_X^2-\theta_0^2)$ instead.
We will analyze two simple cases for $\theta_X$.

As a first case, let us assume 
    detectability at a fraction of the Einstein radius
\be
\theta_X^E = \alpha_X \theta_E \,,
\ee 
where $\alpha_X$ depends on the theory, but not on redshift or lens mass.
The optical depth then reads
\begin{equation} \label{eq:optical_depth_einstein}
 \tau_X^{E} = \frac{3}{2}\Omega_L \alpha_X^2 \int_0^{z_s}dz'\frac{(1+z')^2}{H(z')/H_0}\frac{H_0D_L D_{LS}}{D_S}\,,
\end{equation}
and is independent of mass, which is a known property of lensing probabilities for point-like lenses and sources. Mass dependence often arises from more detailed modeling, e.g. finite source size \cite{Zumalacarregui:2017qqd} or extended lenses producing multiple detectable images \cite{Cusin:2019eyv}. 

For comparison, let us also consider
detectability below a given impact parameter around the lens
\begin{equation}
\theta_X^{\rm ph} = \frac{R_{X}}{D_L}\left(\frac{M}{10^{12}M_\odot}\right)^n
\end{equation}
where $n$ characterizes the mass dependence and $R_{X}$ is detectable radius for typical galactic lenses, which depends only on the parameters of the theory. %, as well as $z_s, z'$ and other lens properties.
The optical depth is then
%, i.e. $R_X$ does not depend on $z',z_s$. Then the optical depth is
\begin{equation} \label{eq:optical_depth_physical}
 \tau_X^{\rm ph} = \Omega_L h \left(\frac{R_X}{22\rm kpc}\right)^2\left(\frac{M/M_\odot}{10^{12}}\right)^{2n-1}\int_0^{z_s}dz'\frac{(1+z')^2}{H(z')/H_0}\,.
\end{equation}
This dependence is general enough to include scalings like the Einstein radius ($n=1/2$, but without the redshift dependence, cf. Eq. \ref{eq:optical_depth_einstein}), the Schwarzschild radius ($n=1$, as in theories with scalar hair) or the Vainshtein radius ($n=1/3$, as in massive gravity or Horndeski theories cf. section \ref{sec:screening}).
In the rest of this section we will assume that all the mass is effectively in lenses of $10^{12}M_\odot$. However, note that for $n<1/2$ the contribution of lighter lenses can be significantly enhanced, cf. Vainshtein radius scaling in section \ref{sec:screening}, Eq. (\ref{eq:vainshtein_radius_def}).

\begin{figure*}
    \centering
    \includegraphics[width=\textwidth]{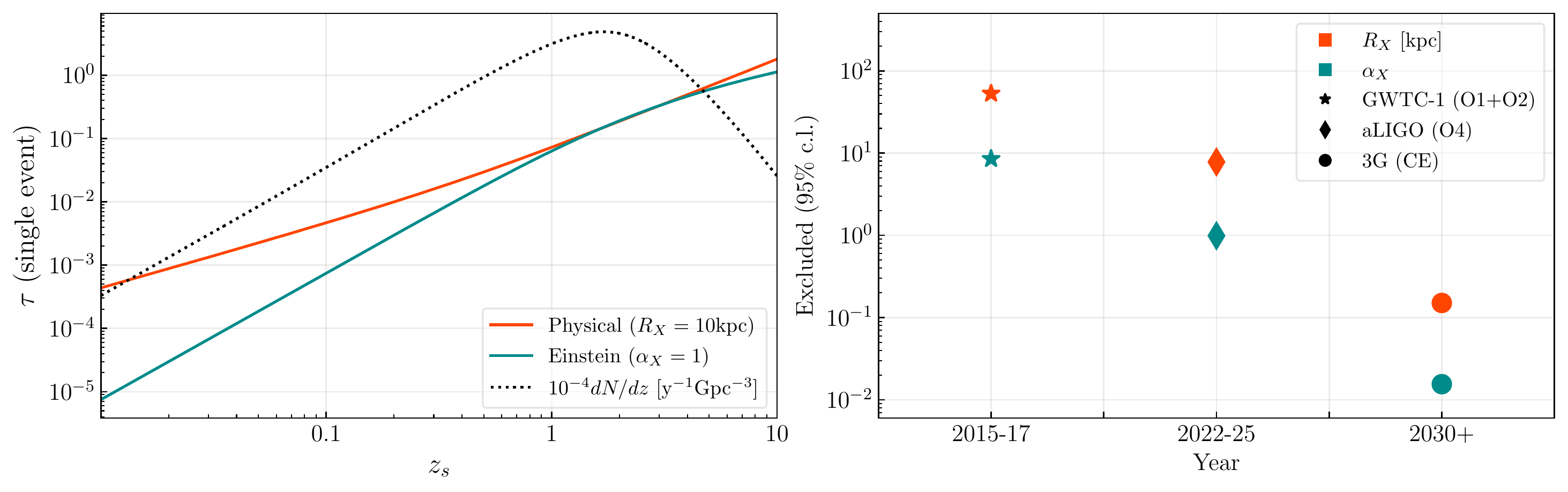}
    \caption{Lensing probabilities assuming detectability at a fraction of the Einstein radius (\ref{eq:optical_depth_physical}) and at fixed physical impact parameter (\ref{eq:optical_depth_einstein}). 
    \textbf{Left panel:} optical depth for a single event as a function of the source redshift, with the (rescaled) event rate shown for comparison (see text). Fiducial model assumes $h=0.7$, $\Omega_L = 0.3$, $T_{\rm obs} = 1$y and $R_0 =30\, {\rm yr}^{-1}{\rm Gpc}^{-3}$. 
    \textbf{Right panel:} constraints on birefringence probabilities, assuming no detection. The stars show the bounds based on GWTC-1, Eq. (\ref{eq:optical_depth_O2_bounds}). Diamonds and circles show the exclusion capacity after 1 year for advanced LIGO design sensitivity (aLIGO) and Cosmic Explorer (CE) respectively, computed using (\ref{eq:optical_depth_total_integral}) for a fiducial binary black-hole population consistent with observations \cite{LIGOScientific:2018jsj} (see details in the text). 
    The horizontal axis indicates the year in which this sensitivities and constraints are expected to be achieved.}
    \label{fig:optical_depth_models}
\end{figure*}

%int_rE=0.141774, int_rV=1.68096
The dependence in the source redshift differs between both cases, as shown in the right panel of figure \ref{fig:optical_depth_models} for a $\Lambda$CDM expansion history. The dimensionless integral in the Einstein scaling case (\ref{eq:optical_depth_einstein}) is an order of magnitude smaller than in the physical scaling (\ref{eq:optical_depth_physical}) for $z\gtrsim 1$. For a source at $z_s=1$, the integral in $\tau_X^E$ is $\sim 0.14$, while the integral in $\tau_X^{\rm ph}$ is $\sim 1.68$. The difference at $z\gtrsim 1$ can be absorbed into the redefinition of the scale $R_X$, but even in that case $\tau_X^{\rm ph}$ is much larger at low redshift due to the projection effect, factor $1/D_L$. The physical scaling optical depth is favoured also at high redshift $z\gtrsim 5$, and it might be probed by LISA massive BH binaries \cite{Klein:2015hvg}.

The cross section models (\ref{eq:optical_depth_einstein},\ref{eq:optical_depth_physical}) can be used to derive constraints from existing GW catalogues.
The detection probability distribution is governed by Poisson statistics
\begin{equation}
 P = \frac{(\tau_{\rm tot})^k}{k!} e^{-\tau_{\rm tot}}\,,
\end{equation}
where $k$ is the number of birefringence detections and the rate (i.e. the mean of the distribution) is given by the total optical depth
\begin{equation}\label{eq:optical_depth_total_sum}
 \tau_{\rm tot} = \sum_i \tau(z_i)\,,
\end{equation}
summed over the $i=1,\cdots,N$ events in a catalogue. The optical depth of each event is evaluated on the mean redshift inferred from the luminosity distance for simplicity (uncertainty of the recovered redshift can be included).
%LIGO optical depths
%RX > 53 kpc excluded (95%)  sum = 2.46
%alphaX > 8.5  excluded (95%) sum = 0.0922
Current constraints can be derived \emph{assuming} non-detection (k=0) in the GWTC-1 of LIGO/Virgo O1+O2 sources  \cite{LIGOScientific:2018mvr}. 
The total optical depth (\ref{eq:optical_depth_total_sum}) evaluated on the models (\ref{eq:optical_depth_einstein},\ref{eq:optical_depth_physical}) can be translated via Poisson statistics into 
\begin{equation}\label{eq:optical_depth_O2_bounds}
\begin{array}{l}
 \alpha_X < 8.5 \\[5pt]
 R_X < 53 \, \text{kpc}
\end{array}
\quad (95\%\, {\rm c.l.})
\end{equation}
 for $\Omega_L = 0.3$, $h=0.7$. 
We note that this limit is subject to a detailed analysis of the waveforms in GWTC-1, to confidently exclude birefringence effects. 
As we will see, future observing runs and next generation detectors can increase these bounds significantly.

In order to estimate the potential of future GW observations we consider the predicted total optical depth
\begin{equation}\label{eq:optical_depth_total_integral}
 \tau_{\rm tot} = \int dz d\vec \lambda \, \tau(z)\frac{dN}{dz}\,,
\end{equation}
where the differential event rate is given by
\begin{equation}
 \frac{dN}{dz} = R(z,\vec \lambda)\frac{T_{\rm obs}}{1+z} \frac{dV_c}{dz} \mathcal{P}_{\rm det}(z,\vec \lambda)\,.
\end{equation}
Here $\vec \lambda$ collectively determines all additional source properties (besides redshift), $\mathcal{P}_{\rm det}$ is the detection probability, $R(z,\vec \lambda)$ is the event rate (per comoving volume) and $ \frac{dV_c}{dz}=\frac{4\pi D_A(z)^2}{(1+z)H(z)}(1+z)^3$ is the comoving volume factor (physical volume times density factor). 
Equation (\ref{eq:optical_depth_total_sum}) is recovered setting $\frac{dN}{dz} = \sum_i \delta(z-z_i)$. 

The predicted total optical depth (\ref{eq:optical_depth_total_integral}) can be used to estimate how future surveys can improve existing bounds (\ref{eq:optical_depth_O2_bounds}). 
We take as a reference model of sources a population of BBHs consistent with GWTC-1 \cite{LIGOScientific:2018jsj}. Specifically we take a power-law distribution of primary masses $p(m_1)\sim m_1^{-1.6}$ between 5 and 45$\Msun$ with a redshift evolution of the merger rate following the star formation rate \cite[Eq. 15]{Madau:2014bja} normalized to $R_0=30\,\text{yr}^{-1}\text{Gpc}^{-3}$. We set the detection threshold at a signal-to-noise ratio of 8 for a single detector. These predictions applied to LIGO O2 sensitivity are in good agreement with the results from GWTC-1, Eq. (\ref{eq:optical_depth_O2_bounds}).

Figure \ref{fig:optical_depth_models} shows the expected bounds on $\alpha_X, R_X$ after a year of observation with advanced LIGO design sensitivity (aLIGO) and Cosmic Explorer (CE) third-generation technology, together with the current bounds (\ref{eq:optical_depth_O2_bounds}). The horizontal axis indicates the expected year when these projections could be achieved. In particular, aLIGO design sensitivity is expected to be achieved during the next observing run O4.
Current constraints can be expected to improve an order of magnitude by O4, and two orders of magnitude after one year of Cosmic Explorer and other 3rd generation ground-based detectors. Note that bounds on the total cross section are quadratic in $\alpha_X, R_X$, so the actual sensitivity increases by $\sim 2,4$ orders of magnitude, respectively. 

The framework introduced in this section applies exclusively to a homogeneous and random distribution of lenses. It is important to note that in certain situations the location of the lens relative to the source might not be random and thus these results may vastly underestimate the probabilities. Examples include when the lens is near the observer (GW events located behind the Galactic Center) or when sources forms very close to the lens (stellar mass BH binaries in the vicinity of a massive black hole) as discussed in section \ref{sec:observational_near_lens}.

%--------
%SECTION: PROPAGATION EIGENSTATES IN HORNDESKI THEORIES
%--------
\section{Propagation eigenstates in Horndeski theories}
\label{sec:horndeski}

As a particular set up, we will concentrate in gravity theories adding just one extra propagating degree of freedom w.r.t. GR. We will restrict to those scalar-tensor theories with covariant second order EoM.
Viable extensions are known \cite{Zumalacarregui:2013pma,Gleyzes:2014dya,BenAchour:2016fzp} but more complex to analyze because of higher derivatives, and some classes induce a rapid decay of GWs into fluctuations of the scalar field \cite{Creminelli:2018xsv,Creminelli:2019nok}.
This naturally leads us to Horndeski's gravity \cite{Horndeski:1974wa}, whose action reads \cite{Kobayashi:2011nu}
\begin{equation}\label{eq:Li}
S[g_{\mu\nu},\phi]=\int\mathrm{d}^{4}x\,\sqrt{-g}\left[\sum_{i=2}^{5}{\cal L}_{i}
\,+\mathcal{L}_{\text{m}}
\right]\,,
\end{equation}
with
\begin{eqnarray}
{\cal L}_{2} & = & G_{2}(\phi,X) \,,
\qquad\qquad {\cal L}_{3} \; = \; -G_{3}(\phi,X)\Box\phi\,,\label{eq:L3}\\
{\cal L}_{4} & = & G_{4}(\phi,X)R
+G_{4,X}(\phi,X)\left[\left(\Box\phi\right)^{2}-\phi_{\mu\nu}\phi^{\mu\nu}\right]\,,\label{eq:L4}\\
{\cal L}_{5} & = & G_{5}G_{\mu\nu}\phi^{\mu\nu}
-\frac{1}{6}G_{5,X}(\phi,X)\Big[\left(\Box\phi\right)^{3}
\nonumber \\ && \qquad\qquad
-3\phi_{\mu\nu}\phi^{\mu\nu}\Box\phi
+2{\phi_{\mu}}^{\nu}{\phi_{\nu}}^{\alpha}{\phi_{\alpha}}^{\mu}\Big]\,.\label{eq:L5}
\end{eqnarray}
This theory has four free function $G_i$ of the filed $\phi$ and its first derivatives $-2X=\phi_\mu\phi^\mu$. Here and for the rest of the paper we adopt the following notation for the covariant derivatives of the scalar field: $\phi_\mu\equiv\nabla_\mu\phi$ and $\phi_{\mu\nu}\equiv\nabla_\mu\nabla_\nu\phi$.

We will divide the analysis of this large class of theories in two. First, we will consider the subclass of theories in which the causal structure of the propagating tensor modes is determined by the background metric. Thus, in these \emph{luminal} theories the phase evolution of GWs is equal to that of light (section \ref{sec:horndeski_luminal}). Then, we will consider \emph{non-luminal} theories in which the tensor modes have a different causal structure (section \ref{sec:horndeski_non-luminal}).

The causal structure of GWs in Horndeski gravity over general space-times in the absence of scalar waves has been studied in \cite{Bettoni:2016mij}. For the subset of luminal theories, the propagation without scalar waves was investigated in \cite{Dalang:2019rke}, while a geometric optics framework including $\vp$ was developed in \cite{Garoffolo:2019mna}. The study of GWs, considered as a background space-time, revealed that scalar perturbations can become unstable in Horndeski theories \cite{Creminelli:2019kjy}, a difficulty that affects even luminal theories such as Kinetic Gravity Braiding, Eq. (\ref{eq:kgb}). In addition, there has been large efforts to study the GW propagation over cosmological and BH spacetimes. We refer the interested reader to the recent review \cite{Ezquiaga:2018btd}.  

Since the GW and scalar wave evolution will in general depend on the propagation direction for an anisotropic background, it is useful to decompose the spatial components of the background tensors in terms of the directions parallel and perpendicular to the propagation trajectory of the GW, defined by the wave vector $k_i$. Specifically, we decompose the spatial gradient of the scalar background as
\be
\phi_i=\phi^{\para}_i+\phi^{\perp}_i\,,
\ee
so that in the transverse gauge
\be
\phi^{i}\nabla_i=\phi_\para^i\nabla_i \quad \text{and}\quad \phi^{i}h_{ij}=\phi_\perp^ih_{ij}\,.
\ee
These identities will be handy in what comes next. 

%---------
%SUBSECTION: LUMINAL THEORIES
%---------
\subsection{Luminal theories}
\label{sec:horndeski_luminal}

Since the evolution equations are coupled in general (even at leading order in derivatives), the first step is to diagonalize them. Depending of the complexity of the theory, the diagonalization can be done covariantly. Indeed, we will see in this section that this is the case for those Horndeski theories with a luminal GW propagation speed. 

Before that, it is useful to recall the case of GR, where one also needs to diagonalize the propagation in order to obtain a wave equation for each polarization. Although in GR there is no additional scalar field, we can effectively treat the trace as an additional degree of freedom. Starting from Einstein's equations, one can see that the tensor EoM of the linear perturbations,
\begin{equation}
\begin{split}
\delta G_{\mu\nu}=&\delta R_{\mu\nu}-\frac{1}{2}h_{\mu\nu}R-\frac{1}{2}g_{\mu\nu}\delta R \\
=&-\frac{1}{2}\Box h_{\mu\nu}+\nabla_{(\mu}\nabla^{\alpha}h_{\nu)\alpha}-\frac{1}{2}\gmn\nabla^\alpha\nabla^\beta h_{\alpha\beta} \\
&-\frac{1}{2}\nabla_{\mu}\nabla_{\nu}h+\frac{1}{2}\gmn\Box h + \Od\lp\nabla h\rp=0\,,
\end{split}
\end{equation}
include a mixing with the trace at leading order in derivative, where $\Od\lp\nabla h\rp$ captures terms linear or lower order in derivatives. The way to diagonalize these equations is to redefine the tensor perturbation to
\be \label{eq:trace-reversed}
\hbmn\equiv\hmn-\frac{1}{2}\gmn h\,
\ee
(well-known as trace-reversed metric perturbations \cite{misner1973gravitation}). In this way, after fixing the transverse gauge on the new perturbations $\nabla^\nu\hbmn=0$, one recovers the standard wave equation
\be
\delta G_{\mu\nu} = -\frac{1}{2}\Box \hbmn + R_{\mu\alpha\nu\beta}\hb^{\alpha\beta}=0\,.
\ee
Note that, at face value, this equation is telling us that the propagation eigenstates of GR are a combination of the tensor perturbations and its trace. In vacuum we can always fix the trace to zero (so that $\hmn=\hbmn$), but in the presence of matter its value has to be computed. 

The fact that in GR only the TT perturbations are non-zero in vacuum can also be easily derived solving the constraint equations. In particular, the 00 Einstein equation tell us that $\Psi=0$, the $0j$ that $w_i=0$ and the spatial trace that $\Phi=0$. We are left then with the $ij$ equations which lead to only two independent equations for $h_+$ and $h_\times$.
 
 Horndeski theories with a luminal GW speed will share with GR the structure of the second order differential operator acting on the tensor perturbations. Such operator corresponds to
\be \label{eq:diff_op}
\bD_{\alpha\beta}^{\  \   \mu\nu}\equiv-\frac{1}{2}\Box\delta_\alpha^\mu\delta_\beta^\nu+\nabla_{(\alpha}\nabla^\mu\delta_{\beta)}^\nu-\frac{1}{2}g_{\alpha\beta}\nabla^\mu\nabla^\nu\,.
\ee
The fact that this operator contains the wave operator plus longitudinal terms makes the GW-cone and light-cone equal, and thus $c_g=c$ in the absence of $\vp$ \cite{Bettoni:2016mij}.

In the following we will generalize this procedure to gravitational theories with luminal GW propagation: generalized Brans-Dicke, kinetic gravity braiding {\cite{Deffayet:2010qz}} and the union of both.

%BRANS-DICKE
\subsubsection{Generalized Brans-Dicke}

A pedagogical exercise is to consider a generalized Brans-Dicke type scalar-tensor theory described by an action
\be \label{eq:brans_dicke}
S=\int d^4x\sqrt{-g}\, \lp G_4(\phi)R+G_2(X)\rp\,,
\ee
which introduces a direct coupling between the scalar field and the second derivatives of the metric through $G_4(\phi)$. 
At leading order in derivatives, the metric EoM for the linear perturbations are given by 
\begin{equation}
\begin{split}
\bD_{\mu\nu}^{\  \  \alpha\beta} \hb_{\alpha\beta}+ G_{4\phi}\lp g_{\mu\nu}\Box\vp-\nabla_\mu\nabla_\nu\vp\rp+ \Od\lp\nabla \hb,\nabla\vp\rp=0\,,
\end{split}
\end{equation}
where for convenience we have already introduced the trace-reversed metric and the differential operator (\ref{eq:diff_op}), and we encapsulate all lower/non-derivative terms which are not relevant for this calculation in $\Od\lp\nabla \hb,\nabla\vp\rp$. 
Thus, there is a mixing of the perturbations, which also occurs in the scalar EoM (see appendix \ref{app:equations} for more details). We can decouple both equations by introducing a new tensor perturbation
\be \label{eq:redefinition_bd}
\hTmn\equiv\hbmn-\frac{G_{4,\phi}}{G_4}\gmn\vp\,
\ee
combining both the trace-reversed and scalar perturbations, which is a well known result in the literature (see e.g. \cite{Fujii:2003pa}). 
After applying the transverse gauge condition on the new field $\nabla^\mu\hTmn=0$, the EoM simplify to 
\begin{align}
-&\frac{1}{2} G_4 \Box \hTmn + \Od\lp\nabla \hT,\nabla\vp\rp=0\,, \label{eq:tensorBD}\\
&\G_s^{\alpha\beta}\nabla_\alpha\nabla_\beta\vp + \Od\lp\nabla \hT,\nabla\vp\rp=0 \label{eq:scalarBD}\,,
\end{align}
where $\G_s^{\alpha\beta}$ is the effective metric for the scalar perturbations
\be
\G_s^{\alpha\beta}=\lp6\frac{G^2_{4,\phi}}{G_4}+2G_{2X}\rp g^{\alpha\beta}-2G_{2XX}\phi^\alpha\phi^\beta\,.
\ee
Therefore, the propagating eigenstates are a combination of the original metric and the scalar perturbations. At this order in derivatives and in the absence of sources, the scalar waves will only be present if they are initially emitted. Moreover, because $\vp$ multiplies $\gmn$, the scalar perturbation will generically contribute to the trace of the tensor perturbations. We can see this explicitly when solving the constraint equations for the non-radiative DoF, obtaining
\begin{align}
&\Psi=-\Phi=\frac{G_{4\phi}}{2G_4}\vp\,,
&w_i=0\,.
\end{align} 
Thus, in Brans-Dicke type theories, the scalar perturbation excites the scalar polarizations of the metric leaving an additional pattern in the GW detector \cite{Will:2014kxa,Maggiore:1999wm}.\footnote{It is to be noted an analogous sourcing of the gravitational (non-radiative) potentials occurs over cosmological backgrounds, see e.g. Ref. \cite[Eq. 3.17-3.21]{Bellini:2014fua} in the limit $k\gg H$.} 
Noticeably, in this theory there is no mixing of the radiative tensorial DoF $h_{+,\times}$ with the scalar $\vp$ ($\G_{+,\times s}=0$), so $h_{+,\times}$ become directly the propagation eigenstates traveling at the speed of light. 

%KGB
\subsubsection{Kinetic Gravity Braiding}

Similarly, we can also diagonalize the propagation equations of Kinetic Gravity Braiding (KGB),
\be \label{eq:kgb}
S=\int d^4x\sqrt{-g}\, \lp G_4\,R-G_3(X)\Box\phi\rp\,
\ee 
a cubic Horndeski theory with a direct coupling between the derivatives of the metric and the scalar field through $G_3(X)$. Note that for simplicity we have fixed $G_4=\mathrm{const}$, although one could easily add a scalar field dependence like in the previous section. Because of this cubic coupling, the metric EoM display a mixing of the scalar and tensor perturbations,
\begin{equation}
\begin{split}
\bD_{\mu\nu}^{\  \  \alpha\beta} \hb_{\alpha\beta}+G_{3,X}\big(\phi_\mu\phi_\nu\Box\vp -2\phi^\alpha\phi_{(\mu}\nabla_{\nu)}\nabla_\alpha\vp \\
+\gmn\phi^\alpha\phi^\beta\nabla_\alpha\nabla_\beta\vp\big)+ \Od\lp\nabla \hb,\nabla\vp\rp=0\,.
\end{split}
\end{equation}
At this order of derivatives, we can diagonalize the equations by changing variables to\footnote{To the best of our knowledge, this metric perturbation diagionalizing KGB equations is novel in the literature.}
\be
\hTmn\equiv\hbmn-\frac{G_{3,X}}{G_4}\phi_\mu\phi_\nu\vp\,. \label{eq:redefG3}
\ee
As in the case of Brans-Dicke, once we apply the transverse condition to the new tensor perturbation $\hTmn$, the EoM reduce schematically to Eq. (\ref{eq:tensorBD}-\ref{eq:scalarBD}) (see details on the form of the effective metric for the scalar field perturbations in appendix \ref{app:equations}). 
Accordingly, the main difference between KGB and Brans-Dicke is that the propagation eigentensor involves the scalar perturbation via the gradients of its background field. In other words, depending on the background, the scalar mode could contribute to other polarizations different from the trace. 

We can see this excitation of non-TT DoF directly by solving the constraint equations. For example, if the scalar background has only temporal components, $\phi_\mu=\phi_0\delta^0_\mu$, the non-radiative DoF read
\begin{align}
&\Psi=\Phi=\frac{G_{3X}\phi_0^2}{4G_4}\vp\,,
&w_i=0\,.
\end{align} 
and $h_+,\times$ propagate independently of $\vp$.  
On the opposite regime, if $\phi_\mu=(0,\phi_i)$, we obtain that
\begin{align}
&\Psi=\frac{G_{3X}\vert\phi_\pa\vert^2}{4G_4}\vp\,, \\
&w_\pe^i=-i\frac{G_{3X}\phi_\pa}{G_4}\frac{\phi^i_\pe}{k}\nabla_0\vp\,,\\
&\Phi=\frac{G_{3X}(\vert\phi\vert^2\Box+3\vert\phi_\pe\vert^2\nabla^0\nabla_0)}{4G_4k^2}\vp\,.
\end{align} 
Moreover, for the radiative DoF, we find that the mixing with the scalar has the same causal structure that the tensor modes,  
\begin{align}
&\G_{hh}=\Box\,, &\G_{+,\times s}=-\frac{G_{3X}\epsilon^{+,\times}_{\mu\nu}\phi^\mu\phi^\nu}{4G_4}\Box\,.
\end{align}
We are then in the $c_h=c_m$ case discussed in section \ref{sec:ch_eq_cm}, meaning that both $h_{+,\times}$ will be propagating eigenstates moving at the speed of light. On the other hand, the scalar eigenstate will be a combination of the original scalar $\vp$ and the tensor modes $h_{+,\times}$

%LUMINAL HORNDESKI
\subsubsection{Luminal Horndeski gravity}

Altogether, the most general luminal Horndeski theory would be a combination of the previous cases
\be
S=\int d^4x\sqrt{-g}\, \lp G_4(\phi)\,R-G_3(\phi,X)\Box\phi+G_2(\phi,X)\rp\,.
\ee 
The dependence in $\phi$ in $G_{2}$ and $G_3$ does not affect the diagonalization of the leading derivative terms in the EoM. Because we are solving for the linear perturbations, the EoM can be diagonalized by a linear combination of the previous field redefinitions, i.e.
\be
\hTmn\equiv\hbmn-\frac{G_{4,\phi}}{G_4}\gmn\vp-\frac{G_{3,X}}{G_4}\phi_\mu\phi_\nu\vp\,. \label{eq:redefG3_2}
\ee
This field redefinition is reminiscent of a disformal transformation \cite{Bekenstein:1992pj,Zumalacarregui:2012us,Zumalacarregui:2013pma,Ezquiaga:2017ner}, e.g. the linearized version of the manipulations presented in Ref. \cite{Bettoni:2015wta}. 
We note that this result agrees with Eq. (40) of \cite{Dalang:2020eaj}.

%--------
%SUBSECTION: NON-LUMINAL THEORIES
%--------
\subsection{Non-luminal theories}
\label{sec:horndeski_non-luminal}

As we increase the order of derivatives of the couplings between the metric and the scalar, we enter on the realm of non-luminal Horndeski theories: theories in which the second order differential operator acting on $\hbmn$ no longer corresponds to the one of GR, $\bD_{\mu\nu}^{\  \  \alpha\beta}$, Eq. (\ref{eq:diff_op}). This induces a different causal structure in the effective GW metric compared to the one that EM waves are sensitive to, leading to $c_g\neq c$ \cite{Bettoni:2016mij}, even in the absence of scalar perturbations $\vp$. These theories involve higher order Horndeski functions with derivative dependence $G_4(X)$ and $G_5(\phi,X)$.

Moreover, in this class of theories, the same couplings that produce an anomalous propagation speed induce a background dependent polarization mixing. Specifically, this mixing can be seen in the EoM from the contraction of perturbed Riemann tensors with first $\phi^\mu$ or second derivatives $\phi^{\mu\nu}$ of the scalar field. Therefore, depending on the scalar field profile the polarizations of the metric may change as they propagate. In practice, this makes the analysis of the propagating DoF difficult in a covariant approach.

%QUARTIC
\subsubsection{Quartic theories}
\label{sec:quartic}

A good example representing this phenomenology is a shift-symmetric quartic Horndeski theory 
\be \label{eq:quartic_horndeski}
\begin{split}
S=\int d^4x\sqrt{-g}\big( G_4(X)\,R&+G_{4,X}\big((\Box\phi)^2 \\
&-\phi_{\mu\nu}\phi^{\mu\nu}\big)+G_2(X)\big)\,,
\end{split}
\ee
where we have added a generalized kinetic term for the scalar. The leading derivative EoM for the tensor and scalar perturbations are then
\be \label{eq:tensorG4}
\begin{split}
G_4 &\bD_{\mu\nu}^{\  \  \alpha\beta} \hb_{\alpha\beta}+G_{4,X}\delta\mathcal{R}_{\mu\alpha\nu\beta}\phi^\alpha\phi^\beta \\
&+\lp G_{4,X}\,\mathcal{C}_{\mu\nu}^{~~\alpha\beta}+G_{4,XX}\,\mathcal{E}_{\mu\nu}^{~~\alpha\beta}\rp\nabla_\alpha\nabla_\beta\vp \\
&+ \Od\lp\nabla \hb,\nabla\vp\rp=0\,, 
\end{split}
\ee
and
\be \label{eq:scalarG4}
\begin{split}
&\mathcal{G}_s^{\alpha\beta}\nabla_\alpha\nabla_\beta\vp+2G_{4,X}\phi^{\mu\nu}\bD_{\mu\nu}^{\  \  \alpha\beta} \hb_{\alpha\beta} \\
&-2 G_{4,XX}\phi^{\mu\nu}\delta\mathcal{R}_{\mu\alpha\nu\beta}\phi^\alpha\phi^\beta + \Od\lp\nabla \hb,\nabla\vp\rp=0\,,
\end{split}
\ee
where $\delta\mathcal{R}_{\mu\alpha\nu\beta}$ is a second order differential operator constructed by a linear combination of the perturbations of the Riemann tensor, $\mathcal{C}_{\mu\nu}^{~~\alpha\beta}$ and $\mathcal{E}_{\mu\nu}^{~~\alpha\beta}$ are background tensors made of second derivatives of the scalar profile and $\mathcal{G}_s^{\alpha\beta}$ is the effective metric for the scalar perturbations which depends on $K_X$ and $G_{4,X}$ (see full definitions in appendix \ref{app:equations}). It is precisely the presence of $\delta\mathcal{R}_{\mu\alpha\nu\beta}$ which induces the non-luminal propagation. Note also that either $G_{4,X}\neq0$ or $G_{4,XX}\neq0$ triggers the mixing of the perturbations in both equations. 

In the following we will concentrate in the simplest theory producing this effect, a quartic theory linear in $X$.%
\footnote{Theories with $G_4 = f(\phi)X$ are equivalent to quintic theories with $G_5(\phi)$ up to a total derivative \cite{Ezquiaga:2016nqo}.} 
It is clear from the equations (\ref{eq:tensorG4}-\ref{eq:scalarG4}) that the dimensionless coupling controlling the mixing is 
\be
\Upsilon\sim\frac{(\ln G_{4})_{,\tilde{X}}}{\mathcal{G}_s}\,\nabla\nabla(\phi/\Mpl)\,,
\ee
where $\tilde{X}=X/\Mpl^2$ and we have introduced $\mathcal{G}_s$, which quantifies the value of $\vert\mathcal{G}_s^{\alpha\beta}\vert$, to ensure canonical normalization of the scalar field. In other words, if $\mathcal{G}_s$ is large, the scalar perturbations decouple from the GW evolution.

We now identify the propagation eigenstates of the quartic theory using two methods: a) perturbative solution for small mixing and b) diagonalization based on a local 3+1 splitting.

%PERTURBATIVE REGION
\paragraph{Perturbative solutions for $\Upsilon\ll1$}
\label{sec:pert_region}

In order to gain some intuition, we will consider first situations in which the GW-scalar mixing is small, $\Upsilon\ll1$, so we can make a perturbative expansion of the propagation equations. Thus we expand the full solution as 
\begin{align}
&\hbmn=\hbmn^{(0)} + \hbmn^{(1)} + \hbmn^{(2)}+\cdots\,, \\
&\vp=\vp^{(0)} + \vp^{(1)} + \vp^{(2)}+\cdots\,, 
\end{align}
solving order by order iteratively. 

Accordingly, at leading order (LO), we have to solve simply 
\begin{align}
&G_4 \Box\hbmn^{(0)}=0\,, \\
&\mathcal{G}_s^{\alpha\beta}\nabla_\alpha\nabla_\beta\vp^{(0)}=0\,,
\end{align}
where we have already applied the transverse condition $\nabla^\mu\hbmn^{(0)}=0$. Therefore, at LO, the equations decouple and we can fix the TT gauge, $\hb^{(0)}=0$. As a consequence, if there is no initial scalar wave $\vp^{(0)}(t_e)$, it will remain zero along the propagation. One can also see that while $\hbmn^{(0)}$ propagate at the speed of light, $\vp^{(0)}$ can have a non-luminal velocity.

At next-to-leading order (NLO), the mixing terms arise in the equations
\begin{align} \label{eq:tensor1G4}
&G_4 \Box\hbmn^{(1)}+G_{4,X}\phi^\alpha\phi^\beta\nabla_\alpha\nabla_\beta\hbmn^{(0)}-2G_{4,X}\,\mathcal{C}_{\mu\nu}^{~~\alpha\beta}\vp^{(0)}_{\alpha\beta}=0\,, \\ \label{eq:scalar1G4}
&\mathcal{G}_s^{\alpha\beta}\nabla_\alpha\nabla_\beta\vp^{(1)} - G_{4,X}\phi^{\mu\nu}\Box \hbmn^{(0)}=0\,,
\end{align}
where we have set $\nabla^\mu\hbmn^{(1)}=0$. 
Note that, since $\hbmn^{(0)}$ is TT, $G_{4,X}\phi^\alpha\phi^\beta\nabla_\alpha\nabla_\beta\hbmn^{(0)}$ is the only non-zero term from $-2G_{4,X}\delta\mathcal{R}_{\mu\alpha\nu\beta}^{(0)}\phi^\alpha\phi^\beta$, where $\delta\mathcal{R}_{\mu\alpha\nu\beta}^{(0)}$ indicates that the perturbations of the Riemann tensors are w.r.t. the zero-th order tensor perturbation $\hbmn^{(0)}$. Consequently, the NLO equations tell us that $\vp^{(1)}$ is only sourced if $\phi^{\mu\nu}_{TT}\neq0$. 
Moreover, one can also see that, when there is no initial scalar wave, the second term of the tensor equation (\ref{eq:tensor1G4}) acts to modify the GW propagation speed. This can be shown explicitly by solving $\hbmn^{(1)}$ with its Green function and noting how the propagator of the total solution $\hbmn^{_\mathrm{NLO}}=\hbmn^{(0)}+\hbmn^{(1)}$ is modified. In the opposite situation, when $\vp^{(0)}(t_e)\neq0$, the different propagation speed of the scalar wave introducing a dephasing in the mixing.  Note however that even in the absence of an initial scalar wave, $\hbmn^{(1)}$ is not necessarily TT.

At next order (NNLO), the equations contain all their possible terms,
\begin{align} \label{eq:tensorNG4}
&G_4 \Box\hbmn^{(n)}-2G_{4,X}\delta\mathcal{R}_{\mu\alpha\nu\beta}^{(n-1)}\phi^\alpha\phi^\beta-2G_{4,X}\,\mathcal{C}_{\mu\nu}^{~~\alpha\beta}\vp^{(n-1)}_{\alpha\beta}=0\,, \\ \label{eq:scalarNG4}
&\mathcal{G}_s^{\alpha\beta}\nabla_\alpha\nabla_\beta\vp^{(n)} - G_{4,X}\phi^{\mu\nu}\Box \hbmn^{(n-1)}=0\,,
\end{align}
so they are valid for any $n>1$ (again $\nabla^\mu\hbmn^{(n)}=0$). 

%3+1
\paragraph{Local, general solution in the 3+1 splitting}

Although the general solution when the mixing is dominant, $\Upsilon\sim1$, is not analytically tractable, we can obtain general solution in a local region of space-time where linearized gravity applies. This is equivalent to going to Riemann normal coordinates. We have to solve the evolution and constraint equations for the 11 DoF of the problem, $s_{ij}$, $\vp$, $\Psi$, $w_i$ and $\Phi$ (see Eq. \ref{eq:3+1_dof}). As before, we will work in the spatially transverse gauge, $\partial^is_{ij}=\partial^iw_i=0$, which it is always possible to choose. Moreover, for clarity in the equations, we will restrict to a static, spatially dependent scalar field background, $\phi_\mu=(0,\phi_i(\mathbf{x}))$. Additional details on the equations for this derivation are given in appendix \ref{app:local_diagonalization}.

Let us focus for the moment on the case of a quartic Horndeski theory in the \emph{absence} of scalar perturbations. In that case the leading derivative EoM are given by (\ref{eq:tensorG4}). Thus, essentially, we need to compute the different components of $\delta G_{\mu\nu}$ and $\delta\mathcal{R}_{\mu\alpha\nu\beta}\phi^\alpha\phi^\beta$. For reference, one should remember that in GR there is only $\delta G_{\mu\nu}$ present. As in GR, the $00$-equation,
\be \label{eq:G4X00}
2G_4\nabla^2\Psi+G_{4X}\lp\phi_i\phi^i\nabla^2\Psi+\phi^i\phi^j \partial_i\partial_j\Psi\rp = G_{4X}\phi^i\phi^j\nabla^2s_{ij}\,,
\ee
provides a constrain equation for $\Psi$. The difference is that $\Psi$ is sourced by $\phi^i\phi^j s_{ij}$, even in vacuum. 

We can proceed similarly for the other equations. For the the $0j$-equations, we obtain the constraint equation for $w_j$ as
\be \label{eq:G4X0j}
\begin{split}
&\nabla^2w_j - \frac{G_{4X}}{G_4}\lp2\phi^k\phi^l\partial_k\partial_{[j}w_{l]}-\phi_j\phi^k\nabla^2w_k\rp \\
&=4\partial_0\partial_j\Psi+2\frac{G_{4X}}{G_4}\big(\phi_j\phi^k\partial_0\partial_k\Psi+\phi_k\phi^k\partial_0\partial_j\Psi \\
&\qquad\qquad\qquad\qquad-2\phi^k\phi^l\partial_0\partial_{[j}s_{l]k}\big)\,.
\end{split}
\ee
In the GR limit we recover the case that $w_i$ is sourced by $\Psi$ and consequently it vanishes in vacuum. Here, the new features are the couplings to the backgrounds as well as the dependence on $s_{ij}$.

Next we move to the trace of the $ij$-equations which yields an equation for the last non-propagating perturbation $\Phi$, i.e
\be \label{eq:G4Xtrace}
\begin{split}
2&\nabla^2\Phi+\frac{G_{4X}}{G_4}\lp\phi_i\phi^i\nabla^2\Phi+\phi^i\phi^j\partial_i\partial_j\Phi\rp =2\nabla^2\Psi-6\partial_0^2\Psi \\
&-\frac{G_{4X}}{G_4}\lp\phi^i\phi^j(\partial_0\partial_iw_j-\partial_0^2s_{ij})+\phi_i\phi^i\lp4\partial_0^2\Psi-2\nabla^2\Psi\rp\rp\,.
\end{split}
\ee
In the GR limit $\Phi$ is only sourced by $\Psi$. Therefore, for the same reason as before, in vacuum both vanish. However, for quartic Horndeski $\Phi$ is sourced by $\Psi$, $w_j$ and $s_{ij}$.

In conclusion, we have solved $\Psi$, $\Phi$ and $w_j$ in the transverse gauge ($\partial^is_{ij}=0$ and $\partial^jw_j=0$) in terms of $s_{ij}$, which are the two transverse-traceless components. We denominate these non-radiative, non-zero perturbations \emph{GW shadows}. We can obtain the equations for $s_{ij}$ plugging in these solutions for the non-propagating perturbations in the spatial tensor equations, cf. (\ref{eq:ij_1}-\ref{eq:ij_2}).

In order to take into account the scalar perturbation $\vp$ we have to both incorporate the new terms in the tensor equations and include the scalar EoM. Because we are expanding over flat space, the second derivatives of the scalar background are purely spatial $\phi_{0\mu}=0$. We also make the further assumption that $G_{4XX}=G_{4XXX}=0$. Then, the new contribution to the tensor equations is 
\be
\begin{split}
G_{4X}\mcA_{\mu\nu}\vp&=G_{4X}\huge(\lp\Box\phi\Box-\phi^{\alpha\beta}\nabla_\alpha\nabla_\beta\rp g_{\mu\nu} -\Box\phi\nabla_\mu\nabla_\nu \\
&-\phi_{\mu\nu}\Box+2\phi_{(\mu\alpha}\nabla^\alpha\nabla_{\nu)}\huge)\vp\,.
\end{split}
\ee
For the $00$-equation, we have
\be
G_{4X}\mcA_{00}\vp=-G_{4X}\lp\Box\phi\nabla_i\nabla^i-\phi^{ij}\nabla_i\nabla_j\rp\vp\,.
\ee
Then, $\Psi$ can be solved in terms of $s_{ij}$ and $\vp$. 
For the $0j$-equations we add
\be
G_{4X}\mcA_{0j}\vp=G_{4X}\lp-\Box\phi\nabla_0\nabla_j+\phi_{jk}\nabla^k\nabla_0\rp\vp.
\ee
Similarly, $w_j$ can be solved in terms of $s_{ij}$ and $\vp$ once $\Psi$ is substituted. 
For the $ij$-equations
\be
\begin{split}
G_{4X}\mcA_{ij}\vp&=G_{4X}\huge(\lp\Box\phi\Box-\phi^{kl}\nabla_k\nabla_l\rp \delta_{ij} -\Box\phi\nabla_i\nabla_j \\
&-\phi_{ij}\Box+2\phi_{(ik}\nabla^k\nabla_{j)}\huge)\vp\,.
\end{split}
\ee
This allow us to compute the spatial trace
\be
G_{4X}\mcA_{ij}\delta^{ij}\vp=G_{4X}\lp-2\Box\phi\nabla_0\nabla_0+\Box\phi\nabla_i\nabla^i-\phi^{kl}\nabla_k\nabla_l\rp\vp.
\ee
From this last equation we can solve $\Phi$ in terms of $s_{ij}$ and $\vp$. 
Finally, we also have the scalar equation
\be
\mcG_s^{\alpha\beta}\nabla_\alpha\nabla_\beta\vp+2G_{4X}\phi^{ij}\delta G_{ij}=0\,.
\ee
Once we solve the constraints, we end up with two independent equations from the $ij$-equations plus the scalar EoM for three DoF, $h_+$, $h_\times$ and $\vp$. Therefore, we have solved the constraint equations. 

For simplicity we present the equations at linear order in $G_{4X}$, where they follow the structure of section \ref{sec:lensing_eigenstates} with coefficients 
\begin{align} \label{eq:G4hh}
G_4\hat{\mcG}_{hh}&=G_4\Box+G_{4X}\vert\phi_\pa\vert^2\nabla_i\nabla^i \,,\\
2G_4\hat{\mcG}_{+s}&=G_{4X}(\phi_{ij}\epsilon_+^{ij})\Box \,,\\
2G_4\hat{\mcG}_{\times s}&=G_{4X}(\phi_{ij}\epsilon_\times^{ij})\Box \,,\\
4G_4\hat{\mcG}_{ss}&=\mcG_s^{\alpha\beta}\nabla_\alpha\nabla_\beta \,. \label{eq:G4ss}
\end{align}
Non-linear terms modify the mixing coefficients $\hat{\mcG}_{\times s}$ and $\hat{\mcG}_{+ s}$ but preserve $\hat{\mcG}_{hh}$. In this way we can solve the propagation diagonalizing the EoM as described in section \ref{sec:lensing_eigenstates}. In the absence of mixing, the propagation speeds for the tensor modes is
\be \label{eq:ch_quartic}
c_{h}^2=1+\frac{G_{4X}\vert\phi_\pa\vert^2}{G_4}\,,
\ee
which also coincides with the speed of the tensorial propagation eigenstate $c_1=c_h$. On the other hand, the scalar speed without mixing reads 
\be \label{eq:cs_quartic}
c_s^2=1-\frac{G_{2XX}\vert\phi_\pa\vert^2}{G_{2X}}
\ee
in the limit where $G_{4XX}=G_{4XXX}=0$ (a more general expression can be derived from the full equations in \cite{Bettoni:2015wta}). One should note that inhomogeneous GW speed (\ref{eq:ch_quartic}) generalizes the result of  \cite{Brax:2015dma,Bettoni:2016mij} where $\vp$ was set to 0 and a TT-gauge was assumed without solving the constraint equations. This result agrees with the radial and angular speed obtained from the calculation of the small-scale perturbations around a BH in Horndeski gravity \cite{Kobayashi:2012kh} and generalizes that result to arbitrary propagation direction.

Finally, we have to remember that although non-propagating, $\Psi$, $\Phi$ and $w_j$ cannot be set to zero. At leading order in $G_{4X}$ they read (assuming propagation in $z$ direction)
\begin{align}
2G_4\Psi&=G_{4X}\lp\phi^{i}\phi^js_{ij}+(\phi^\pe)_{i}^{~i}\vp\rp \\
kG_4w_x&=2iG_{4X}\nabla_0\lp\phi_x\phi_zh_++\phi_y\phi_zh_\times+\phi_{xz}\vp\rp \\
kG_4w_y&=2iG_{4X}\nabla_0\lp\phi_x\phi_zh_\times-\phi_y\phi_zh_++\phi_{yz}\vp\rp \\
2k^2G_4\Phi&=G_{4X}\Huge(\nabla_0\nabla_0\lp2\phi^i\phi^js_{ij}+\delta^{ij}\lp\phi_{ij}-\phi^\pe_{ij}/3\rp\vp\rp  \nonumber\\
&~~~~~~+k^2\phi^i\phi^js_{ij}\Huge)\,.
\end{align}
This suggests that all the tensor polarizations will be excited and that the fact that there are only 3DoF can be seen from the correlations among the different polarizations. 
GW detectors can in principle detect these GW shadows.

%QUINTIC
\subsubsection{Quintic theories}
\label{sec:quintic_theories}

Quintic Horndeski theories also feature GW-scalar mixing at leading order in derivatives that cannot be diagonalized covariantly. In fact, the interactions have an increased level of complexity. In addition to the operators in (\ref{eq:tensorG4}-\ref{eq:scalarG4}), there will be, for example, contractions of the perturbations of the Riemann tensor with second derivatives of the scalar background $\phi^{\mu\nu}$. Because the scalar second derivative tensor could have different projections into the GW polarizations $e^+_{\mu\nu}$ and $e^\times_{\mu\nu}$, propagation effects are subject to polarization dependence. In particular, even in the absence of scalar waves, it is possible for the GW speed to depend on the polarization in a generic quintic Horndeski model. For instance, operators like
\be
\phi_{i}\phi_j\,\phi^{kl}\Box h_{kl} = 2\,\phi_i\phi_j \, \lp\phi_+\Box h_+ + \phi_\times \Box h_\times\rp
\ee 
would introduce such a birefrengent effect. 

An interesting exception is scalar Gauss-Bonnet gravity (sGB) \cite{Nojiri:2005vv}, where due to the symmetry of the theory the tensor speed does not depend on the polarization \cite{EzquiagaBravo:2019xyh}. 
This theory is the described by the Lagrangian
\be
\Lag=\frac{R}{2}-\frac{1}{2}\nabla_a\phi\nabla^a\phi-V(\phi)+f(\phi)\mcG\mcB\,,
\ee
where $\mcG\mcB=R^2-4R_{ab}R^{ab}+R_{abcd}R^{abcd}$ is the Gauss-Bonnet invariant. 
After a bit of calculus, one can show that in the absence of scalar waves, the leading order equations for sGB are the same that for a quartic theory if one replaces
\begin{align}
    &\phi_\mu\phi_\nu/\Mpl^2\to f_{\mu\nu}\equiv f_{\phi\phi}\phi_\mu\phi_\nu+f_{\phi}\phi_{\mu\nu}\,, \\
    &\Mpl^{2}\frac{G_{4X}}{G_4}\to\tilde{G}\equiv\frac{16}{\Mpl^2-16f_{\alpha\beta}g^{\alpha\beta}}\,.
\end{align}
Then, locally and at leading order, one obtains the propagation velocity
\begin{equation}
    c_h^2=1+16\frac{f_{\pa}}{\Mpl^2}=1+16\frac{(f_\phi \phi_{uu}+f_{\phi\phi}\vert\phi_\pa\vert^2)}{\Mpl^2}\,,
\end{equation}
which is the same for both polarizations. It is to be noted that here $\phi_{uu}$ corresponds to the projection of the second derivatives of the scalar field background in the direction of propagation. Therefore, the novelty in the propagation speed of GWs in sGB compared to a quartic Horndeski theory is precisely this dependence in the second derivatives. 

We can go one step further and compute the mixing of the GWs with scalar waves at leading order in derivative in a vacuum solution ($R^\text{B}=R^\text{B}_{\mu\nu}$=0). The EoM would look like
\begin{align}
    &\delta G_{\mu\nu}+\tilde{G}\delta \R_{\mu\alpha\nu\beta}f^{\alpha\beta}+4f_\phi R^\text{B}_{\mu\alpha\nu\beta}\varphi^{\alpha\beta}=0\,, \\
    &\Box\varphi+2f_\phi R_{\text{B}}^{\mu\alpha\nu\beta}\delta R_{\mu\alpha\nu\beta}=0\,,
\end{align}
where $\delta G$ and $\delta R_{\mu\alpha\nu\beta}$ are the perturbations of the Einstein and Riemann tensor respectively defined in appendix \ref{app:equations}. From these equations we can see that the main difference of the mixing in sGB and quartic Horndeski is that in the former the mixing is through the curvature background while in the latter this happens through the scalar field background. We leave the analysis of the detectability of the mixing of GWs and scalar waves in sGB for future work.

%-------
%SECTION: SCREENING
%-------
\section{Probing GW propagation in screened regions}
\label{sec:screening}

In this section we will present detailed GW lensing predictions for a concrete Horndeski theory featuring Vainshtein screening. We first introduce the theory Lagrangian and parameters, as well as some quantities of interest. 
In section \ref{sec:screening_local} we present the local solutions of the scalar field around spherical lenses, including screening phenomena.
Section \ref{sec:screening_cosmo} briefly describes the cosmological behavior and limits imposed by  compatibility with the GW speed on the cosmological background following GW170817.
In section \ref{sec:screening_time_delays} we present detailed predictions for the multi-messenger and birefringent time delays for point-like lenses.
Section \ref{sec:screening_shadows} explores the emergence of GW shadows for signals propagating in a screened region.
Finally, Section \ref{sec:screening_obs_prospects} discusses the prospects to further probe Horndeski theories using GW lensing and birefringence.

To exemplify this modified GW propagation due to screening, let us come back to a quartic Horndeski theory (see section \ref{sec:quartic}).
We will consider a linear coupling to the curvature of the form
\be \label{eq:example_theory}
\Lag=\Lag_\text{shift-sym} + \pFphi{\phi}{\Mpl}R\,,
\ee
where the shift-symmetric quartic theory $\Lag_\text{shift-sym} $ is given by (\ref{eq:quartic_horndeski}) in which the free functions $G_i$ depend only on the derivatives of the scalar. This linear coupling can be thought as the leading order term of an exponential coupling $e^{\pFphi\phi/\Mpl}$, which in the Einstein frame corresponds to a linear coupling to the trace of the energy-momentum tensor. 

For concreteness, we will consider a polynomial expansion in the Horndeski parameters
\begin{align} 
G_{2}&=\Mpl^2\lp2\pTX\tilde{X}+\pTXX\frac{\tilde{X}^2}{\LT^2}\rp\,,\\
G_4&=\frac{\Mpl^2}{2} \lp1+ 2\pFphi\,\tilde{\phi}+ 2\sum_{n=1}^{N} \pFXn\lp\frac{-\tilde{X}}{\LF^2}\rp^n\rp\,. \label{eq:G4poly}
\end{align}
Note that we are measuring the field in Planck units $\tilde{\phi}=\phi/\Mpl$ so that $\tilde{X}=X/\Mpl^2$.  
Each of these terms have an associated energy scale $\Ln$ which determines the length scale at which non-linearities become relevant. 
We can define the non-linear length scale 
\be\label{eq:rFdef}
(\rF)^3=\frac{\rS}{2\LF^2}\,,
\ee
associated to the quartic theory, and 
\be
(\rT)^2=\frac{\rS}{\LT}\,,
\ee
associated to the scalar kinetic interaction. Here $r_s=2GM$ is the Schwarzschild radius.

\subsection{Local background} \label{sec:screening_local}

Screening mechanisms suppress fifth forces around massive objects, so that GR holds in their vicinity. This is achieved in different ways depending on the underlying theory \cite{Joyce:2014kja}, but typically it is caused by a particular background configuration preventing the propagation of scalar modes (fifth forces). These backgrounds can be induced by the local matter density or curvature profile, depending on the screening mechanism. 
Screened environments are natural set-ups for GW lensing beyond GR, since they introduce non-trivial background profiles around massive objects that could modify the GW propagation. GW lensing effects beyond GR are thus expected to be different for different types of screening mechanisms. 

For the quartic theory under consideration, screening is caused by non-linear derivative self-interactions of the scalar field. 
Screening becomes effective within a scale known as the \textit{Vainshtein radius}:
\be \label{eq:vainshtein_radius_def}
\rV \equiv \pFphi^{1/3}\rF=\left(\frac{\pFphi GM}{\LF^2}\right)^{1/3}\,,
\ee
(assuming $\pFXn=1$ in the last equality). 
Whenever the coupling to matter $\pFphi$ is of order one, the non-linear scale (\ref{eq:rFdef}) corresponds to the Vainshein radius.
The linearized field equation is valid for $r\gg \rV$: in that unscreened region the scalar field mediates a force $\sim \pFphi^2$ times that of gravity.

%-FIGURE UPSILON-
\begin{figure*}[t!]
\centering 
\includegraphics[width = 0.48\linewidth,valign=t]{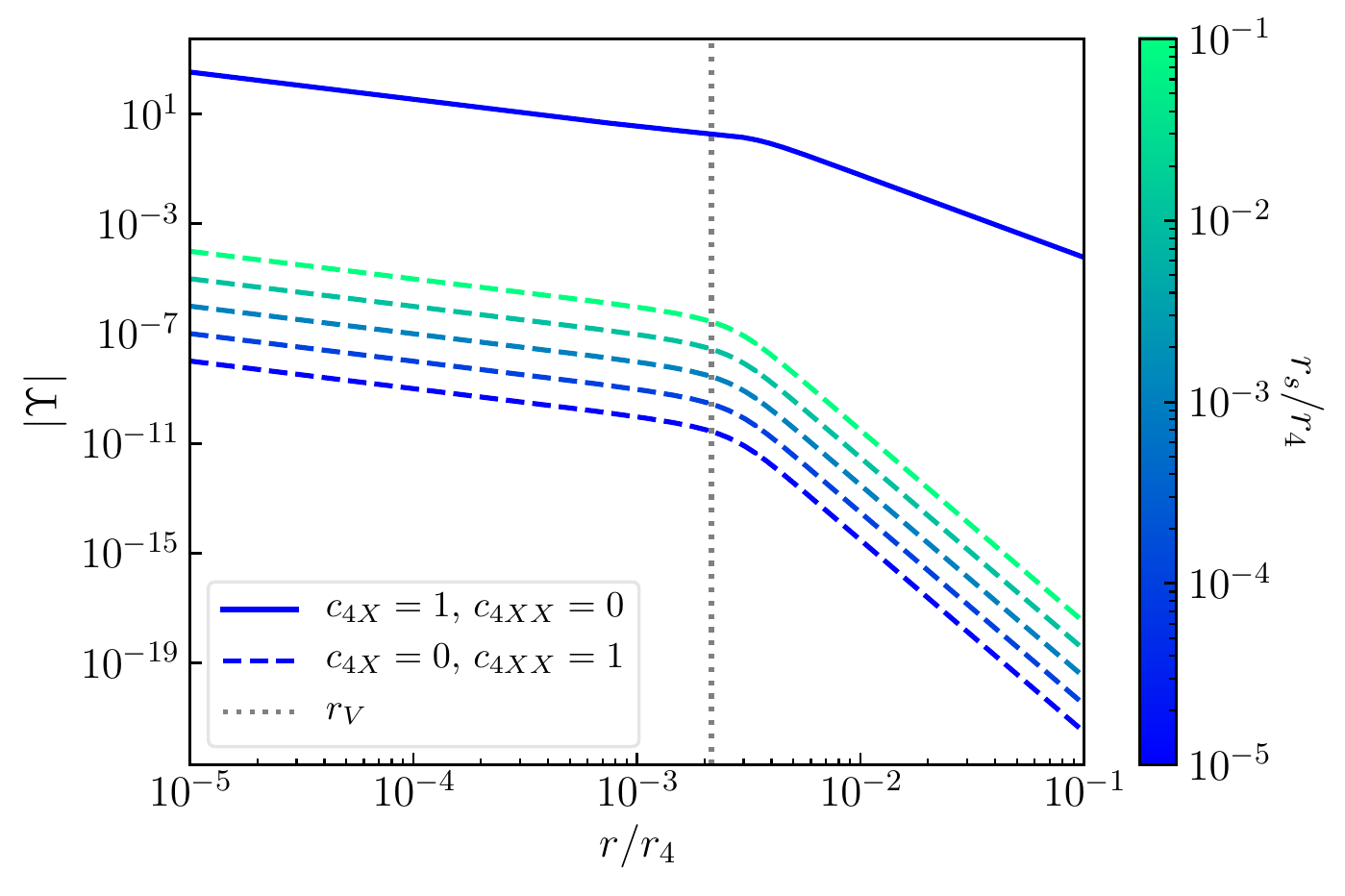}
\includegraphics[width = 0.48\linewidth,valign=t]{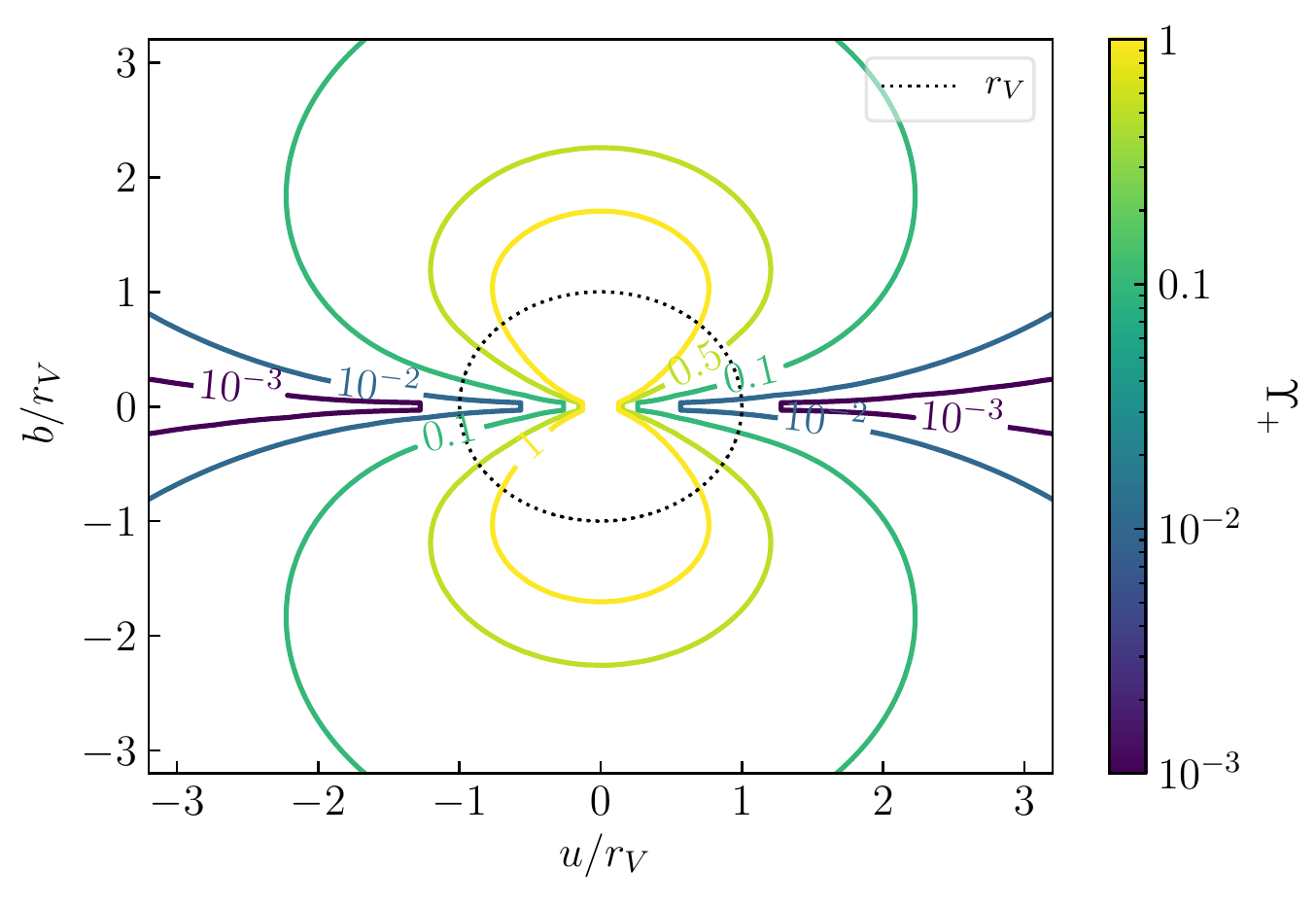}
 \caption{GW mixing amplitude $\vert\Upsilon\vert$ (left) and plus polarization contribution $\Upsilon_+$ (right) for a GW propagating in $\hat u$ direction in a quartic Horndeski theory with a standard scalar kinetic term ($\pTX=1$ and $\pTXX=0$) and $\pFphi=10^{-8}$ (so that a theory with $\LF=H_0$ is consistent with GW170817, cf. section \ref{sec:screening_cosmo}). For the polarization plot we further fix $\rS/\rF=0.1$ and $\pFX=1$. The distances are normalized with respect to the scale of the quartic theory $\rF$ and the Vainshtein radius $\rV$.}
 \label{fig:upsilon}
\end{figure*}
%------------

It will be convenient to measure distances in units of the non-linear scale of the quartic theory: $\tilde{r}=r/\rF$. In this units, following \cite{Narikawa:2013pjr}, we can obtain the screening background from the dimensionless quantity $x(\tilde{r})$, whose algebraic equation for this theory is given by\footnote{To link with the notation of \cite{Narikawa:2013pjr} one can set $\mu=\beta=0$, $\xi=\pFphi$, $\eta=2\tilde{c}_2$, $\alpha=-\pFX$ and $\nu=\pFXX$, as well as $A(r)=\tilde{M}(r)/\tilde{r}^3$.}
\be \label{eq:screening_back}
\begin{split}
 \lp\pTX+3\pFphi^2\rp\, x -6\pFX&\pFphi\,x^2 + \lp\pFXX+2\pFX^2\rp\,x^3  \\
&=- \pFphi\tilde{M}(\tilde{r})/\tilde{r}^{3}\,,
\end{split}
\ee
where $\tilde{M}(r)$ accounts for the mass enclosed in a sphere of radius $r$, i.e. $\tilde{M}(r)\equiv 4\pi M^{-1}\int_0^{r}(-T_t^t)r^2dr$. 
To isolate the dependence on the source mass distribution, we make the definition
\be
\frac{\partial\tilde\phi}{\partial\tilde{r}}=\frac{1}{2}\,\tilde{r}\,\tilde{r}_s\,x(\tilde{r}) \equiv \tilde{r}_s\,\frac{\partial\bar{\phi}}{\partial\tilde{r}}\,.
\ee
This is a convenient rewriting of the local scalar field background because all the dependence in the lens mass is isolated in the prefactor $\tilde{r}_s$, while $\partial\bar\phi/\partial\tilde r$ is a profile depending only on the parameters of the theory. 
For example, outside of the source but well within the screening radius, $r\ll \rV$, the profile becomes constant
\be \label{eq:back_screening}
\left.\frac{\partial\bar{\phi}}{\partial\tilde{r}}\right\vert_{r\ll \rV}\rightarrow-\frac{1}{2}\lp\frac{\pFphi}{\pFXX+2\pFX^2}\rp^{1/3}
\ee
and far from the source we recover the decay with the inverse square distance
\be
\left.\frac{\partial\bar{\phi}}{\partial\tilde{r}}\right\vert_{r\gg \rV}\rightarrow-\frac{1}{2}\frac{\pFphi}{\pTX+3\pFphi^2}\,\frac{1}{\tilde{r}^2}\,.
\ee
As it is evident from the above equation, $\pFphi$ indeed weights the coupling to matter. One should note that by differentiating (\ref{eq:screening_back}) along the radial direction one can also obtain an algebraic equation for $\partial^2\bar\phi/\partial\tilde r^2$ as a function of the theory parameters and $\partial\bar\phi/\partial\tilde r$. This is useful for instance to compute the second derivative background limit within the screening region.

We have seen previously that the coupling of the scalar perturbation to the tensorial radiative modes is supported by the second derivatives of the scalar background (recall equation \ref{eq:scalarG4}). For a radial scalar configuration the second order partial derivatives read 
\begin{equation}
 \phi_{;ij} = \phi''\frac{x_i x_j}{r^2} + \frac{\phi'}{r}\left(\delta_{ij}-\frac{x_i x_j}{r^2}\right)\,,
\end{equation}
using $r=\sqrt{x^2+y^2+z^2}$. 
The projection to spherical coordinates ($x=r \sin(\theta)\cos(\phi)$, $y=r \sin(\theta) \sin(\phi)$, $z=r \cos(\theta)$) yields the following projections (see Fig. \ref{fig:diagram_prop})
\begin{eqnarray}
 \phi_+ &=& \phi_{xx}-\phi_{yy} = \left(\phi'' - \frac{\phi'}{r}\right)\cos(2\phi) \sin(\theta)^2 \,,
 \\
\phi_\times &=& 2\phi_{xy}= \left(\phi'' - \frac{\phi'}{r}\right)\sin(2\phi)\sin(\theta)^2\,.
\end{eqnarray}
Applying this general projection to the quartic theory, we can then read from the entries of the mixing matrix in eqs. (\ref{eq:G4hh}-\ref{eq:G4ss}) to obtain the scalar-tensor mixing coefficients\footnote{Note that these mixing coefficients directly connect with the simplified notation of equation (\ref{eq:mixing_coefficients}) if one defines $M_\phi=\vert\Upsilon\vert\sin^2\theta$.}
\begin{align}
\Upsilon_+&=\vert\Upsilon\vert \cos(2\phi) \sin(\theta)^2 \,,\\
\Upsilon_\times&=\vert\Upsilon\vert \sin(2\phi) \sin(\theta)^2\,,
\end{align}
where the modulus reads
\be \label{eq:ups_modulus}
\begin{split} 
\vert\Upsilon\vert =& \frac{4\sum_{n=1}^{N}n\,\pFXn\,\lp\frac{\rS}{\rF}\rp^{n-1}\lp\frac{\partial\bar{\phi}}{\partial\tilde{r}}\rp^{2(n-1)}\lp\frac{\partial^2\bar{\phi}}{\partial\tilde{r}^2}-\frac{1}{\tilde{r}}\frac{\partial\bar{\phi}}{\partial\tilde{r}}\rp}{1+4\pTX\lp\frac{\partial\bar{\phi}}{\partial\tilde{r}}\rp^2\lp\frac{\rT}{\rF}\rp^4}\,.
\end{split}
\ee
This will be the quantity determining how much mixing between polarizations there is when crossing a screened region, which is controlled mostly by the ratio $r_s/\rF$. Note that whenever the kinetic screening dominates over the Vainshtein mechanism, $\rT\gg\rF$, the tensor-scalar mixing $\Upsilon$ will be suppressed. 

The spatial dependence of the mixing modulus is shown in the left panel of Fig. \ref{fig:upsilon} for a quartic theory with a standard scalar kinetic term ($\pTX=1$ and $\pTXX=0$). One should note that the linear theory ($\pFXn=0$ for $n\neq 1$) represented in solid lines is independent of the non-linear scale $\rF$ represented in the color bar. This can be seen directly fixing $N=1$ in the general formula for $|\Upsilon|$ given in (\ref{eq:ups_modulus}). On the contrary, for a quadratic in $X$ quartic theory (like Covariant Galileons), the mixing is sensitive to the non-linear scale and highly suppressed because if screening is efficient we are always in the regime $\rS\ll \rF$. Taking into account the polarization information, in the right panel of Fig. \ref{fig:upsilon} we present the spatial dependence of the $+$ polarization mixing for a GW propagating in the $\hat z$-direction. The mixing is larger perpendicular to the propagation direction.

Interestingly, the quartic theory linear in $X$ can be mapped (see e.g. \cite{Ezquiaga:2016nqo}) to an Einstein-Hilbert action plus a modified gravity term $\Lag_{MG}\sim G_{\mu\nu}\phi^\mu\phi^\nu$. A theory which has been studied in \cite{Gubitosi:2011sg}. 
While theories with $n>1$ suppress $|\Upsilon|$ they produce larger screening regions once other constraints are imposed, which may overcome the reduced mixing $|\Upsilon|$. However, including $n>1$ requires additional terms in the equations for GW propagation (for details see Eq. \ref{eq:full_scalar_metric}), 
difficulting the analysis.
We will focus on the $n=1$ theory for this first study, leaving the general case for future work.

%COSMOLOGY
\subsection{Cosmological background} \label{sec:screening_cosmo}

Before moving towards the GW lensing observables and their detectability, it is important to consider which region of the parameter space is still viable given present data. In particular, the almost simultaneous arrival of GW and EM radiation from the binary neutron-star merger GW170817 sets the most stringent constraints on the cosmological solutions of the quartic theory under consideration. 

In the theory under consideration (\ref{eq:example_theory}), the cosmological evolution of the scalar field velocity is given approximately by 
\begin{equation}
 \dot\phi + 2 \frac{\pT}{\LT^2}\dot\phi^3 \sim \pFphi H(a)\,.
\end{equation}
This relation is exact for an exponential coupling $G_4 = \exp(\pFphi \phi)$ in a matter-dominated universe (cf. Ref. \cite[Eq. 42]{Zumalacarregui:2020cjh}), but reduces to the theory at hand when $\pFphi \ll 1$ and  the contributions of $G_{4,X}$ are negligible compared to the $G_2$ terms. It also provides a good order-of-magnitude description for the late universe in the presence of a cosmological constant. If the canonical kinetic term dominates, the scalar is cosmologically unscreened and its velocity reads
\begin{equation}
\dot\phi \sim \pFphi H \quad \text{if}\quad (\dot\phi \ll \LT/\sqrt{\pT})\,,
\label{eq:cosmo_solution_unscreened}
\end{equation}
whereas $\dot\phi \sim \left(\pFphi H \LT^2/2\pT\right)^{1/3} $  if the non-canonical term dominates ($ \dot\phi \gg \LT/\sqrt{\pT}$), corresponding to the cosmologically screened regime. 
In what follows we will assume the unscreened solution (\ref{eq:cosmo_solution_unscreened}).

For a quartic theory (\ref{eq:G4poly}) the cosmological change in GW speed at $z\sim 0$ reads
\begin{equation}
 \alpha_T = 4G_{4,X}X = \sum_n 4n \pFXn\left(\frac{-\pFphi^2 H_0^2}{2\LF^2}\right)^n\,
\end{equation}
where the last equality assumes the unscreened cosmological solution (\ref{eq:cosmo_solution_unscreened}).
Assuming that only one among the $\pFXn$ coefficients is non-zero
\begin{equation}
 \alpha_T \sim \left(\frac{\pFphi H_0}{\LF}\right)^{2n}\,,
\end{equation}
where we have set $c_{4Xn}=1$, as it is redundant with $\LF$ in this case. 
Then the GW170817 constraint $|\alpha_T|\lesssim 10^{-15}$ gives a relation between the two theory parameters. 
For a quartic theory linear in $X$ ($n=1$), this bound can be satisfied for sufficiently small matter couplings compared to the scale of the theory, i.e.
\be\label{eq:cosmo_alpha_T_limit}
\pFphi \lesssim 10^{-8}\LF/H_0\,.
\ee
The suppression of modified gravity effects ensures that the approximations used to estimate the cosmological evolution  (\ref{eq:cosmo_solution_unscreened}) remain valid if $\LF\gtrsim H_0$. If $\LF\lesssim H_0$, quartic Horndeski gravity breaks down as an effective field theory at energy scales comparable to GWs with typical LIGO/Virgo frequencies \cite{deRham:2018red}.

As we will see below, GW lensing and birefringence effects can extend constraints based on the GW speed over the cosmological background, Eq. (\ref{eq:cosmo_alpha_T_limit}). The reason behind it is that the scalar field gradient sourced by a massive lens is significantly larger than the cosmological time variation $\phi'/\dot\phi \gg 1$ in a region much larger than the Vainshtein radius. We will use this fact to approximate $\dot\phi\sim 0$ to compute GW propagation, i.e. considering static lenses hereafter.

%TIME DELAYS
\subsection{Time delays} \label{sec:screening_time_delays}

One of the main GW propagation observables is the time delay between different propagation eigenstates and with respect to a possible EM counterpart. For that, we need to compute first the propagation speeds. 
Assuming that the GW propagates in the $\hat{u}$ direction, using Eq. (\ref{eq:ch_quartic}), we obtain the propagation speed of the tensor modes in the absence of scalar waves
\be
c_h^2=1-2\cos^2\theta\frac{\Mpl^2}{G_4}\sum_n n\,\pFXn\lp\frac{\rS}{\rF}\rp^n\lp\frac{\partial\bar\phi}{\partial\tilde{r}}\rp^{2n}\,,
\ee
which tends to the speed of light when $\rS\ll \rF$ and/or $\theta\rightarrow\pi/2$. This velocity also corresponds to the one of the purely tensorial propagation eigenstates, $c_1=c_h$, when there is mixing. 
In the absence of mixing, the scalar speed is given by Eq. (\ref{eq:cs_quartic}) to arrive at
\be
c_s^2=1+2\cos^2\theta\frac{\Mpl^2}{G_2}\pTXX\lp\frac{\rT}{\rF}\rp\lp\frac{\partial\bar\phi}{\partial\tilde{r}}\rp^{2}\,,
\ee
which tends to the speed of light when $\rT\ll\rF$ or $\pTXX\rightarrow0$. 

%-FIGURE SPEEDS-
\begin{figure}[t!]
\centering 
\includegraphics[width = 0.95\columnwidth,valign=t]{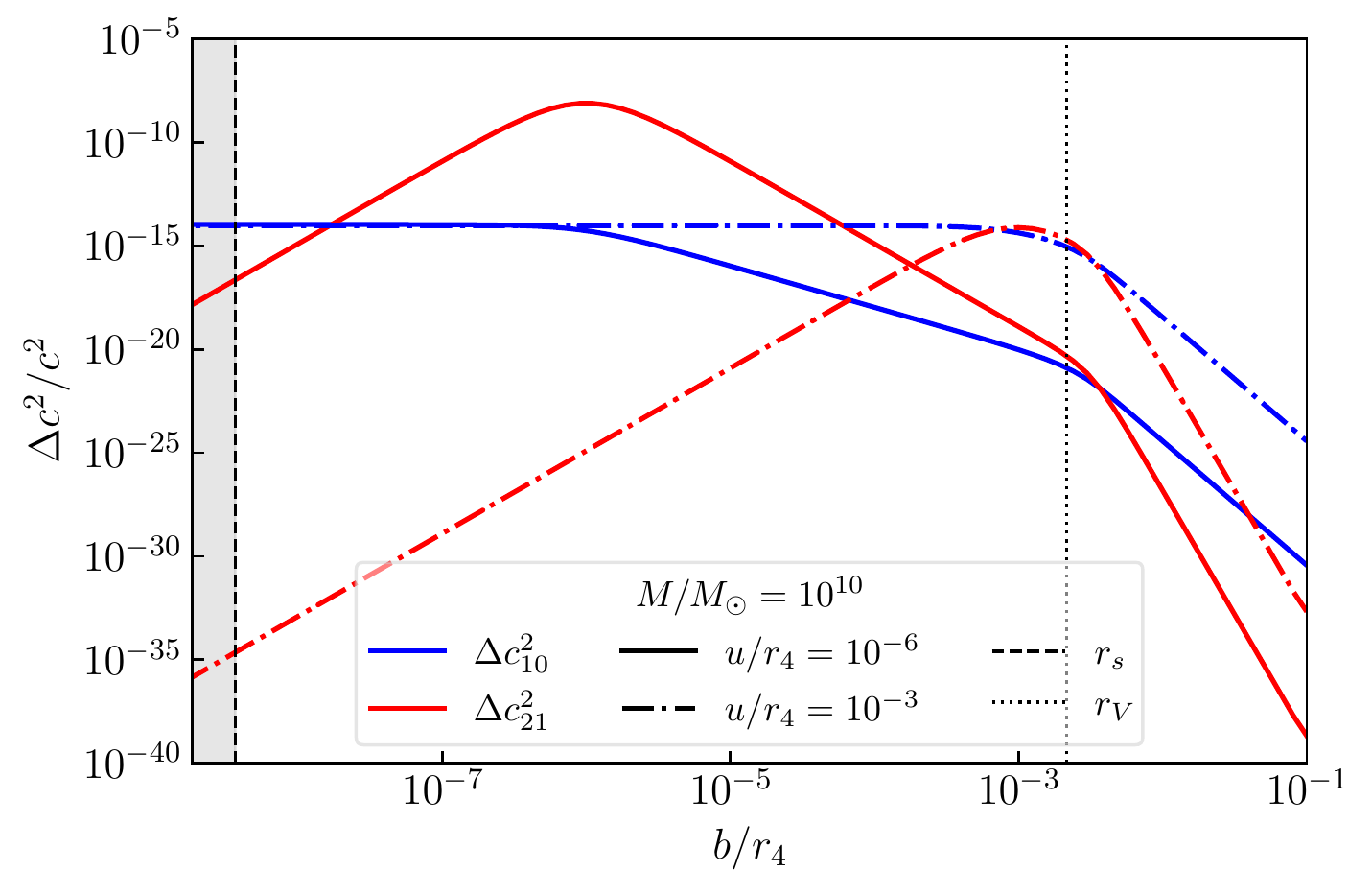}
 \caption{Difference in the speed between the metric eigenstate and light (blue) and the two mostly-metric eigenstates (red) as a function of the impact parameter $b$ in units of the scale of the quartic Horndeski theory $\rF$. 
 The solid and dashed-dotted lines evaluate the speed at different points of the trajectory $u$. 
 For this example we have chosen $\pFphi=10^{-8}$, $\LF=H_0$, $\pFX=-1$ and $\pTX=1$.}
 \label{fig:speed}
\end{figure}

%-FIGURE DELAY-
\begin{figure}[t!]
\centering 
\includegraphics[width = 0.95\columnwidth,valign=t]{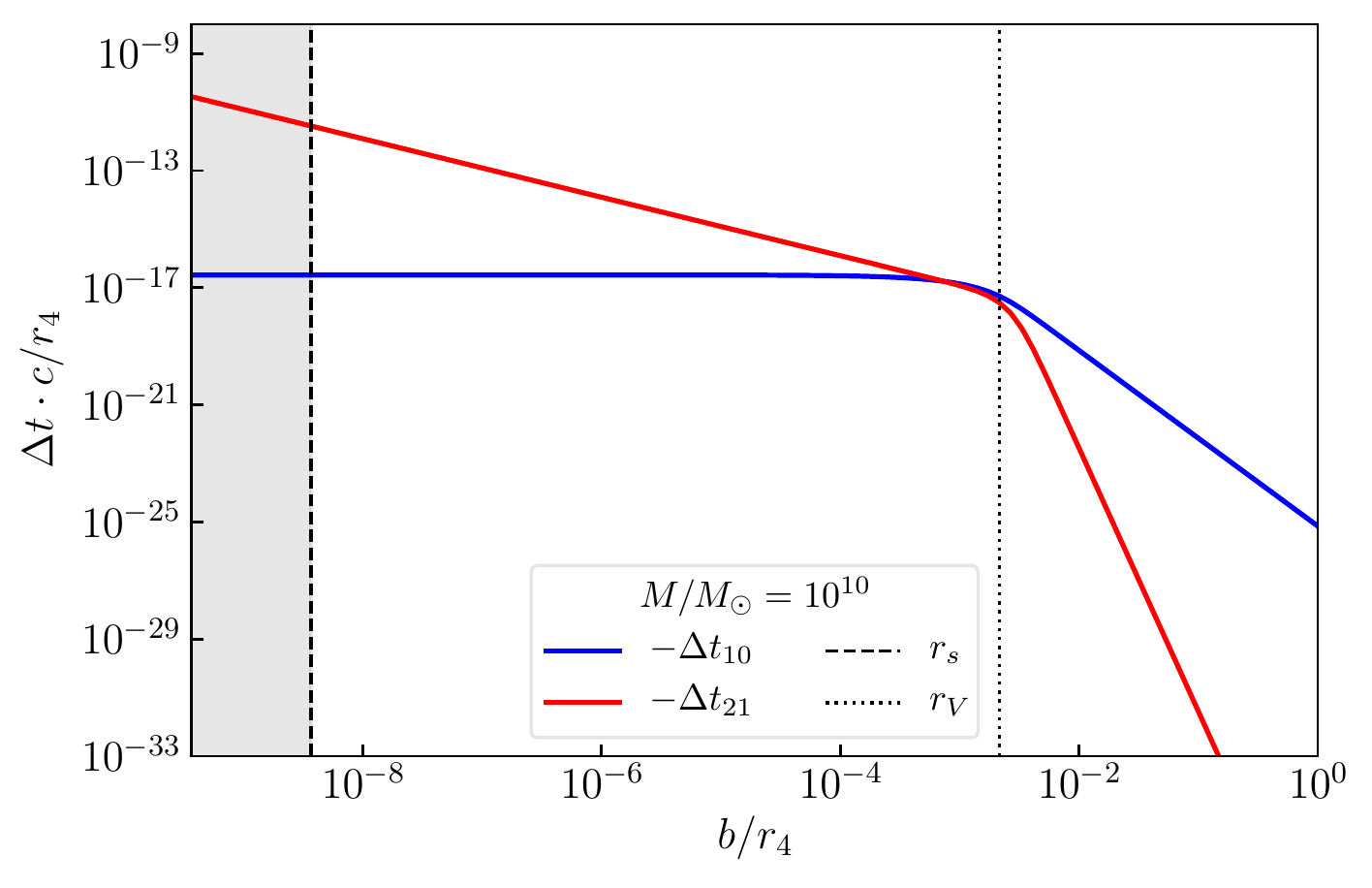}
 \caption{Shapiro crossing time delay induced by a point mass of $M=10^{10}M_\odot$ in a theory with $\pTX=-\pFX=1$, $\pTXX=\pFXX$ and $\pFphi=10^{-8}\Lambda/H_0$ as a function of the impact parameter $b$ in units of the non-linear scale of the quartic Horndeski theory $\rF$.}
 \label{fig:delay_b}
\end{figure}

%-FIGURE CONTOUR PLOTS-
\begin{figure*}[t!]
\centering 
\includegraphics[width = 0.9\linewidth,valign=t]{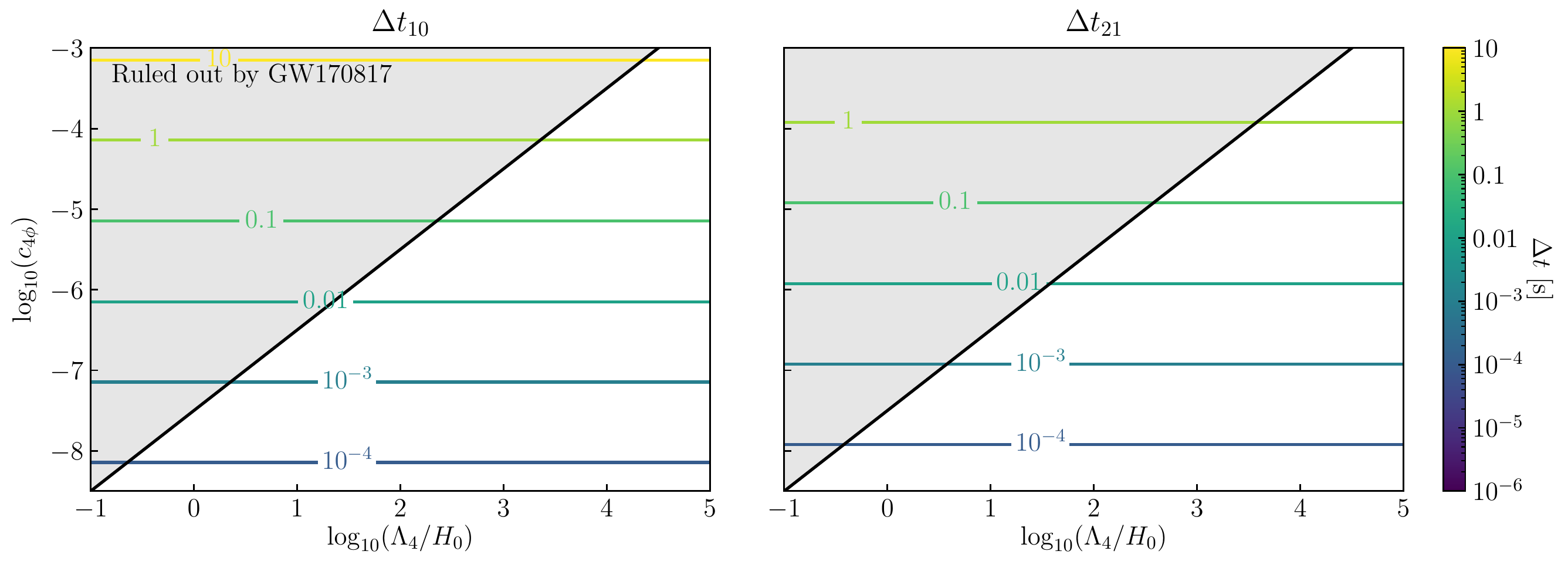}
 \caption{Shapiro time delays between the purely tensor eigenstate with speed $c_1=c_h$ w.r.t. the speed of light $c$, $\Delta t_{10}$, (left) and among the two mostly tensor polarizations, $\Delta t_{21}$, (right) as a function of the parameter space of the coupling to matter $\pFphi$ and the energy scale of the theory $\LF$. We plot the time delay accumulated after crossing the screened region of a point lens of $10^{10}M_\odot$ with an impact parameter $b=\rV$. The gray shaded region is the part of the parameter space already ruled out by the cosmological time delay in confrontation with GW170817.}
 \label{fig:contour_delay}
\end{figure*}
%------------

When there is mixing but $\Upsilon$ is small, the speeds of the propagation eigenstates are %controlled by  
\begin{align}
c_1^2&=c_h^2 \\
c_2^2&=c_h^2+\vert\Upsilon\vert^2\sin^4\theta \frac{(\Delta c_h^2)^2}{\Delta c_{hs}^2}+\cdots\\
c_3^2&=c_s^2-\vert\Upsilon\vert^2\sin^4\theta \frac{(\Delta c_s^2)^2}{\Delta c_{hs}^2}+\cdots
\end{align}
where we have defined the difference in the speed w.r.t. the speed of light $\Delta c_{i}^2=c_{i}^2-c^2$ and among different eigenstates $\Delta c_{IJ}^2=c_I^2-c_J^2$. The dots refer to higher order terms in the expansion in $|\Upsilon|$.

We can now compute the associated time delays between different signals. As discussed in section \ref{sec:lensing_pheno}, there will be two contributions: the Shapiro and the geometrical time delay. We discuss them separately before commenting on time delays between multiple images in strong lensing.

\subsubsection{Shapiro time delay}

The Shapiro time delay between the tensorial eigenstate and an EM counterpart, in the limit of small velocity difference $\Delta c_h^2/c^2\ll1$, is 
\be
\Delta t_{10} = \int du\lp\frac{1}{c_h}-\frac{1}{c}\rp=\int du\lp-\frac{\Delta c_h^2}{2c^3}+\cdots\rp\,,
\ee
where, again, $u$ is the GW propagation direction. 
On the other hand, the difference between the two mostly tensorial polarizations, in the limit of $\Delta c_{21}^2/c_h^2\ll1$, is
\be \label{eq:delta_t_21}
\begin{split}
\Delta t_{21} &= \int du\lp\frac{1}{c_2}-\frac{1}{c_1}\rp=\int du\lp-\frac{\Delta c_{21}^2}{2c_h^3}+\cdots\rp \\
&=\int du\lp-\frac{\vert \Upsilon\vert^2\sin^4\theta(\Delta c_{h}^2)^2}{2\Delta c_{hs}^2c_h^3}+\cdots\rp\,.
\end{split}
\ee
We see then that for a small mixing $\vert\Upsilon\vert\ll1$ the time delay between the mostly tensorial eigenstates will be suppressed compared to the time delay of the fastest mode and the speed of light. We can observe this directly in Fig. \ref{fig:speed}, where we present the difference in the speed and associated time delays. Now, because the multi-messenger time delay $\Delta t_{10}$ scales with the scalar background $\partial\bar\phi/\partial\tilde r$, which becomes constant in the inner screened region, the delay saturates at impact parameters smaller than the Vainshtein radius. We find this precisely in the blue line of Fig.~\ref{fig:delay_b}. 
On the hand, for the tensorial polarization delays, because the delay is also proportional to $\vert\Upsilon\vert^2\sin^4\theta$, the delay increases as a function of the impact parameter. This is shown in the red line. We then conclude that for impact parameters much smaller than the screening radius the delay between the tensorial eigenstates $\Delta t _{21}$ becomes more constraining that the multi-messenger delay $\Delta t_{10}$. Nonetheless, such close encounters are less probable (cf. section \ref{sec:observational_probabilities}).

Going to the particular quartic theory studied in this section, we can see that the multi-messenger time delay scales as
\be
\frac{d t_h}{dz}\sim 2\cos^2\theta{\sum_n n\,\pFXn\lp\frac{\rS}{\rF}\rp^n\lp\frac{\partial\bar\phi}{\partial\tilde{r}}\rp^{2n}}\,.
\ee
Since the scalar field profile decays rapidly outside of the screened region, determined by $\rV\sim \pFphi^{1/3}\rF$, the order of magnitude of the delay will be given essentially by $\rV$ times the ratios $(\rS/\rF)^n(\partial\bar\phi/\partial\tilde r)^{2n}$, where we can find the scaling of the scalar background in (\ref{eq:back_screening}). For a theory with $G_4$ linear in $X$ we can compute the order of magnitude of the maximum time delay
\be
\begin{split}
\left.\Delta t_h\right\vert_{\text{max}}&\lesssim \frac{\pFphi}{\pFX^{1/3}
}\frac{2GM_L}{c^3} \\
&\sim 1\,\text{s}\lp\frac{1}{\pFX}\rp^{1/3}\lp \frac{\pFphi}{10^{-4}}\rp \lp\frac{M_L}{10^{10}M_\odot}\rp\,.
\end{split}
\ee
The time delay thus increases with the coupling to matter $\pFphi$ and the lens mass. We could also integrate analytically for $u,b\ll \rV$, since we know the solution of $\partial\bar{\phi}/\partial\tilde{r}$, to obtain
\be
\left.\Delta t_h\right\vert_{u,b\ll \rV}\simeq \lp\frac{\rS}{\rF}\rp\lp\frac{\pFphi^{2}}{4\pFX}\rp^{1/3}\lp u - b\cdot\text{atan}\lb\frac{u}{b}\rb\rp\,,
\ee
where the integration is performed from $-u$ to $u$. 

This order of magnitude calculation can be compared with the explicit calculation that we present in the left panel of Fig. \ref{fig:contour_delay} as a function of the parameter space $\pFphi$ and $\LF$ for a super-massive black hole (modeled as a point lens) of mass $10^{10}M_\odot$. We emphasize that our results can be easily adjusted to other masses.  
It is important to note though that for larger masses (galactic order of magnitude) one would expect the mass to be distributed in a halo, so that the point lens approximation is  broken. Introducing a realistic mass distribution would reduce the mass contained in the inner screened region, reducing the induced time delay. We will elaborate more on this later in section \ref{sec:screening_obs_prospects}. 
Finally, let us mention that for the multi-messenger time-delay $\Delta t_{10}$ there would be an astrophysical uncertainty of order $1-10$ seconds. This means that in practice one can only rule out modify gravity theories with larger delays. 

We can also compute explicitly the time delay between the polarization eigenstates. Starting from (\ref{eq:delta_t_21}) and noting that for our fiducial theory $c_s=c$ (so that $\Delta c^2_{hs}=\Delta c^2_h$), we find
\begin{equation}
\frac{dt_{21}}{dz}\sim\vert \Upsilon\vert^2\sin^4\theta\frac{dt_h}{dz}\,.
\end{equation}
This means that $\Delta t_{21}$ will be suppressed with respect to $\Delta t_h$. 
From Fig. \ref{fig:delay_b} we see that $\Delta t_{21}$ increases inversely proportional to the impact parameter $b$. We can also compute the time delay analytically close to the lens
\be
\left.\Delta t_{21}\right\vert_{u,b\ll \rV}\simeq \frac{1}{4}\pFX^3\lp\frac{\pFphi}{\pFXX+2\pFX^2}\rp^{4/3}\,\frac{\rS}{\rF}\cdot\frac{\text{atan}(z/b)}{b/(\rF)^2}\,,
\ee
which in this case will dominate the overall integral since the major part of the delay is accumulated close to the lens. Nonetheless, one should remember that smaller values of $b$ are less probable to occur. For that reason we fix $b=\rV$ to compute the time delays in the right panel of Fig. \ref{fig:contour_delay}. Thanks to the $\Delta t_{21}\sim 1/b$ scaling, this plot can easily be adapted to other choices of the impact parameter. 
One should remember that the detectability of the birefringence time delay is only limited by the time resolution of GW detectors that can be considered to be $\sim$ms.

%-FIGURE DEFLECTION ANGLE-
\begin{figure}[b!]
\centering 
\includegraphics[width = 0.95\columnwidth,valign=t]{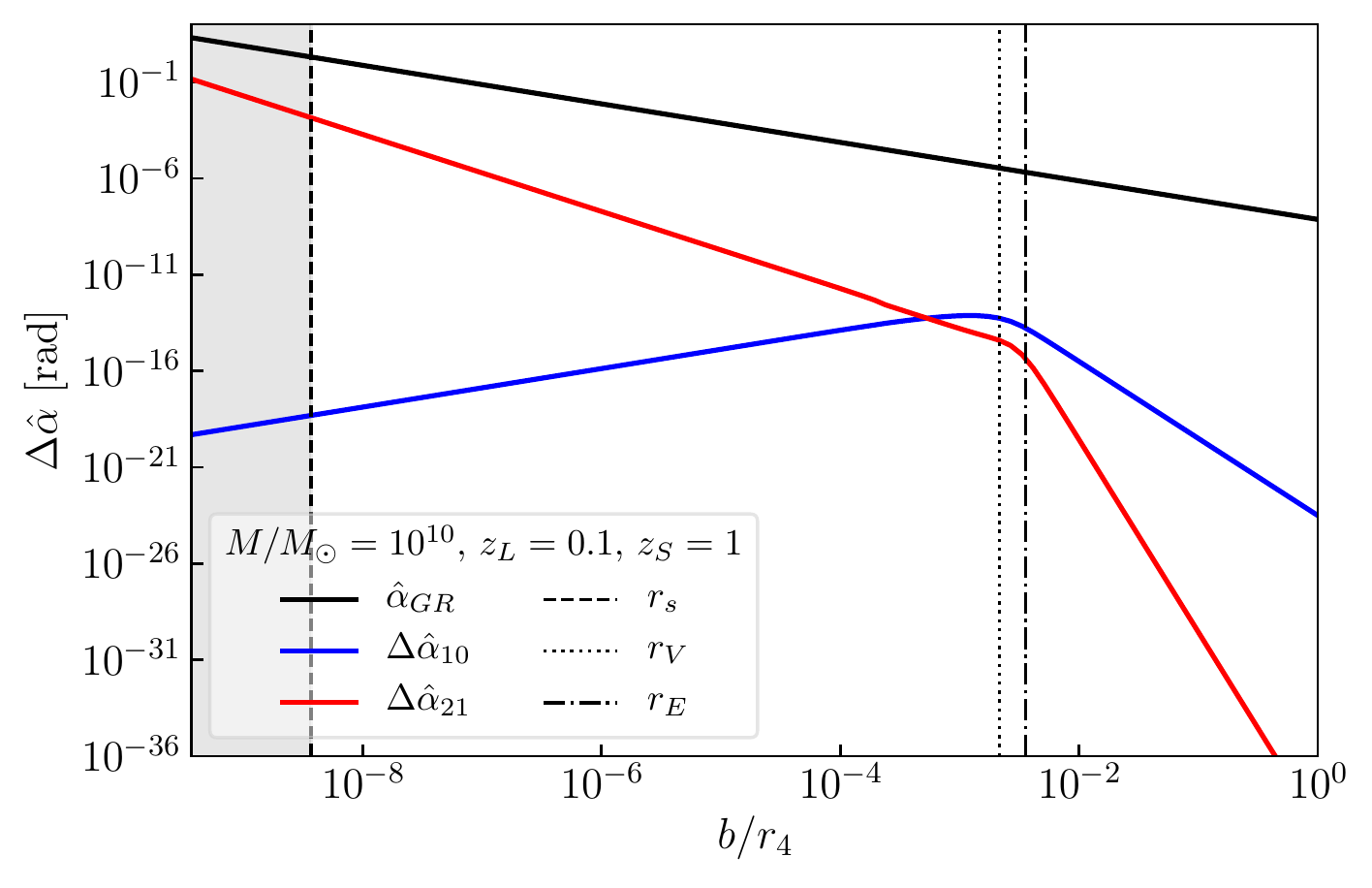}
 \caption{Deflection angle induced by a point mass of $M=10^{10}M_\odot$ in a theory with $\pTX=-\pFX=1$, $\pTXX=\pFXX$ and $\pFphi=10^{-8}\LF/H_0$ as a function of the impact parameter $b$ in units of the scale $\rF$ of the theory.}
 \label{fig:alpha_b}
\end{figure}
%-------------

\subsubsection{Geometrical time delay}

%-FIGURE SHAPIRO VS GEOM-
\begin{figure*}[t!]
\centering 
\includegraphics[width = 0.48\linewidth,valign=t]{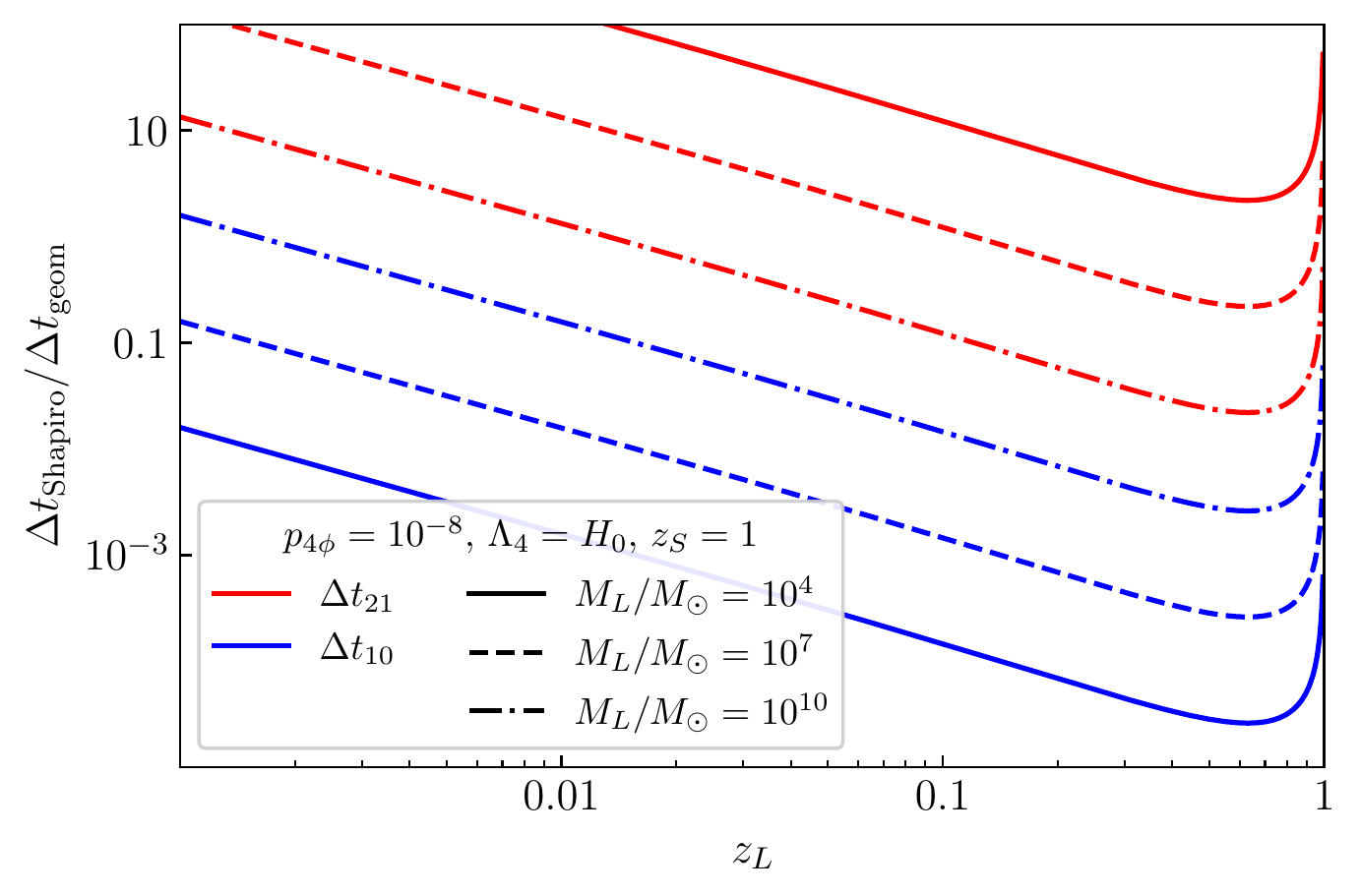}
\includegraphics[width = 0.48\linewidth,valign=t]{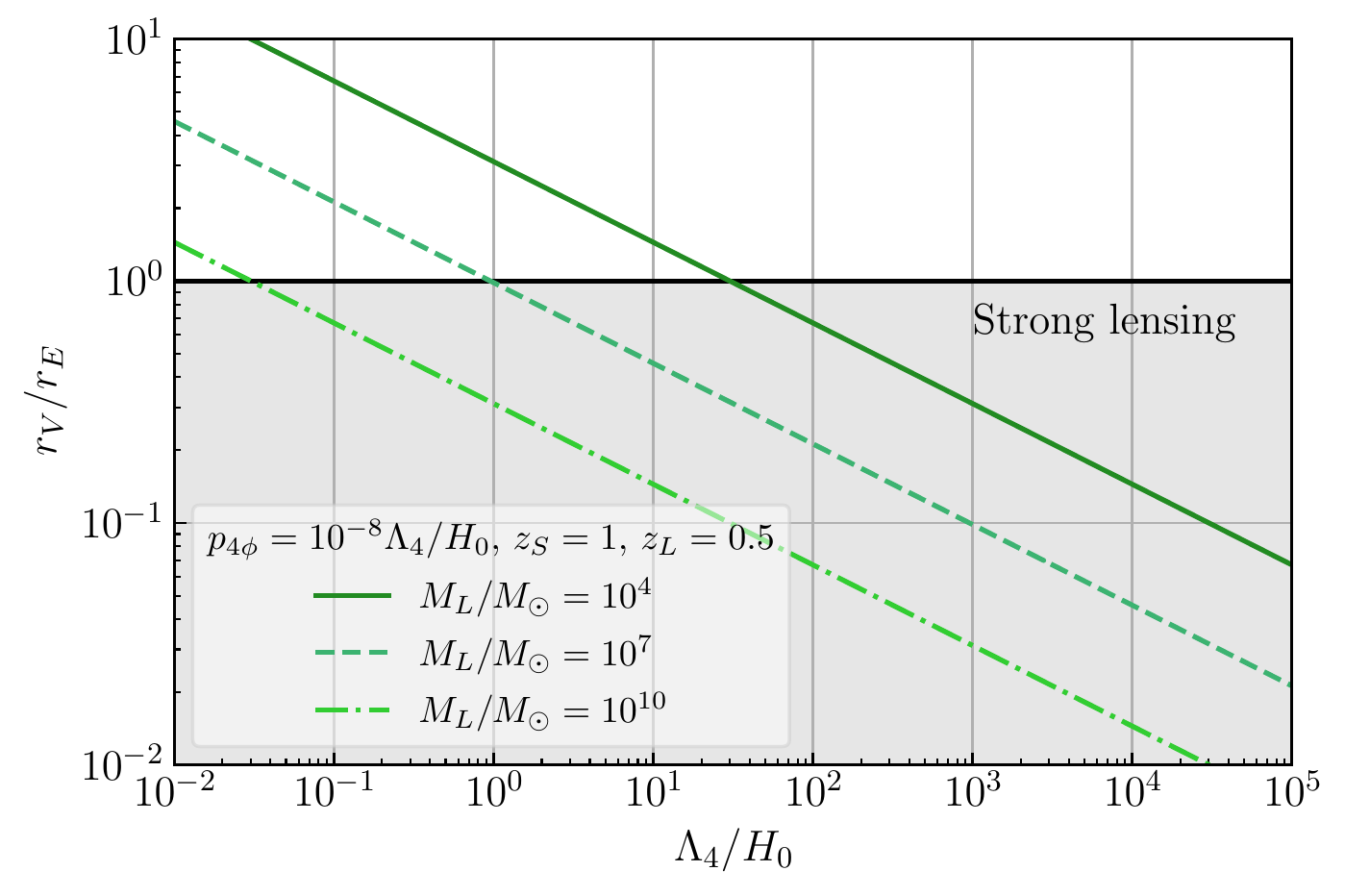}
 \caption{On the left, comparison of the Shapiro and geometrical time delays (evaluated at the Vainshtein radius) as a function of the lens redshift for a fixed source at $z_S=1$. On the right, ratio of the Vainshtein and Einstein radius as a function of the scale of the quartic Horndeski theory. In both panels we consider different point lens masses.
 }
 \label{fig:shapiro_vs_geom}
\end{figure*}
%------------

In order to compute the geometrical time delay (\ref{eq:geom_delay}), we first need to obtain the deflection angle associated to each propagation eigenstate, which can be obtained from their propagation speed (\ref{eq:deflec_angle}). In particular, the deflection angle between the tensor propagation eigenmode and the speed of light is given by
\be
\begin{split}
\Delta \hat \alpha_{10}&=-\frac{1}{2}\int \frac{du}{\rF} \sin\theta\,\frac{\partial\Delta c_{10}^2}{\partial \tilde{r}} \\
& = 4\pFX\lp\frac{\rS}{\rF}\rp\int \frac{du}{\rF} \sin\theta \cos^2\theta\frac{\partial\bar\phi}{\partial\tilde r}\frac{\partial^2\bar\phi}{\partial\tilde r^2}\,,
\end{split}
\ee
where in the second line we have specialized for a quartic theory linear in $X$. 
This corresponds to the blue line in Fig. \ref{fig:alpha_b}, where one can see that inside the screened region the deflection angle difference decreases with the impact parameter.

We can similarly compute the deflection angle between the two mostly tensorial propagation eigenstates
\be
\begin{split}
\Delta \hat\alpha_{21}&=-\frac{1}{2}\int \frac{du}{\rF} \sin\theta\,\frac{\partial\Delta c_{21}^2}{\partial \tilde{r}} \\
&=-\frac{1}{2}\int \frac{du}{\rF} \sin^5\theta\lp2\vert\Upsilon\vert\frac{\partial\vert\Upsilon\vert}{\partial\tilde r}\Delta c_{10}^2+\vert\Upsilon\vert^2\frac{\partial\Delta c_{10}^2}{\partial \tilde{r}}\rp\,\,.
\end{split}
\ee
This corresponds to the red line in Fig. \ref{fig:alpha_b}. In this case, the deflection angle is dominated by the behavior close to the lens. This can be approximated analytical solving the integral in the limit $u,b\ll \rV$. For our fiducial theory we obtain 
\be
\Delta\hat\alpha_{21}=\frac{16\cdot2^{2/3}}{105}\pFphi^{2/3}\pFX^{1/3}\lp\frac{\rS}{\rF}\rp\lp\frac{\rV}{b}\rp^{2}\,.
\ee
From this expression the most important feature is that it scales with the inverse of the square of the impact parameter. This growth is faster than the typical $1/b$ induced by a gravitational potential, as can be seen comparing with the black solid line in Fig. \ref{fig:alpha_b}. One can see though that for impact parameters of the order of the Vainshtein radius the difference in the deflection angle is small compared to the effect of the point mass potential.

Using (\ref{eq:geom_delay}) we can translate the difference in the deflection angles into the geometrical time delay. As we have seen in Fig. \ref{fig:alpha_b}, the difference in the deflection angle will be much smaller than the deflection angle induced by the gravitational potential (except very close to the lens where the difference reduces). Therefore we can use the approximate expression for the geometrical time delay given in equation (\ref{eq:geom_delay_small_deflection}) that makes use of this hierarchy in the order of magnitude of the deflection angles.

The mass of the lens and its relative location in the line of sight determine the relative importance of the Shapiro and geometric contributions to the total time delay.
In Fig. \ref{fig:shapiro_vs_geom}  we present the ratio of both time delays as a function of the lens redshift. The geometrical time delay dominates for lenses halfway to the source, while Shapiro dominates when $z_L\to 0, z_S$. 
With fixed $b\sim r_V$, this is driven by the proportionality with the universal deflection angle $\alpha_0\propto r_E^2/b$ in Eq. (\ref{eq:geom_delay_small_deflection}), as $r_E^2\propto D_L D_{LS}$ is reduced when the lens is near the source or the observer.

The Shapiro-to-geometric contribution depends differently on lens mass in multi-messenger and birefringent delays (different line styles, left panel of Fig. \ref{fig:shapiro_vs_geom}). 
The Shapiro contribution to $\Delta t_{21}$ is reduced with increasing mass, while the opposite is true for $\Delta t_{10}$, independently of the lens redshift.  
The mass dependence can be understood from the right panel of Fig. \ref{fig:shapiro_vs_geom}, where we present how the Vainshtein radius compares to the Einstein radius as a function of the scale of the theory for fixed $\alpha_T=10^{-16}$. 
For $\LF=H_0$, as chosen in the left panel, a lens with $10^{10}\Msun$ will have the Vainshtein radius well within the strong lensing region, while a $10^5\Msun$ lens will have $\rV>r_E$. Whenever the impact parameter is smaller than $r_E$, the geometrical time delay will be large. 
On the other hand, the multi-messenger Shapiro delay scales with the Vainshtein radius and decreases faster than the geometrical one when the lens mass is reduced. 
Finally, the birefringent Shapiro delay is mostly accumulated near the lens and thus is less affected by the reduction of the lens mass than the analogous geometrical delay.

\subsubsection{Multiple image time delays}

As introduced in section \ref{sec:observational_multiple_images}, in the regime of strong lensing there will be multiple images with an associated delay between them. At the same time, each of this images will be subject to the effects of Shapiro and geometrical time delay for the propagation eigenstates. It is therefore appropriate to ask how this multiple image time delays compare to the delay between the propagation eigenstates. 

For the example screening theory that we are considering here, we have seen in the right hand plot of figure \ref{fig:shapiro_vs_geom} that indeed the Vainshtein radius falls inside the Einstein radius for a sector of the parameter space. At small impact parameters, the time delay between the images scales as $\Delta t_{\pm}\sim t_M\cdot b/r_E$ while the Shapiro time delay between the mostly tensor polarization scales as $\Delta t_{21}\sim t_M\cdot r_V/b$. Therefore, depending on the value of $b$, the time delay between the images could be larger than the one between the eigenstates or viceversa. Note that in terms of statistics $\Delta t_\pm > \Delta t_{21}$ is much more probable, as it corresponds to larger impact parameters. Therefore, if a pair of GWs were identified as strongly lensed images of the same event, this would be a perfect candidate to look for additional lensing effects due to screening. 

%MIXING
\subsection{Polarization mixing and GW shadows} \label{sec:screening_shadows}

At leading order in the GW propagation, the other main observable is the appearance of additional polarizations beyond the transverse-traceless tensorial modes. 
We have shown that there are two ways in which extra polarization can arise: \emph{(i)} by a direct mixing between the tensor-scalar perturbations through $\Upsilon$ and \emph{(ii)} by background profiles inducing non-radiative polarizations, what we have called GW shadows. 

\begin{figure}[t!]
\centering 
\includegraphics[width = 0.95\columnwidth,valign=t]{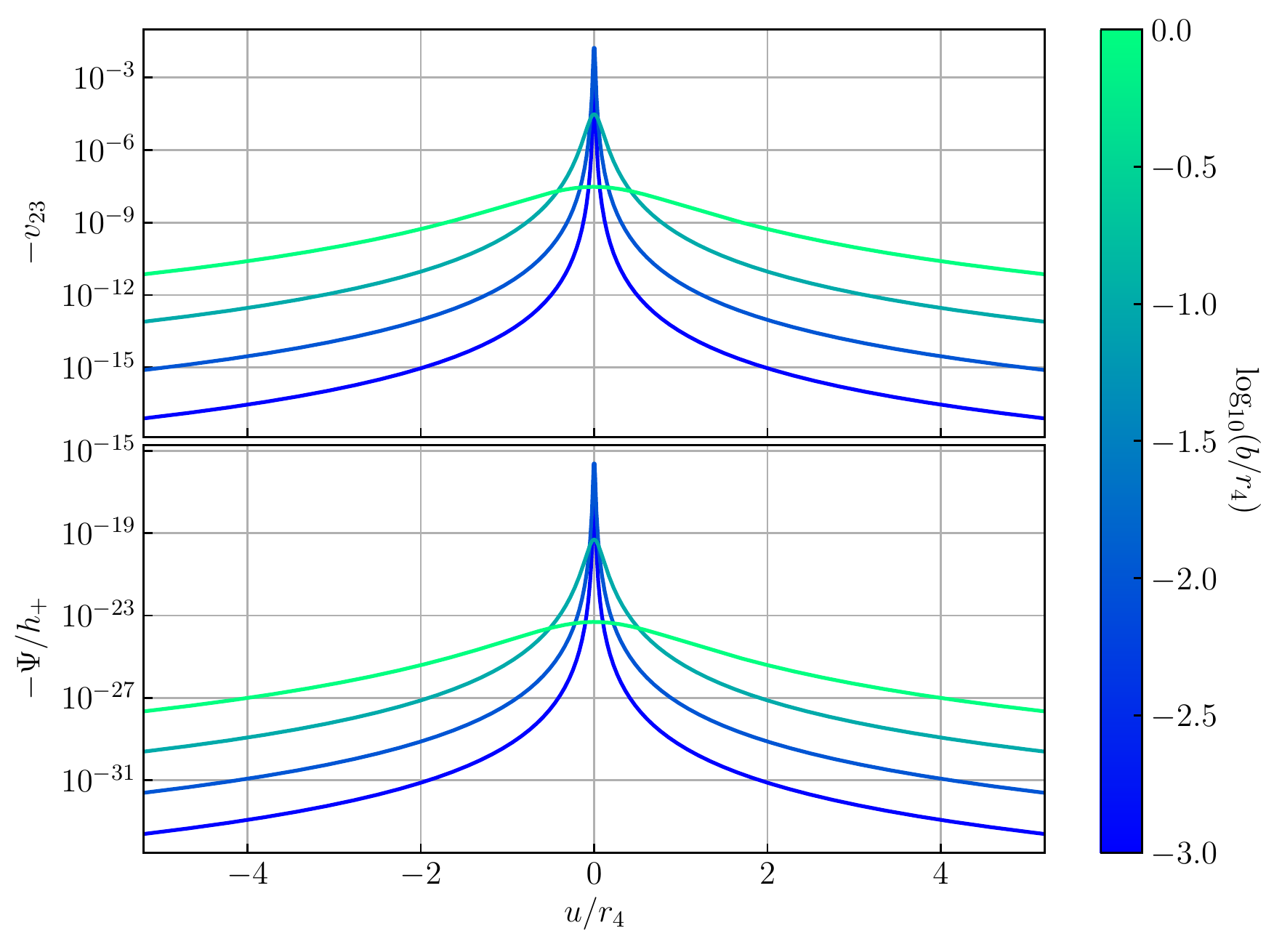}
 \caption{On the top tensor-scalar mixing $v_{23}$ as a function of the propagation direction $u$ normalized by the scale $\rF$. On the bottom \emph{shadow} scalar polarization $\Psi$ induced by an incoming $h_+$ polarized GW in the $u$ direction.}
 \label{fig:add_polarizations}
\end{figure}

For the propagating scalar polarizations, we can quantify the coupling of the tensor-scalar mixing from the mostly-tensor eigenvector. In particular, the third entry $v_{23}$ in the mixing matrix (\ref{eq:change_basis}) informs us of the amplitude of the scalar mode that would be generated even if initially there is no scalar mode. In the top panel of Fig. \ref{fig:add_polarizations} we plot this tensor-scalar mixing. For a linear quartic theory with a standard scalar field kinetic term, the mixing simplifies to 
\be
v_{23}=-\vert\Upsilon\vert\sin^2\theta\,.
\ee 
We can observe that the amplitude of the scalar perturbation can only be sizable near the lens.

With respect to the shadows polarizations, we can take as an example the non-propagating polarization $\Psi$. In particular, we consider a $+$ polarized GW propagating in the $z$ direction. Then, the amplitude is given by
\be
\Psi\sim\frac{G_{4\tilde{X}}}{G_{4}}\lp\frac{\partial\bar\phi}{\partial\tilde r}\rp^2\sin^2\theta\,h_+\,,
\ee
where the approximate equality accounts for the fact that we are neglecting the contribution from $\vp$ (that as we have just seen is small if initially $\vp$ is not sourced). 
Because of the $\sin^2\theta$ proportionality, the amplitude of $\Psi$ evolves similarly to $v_{23}$, as shown in the lower panel of Fig. \ref{fig:add_polarizations}, however the amplitude is many orders of magnitude smaller. Therefore, for this class of theories compatible with GW170817 detecting GW shadows seems out of reach.

%-FIGURE ADDITIONAL POLARIZATIONS-

\subsection{Observational prospects} \label{sec:screening_obs_prospects}

To conclude this section we will discuss the observational prospects of detecting these novel lensing effects beyond GR. The first question would be in which systems these effects would be relevant. In our calculations we have done two important assumptions: we have worked at leading order in geometric optics and modeled the lenses as point masses. Both effects limit the lenses available to test quartic Horndeski theories. We will comment on the implications of these assumptions and then discuss the potential of GW birefringence to probe our example theory.

\begin{figure}[t!]
\centering 
\includegraphics[width = 0.95\columnwidth,valign=t]{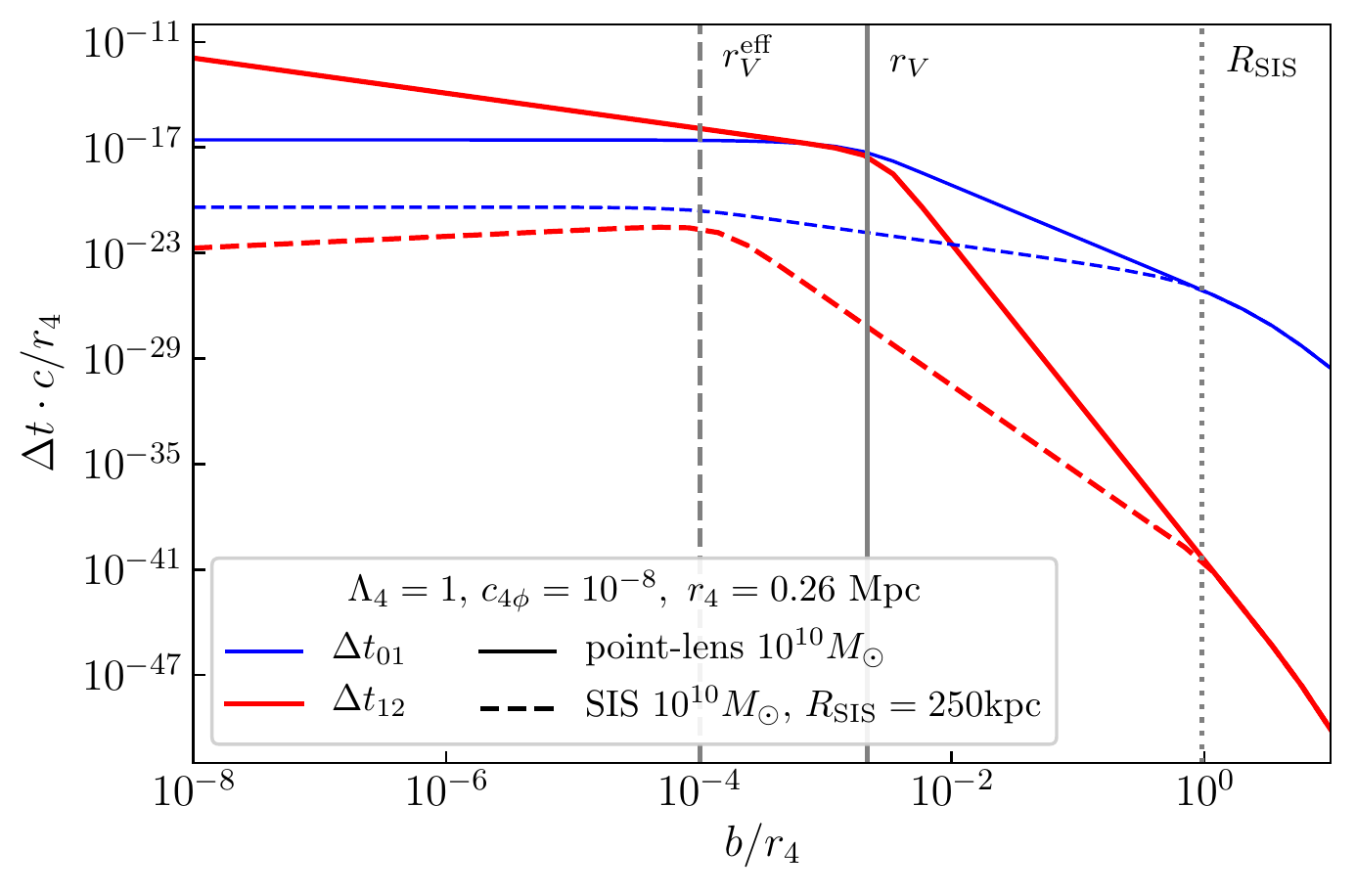}
 \caption{Effect of extended lenses. Lines show the Shapiro time delays as a function of the impact parameter for a point lens (solid) and a singular isothermal profile (dashed) truncated at $R_{\rm SIS}=250$kpc (\ref{eq:sis}), both with the same mass $10^{10}M_\odot$. The birefringence time delay $\Delta t_{12}$ (thick red) is more suppressed than the multi-messenger delay $\Delta t_{01}$ (thin blue). Vertical lines denote the nominal Vainshtein radius (solid), the SIS effective Vainshtein radius (dashed) and the size of the lens (dotted), see text.}
 \label{fig:beyond_point_mass_profile}
 %\end{figure}
 \vspace{5pt}
 %\begin{figure}
\includegraphics[width = 0.95\columnwidth,valign=t]{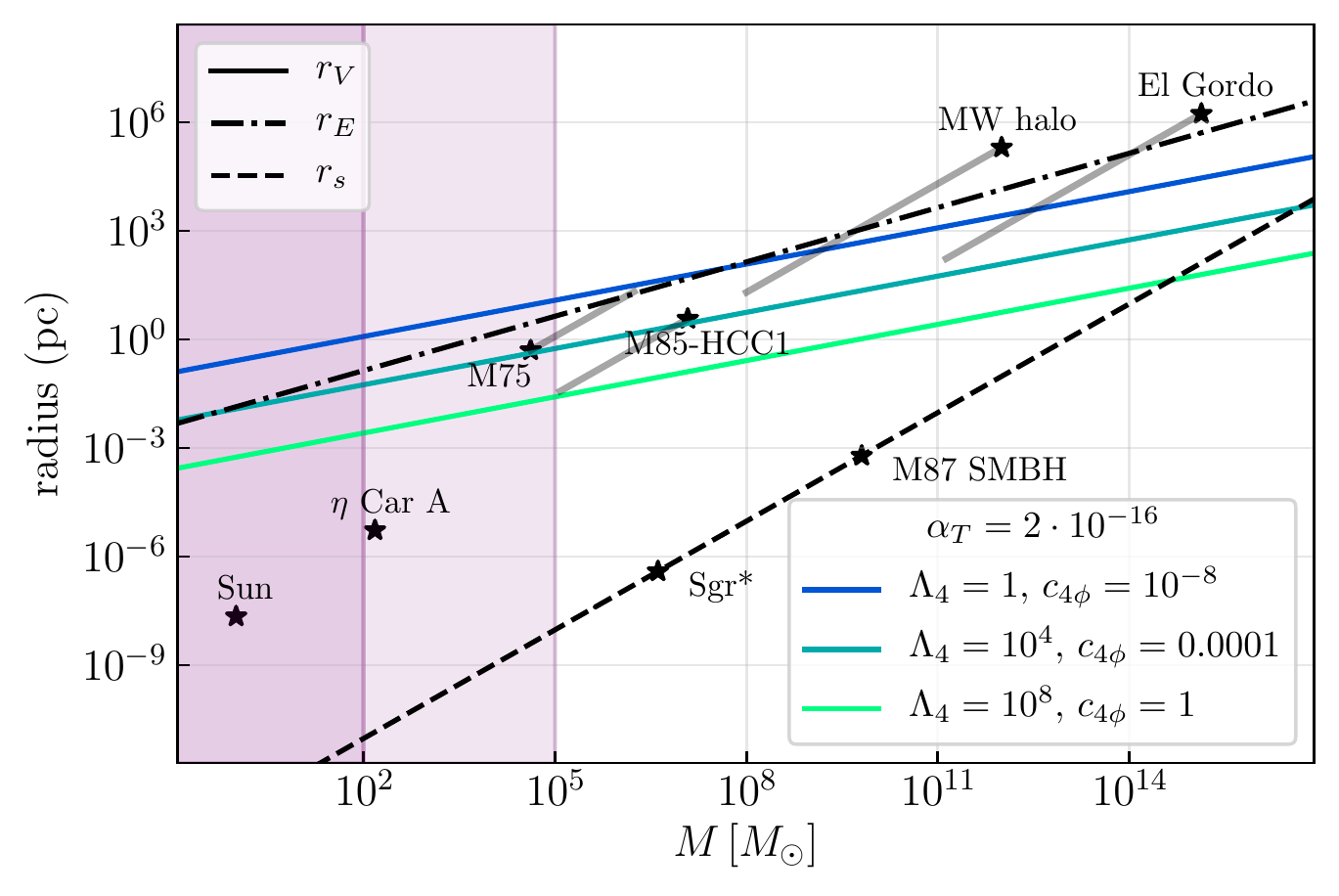}
 \caption{Masses and sizes of prototypical lenses. The Vainshtein radius is shown for example theories compatible with GW170817. The Einstein radius assumes that the lens and the source are at cosmological distances $r_E\sim 0.03\text{pc}\sqrt{M/M_\odot}$.
 Shaded regions correspond to the onset of wave effects (\ref{eq:diffraction_limit}) for GWs frequencies $\sim$ kHz, Hz, where geometric optics is not applicable.
 Markers show the physical size of known objects (see text).  
 Gray lines show the enclosed mass of extended objects assuming an isothermal profile (\ref{eq:sis}).}
 \label{fig:beyond_point_mass_lenses}
\end{figure}

%geometric optics
Working in the geometric optics regime imposes a lower limit on the frequency of GWs for which the short-wave expansion  (\ref{eq:wkb_metric},\ref{eq:wkb_scalar}) applies. The exact limit depends on the background solution around the lens and theory-specific lower-order corrections to the propagation equations. 
Even neglecting beyond GR corrections, the frequency range is restricted by the diffraction limit in GR, below which lensing magnification becomes very inefficient. The GR diffraction limit, Eq. (\ref{eq:diffraction_limit}), corresponding to GW wavelengths larger than or comparable to the Schwarzschild radius of the lens and is shown in figure \ref{fig:beyond_point_mass_lenses} for $f\sim$ Hz \& kHz.

The diffraction limit excludes stellar-mass lenses to test birefringence using a short-wave expansion. This would be excellent lens candidates, as most stellar objects can be considered point-like, i.e. their sizes are much smaller than their Vainshtein radii, even for theories compatible with GW170817, cf. figure \ref{fig:beyond_point_mass_lenses}. 
Note that the validity of geometric optics is a limit on the framework, indicating the need of a wave-optics description. In particular, it does not mean that birefringence or time delays cease to exist. If a description similar to the wave-optics integral is valid at low frequencies (Eq. 3 of \cite{Ezquiaga:2020spg}), the birefringence time delay should leave an imprint on the waveforms, even if gravitational magnification is negligible.

%-FIGURE BEYOND POINT MASS-

%radial dependence
The point-mass assumption is a good description for impact parameters larger than the lens size. Because birefringence is suppressed beyond $\rV$, ideal lenses should be smaller than their own Vainshtein radii.
Effects on general lenses can be computed given their mass distribution.
By Gauss' theorem, the scalar field profile around a spherically symmetric lens is sourced by the enclosed mass at a given radius, i.e. $\tilde M(r)$ in Eq. (\ref{eq:screening_back}). 
We will model extended sources as truncated singular isothermal spheres (SIS)
\begin{equation}\label{eq:sis}
 \rho_{\rm SIS} \sim r^{-2} \quad (r\leq R_{\rm SIS})\,,
\end{equation}
and $\rho_{\rm SIS}=0$ for $R>R_{\rm SIS}$. The truncation at $R_{\rm SIS}$ ensures a finite total mass, but does not affect the results for low impact parameter.
The SIS profile is widely used as a model for simple gravitational lenses. 
Note that the matter density diverges at the center.%
\footnote{A regular value of the central density will suppress beyond-GR effects near the center.
In the case of a homogeneous density $\rho(r)\sim \text{const}$, $\Upsilon\sim 0$ and birefringence effects vanish entirely.}

% The effect of extended masses is to spread the 
The reduced enclosed mass at low radii flattens the derivatives of the scalar field, lowering the time delays in extended lenses.
Figure \ref{fig:beyond_point_mass_profile} shows the Shapiro time delays for two lenses with $10^{10}M_\odot$: one with a point-like distribution and another one with a SIS profile truncated at $R_{\rm SIS} \approx 250$kpc. 
The delay between gravitational polarizations $\Delta t_{12}$ is more affected than the multi-messenger time delay $\Delta t_{01}$ due to the different dependence on the scalar field derivatives, including the shear via $\Upsilon$. This reduces both the slope and the amplitude of $\Delta t_{12}$.

The maximum time delay in finite lenses occurs at a parameter impact smaller than the nominal Vainshtein radius. The reason is that only the total mass with a radius $r$ contributes to the scalar field profile. This motivates the definition of an \textit{effective Vainshtein radius} $\rV^{\rm eff}$ satisfying
\begin{equation} \label{eq:extended_effective_r_V}
 \rV(M(\rV^{\rm eff})) = \rV^{\rm eff}\,,
\end{equation}
where the dependence on theory parameters has been omitted. For the truncated SIS, the mass dependence $M\propto r$ results in $\rV^{\rm eff} = \rV^{3/2}/R_{\rm SIS}^{1/2}$. For a point-lens the effective and nominal Vainshtein radius are equal since the enclosed mass $M$ is constant. Note that non-singular lenses may have no solutions to Eq. (\ref{eq:extended_effective_r_V}), indicating that no screening occurs.

%possible sources
The requirement of lenses being smaller than their effective Vainshtein radii limits the type of objects (or portions thereof) that can contribute significant time delays.
Figure \ref{fig:beyond_point_mass_lenses} shows the sizes and masses of known astronomical objects that could act as lenses. These are, in order of increasing mass, the Sun, a large star $\eta$ Car A, a dense globular cluster M75, the massive black hole Sgr* in the center of the milky way, the very dense dwarf galaxy M85-HCC1, the super-massive black hole in M87, the Milky Way halo and the Galaxy Cluster ``El Gordo''. 
The mass profiles of extended objects have been extrapolated inward assuming a SIS distribution (\ref{eq:sis}) using the total mass and size (or outward using the central density in the case of M75). This extrapolation suggests that some portion of extended lenses will be within its own effective Vainshtein radius, at least for theories compatible with GW170817 with low $\LF$. 

%SMBH and the conformal coupling
Super-massive black holes (SMBHs) appear as the optimal lenses to further constrain quartic Horndeski theories. But because black hole solutions have vanishing Ricci curvature, SMBHs would not source the field via the conformal coupling (\ref{eq:example_theory}) in the specific theory under consideration. SMBHs could still provide an effective lens if they lead to the accumulation of dark matter around the black hole with sufficient density. In such scenarios, a ``dark matter spike'' could encompass a mass comparable to that of the central black hole in a very small central region of radius $r\sim 0.1(M_{\rm SMBH}/10^6M_\odot)$pc \cite{Gondolo:1999ef}, sourcing the scalar field profile at the level to render the lens efficient. 
A coupling to the SMBHs may be induced by the cosmological evolution of the field, as it has been shown to occur for cubic Galileons (unconstrained by GW170817) \cite{Babichev:2012re,Babichev:2016fbg,Brax:2020ujo}.

\begin{figure}[t!]
\centering 
\includegraphics[width = 0.95\columnwidth,valign=t]{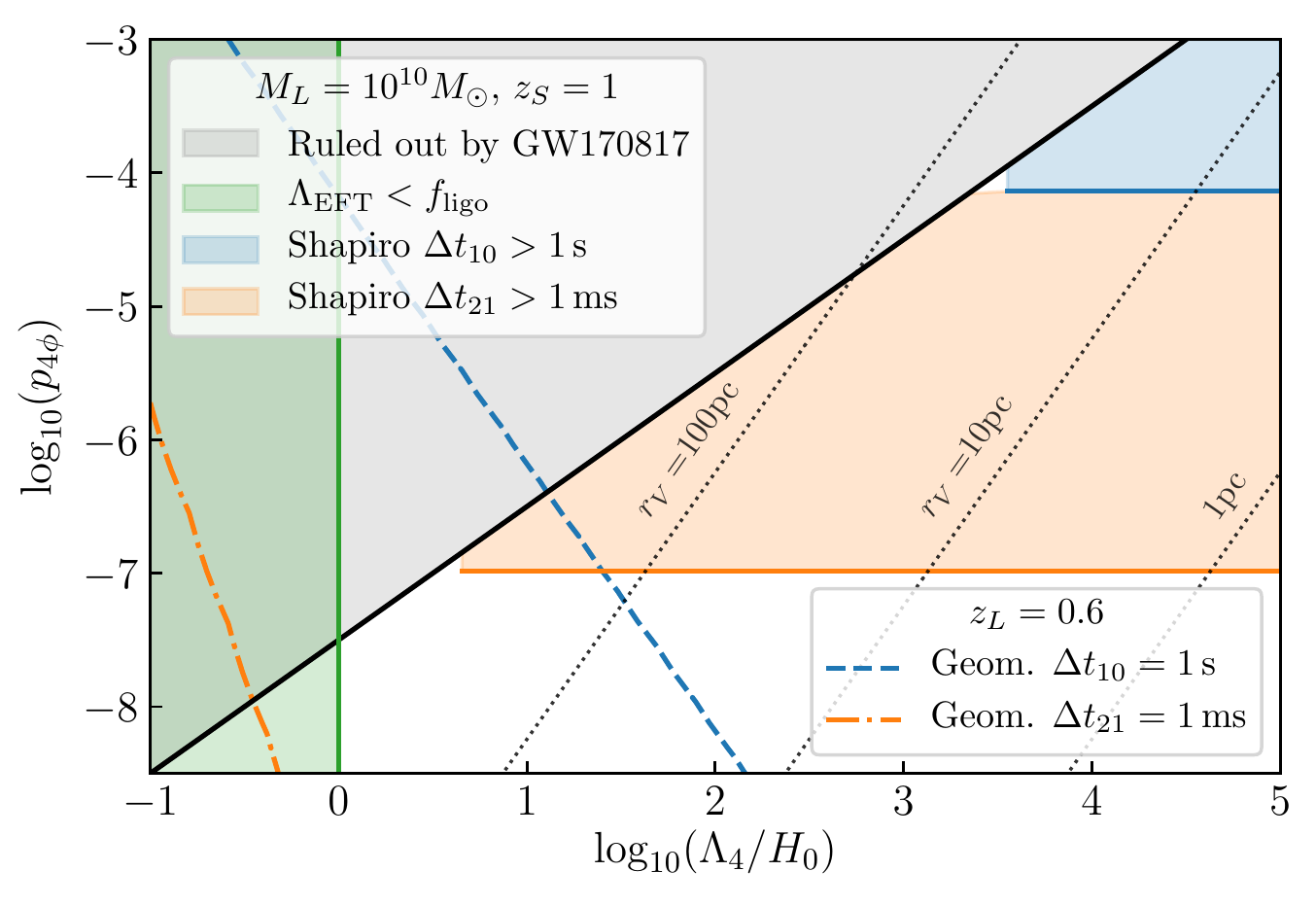}
 \caption{Summary of the constraints on the parameter space of the quartic theory $\pFphi$ and $\Lambda_{4}$. Theories that predict a cosmological time delay $\alpha_T\gtrsim10^{-15}$ are ruled out by GW170817, although part of this sector has a effective field theory (EFT) scale $\Lambda_\text{EFT}^3=\Mpl\LF^2$ smaller than LIGO frequencies: $\Lambda_\text{EFT}\lesssim f_\text{ligo}\lesssim10^{3}$Hz. The inhomogeneous time delay induced by the screening probes a new sector with the multi-messenger time delay $\Delta t_{10}$ and the delay between the propagation eigenstates $\Delta t_{21}$. We take as a reference a point lens of $10^{10}M_\odot$.}
 \label{fig:summary_plot}
\end{figure}

We summarize the parameter space of the quartic Horndeski theory that could be constrained with lensing time delays in figure \ref{fig:summary_plot}. As an order of magnitude estimate, we consider testable multi-messenger time delays $\Delta t_{10}>1$s and delays between the polarizations $\Delta t_{21}>1$ms. It is clear from the plot that a large new sector of the parameter space $\pFphi$, $\LF$ could be probed beyond current constraints from GW170817. For reference we also highlight the parameter space in which the scale of the effective field theory cutoff is smaller than LIGO frequencies \cite{deRham:2018red}. Moreover, one can also see that the birefringent Shapiro time delay can constrain a larger portion of the theory than the multi-messenger delay, as can be seen comparing the orange and blue shaded regions respectively. 

%-FIGURE SUMMARY PLOT-
%---

We also include the geometrical time delay induced by the modified deflection angle with dashed and dashed-dotted lines for the multi-messenger and birefrengent delays respectively. We have chosen the redshift of the lens to give the maximum delay and, in this case, it can be more constraining than the Shapiro delay. One should note that while the geometrical time delay is subject to the lens-source-observer geometry, the Shapiro delay only cares about how close to the lens the GW passes. This means that for example if the lens is very close to the observer or source, the Shapiro time delay will dominate over the geometrical. This is interesting because from the sky localization of the source we can then ask for instance whether the GW has traveled close to the center of the Milky Way or Andromeda and quantify what would be the associated Shapiro delay. 

The possibility of detecting Shapiro time delays via birefringence allows novel tests of GR via GW lensing. One such possibility is the case of a binary merging in the environment of a SMBH, discussed in section \ref{sec:observational_near_lens}. 
If it turns out that there is non-negligible population of BBHs merging near an AGN, these would be ideal sources to constrain this type of modify gravity theories. 
For example, if the EM flare associated to GW190521 \cite{Abbott:2020tfl} was confirmed as an indication of this type of systems \cite{PhysRevLett.124.251102}, this would imply that the binary would have merge very close to the SMBH, around $20-300\,r_s$ \cite{Bellovary:2015ifg}. In this interpretation the mass of the SMBH near GW190521 would be $\sim10^8\Msun$ so that the BBH would be located at $\sim0.2-3\cdot10^{-3}(M_L/10^8\Msun)$\,\text{pc}. 
For reference, we include in figure \ref{fig:summary_plot} with dotted lines where the Vainshtein radius is placed in the $(\pFphi,\LF)$ parameter space for $r_V/r_s=10^3,10^4,10^5$. Therefore, BBHs merging in the accretion disk of an AGN would probe the whole birefringent Shapiro time delay parameter space region in orange.

%--
%CONCLUSIONS
%--
\section{Conclusions and prospects} \label{sec:conclusions}

%sketch of motivation
Gravitational lensing of gravitational waves (GWs) is sensitive to the propagation of GWs around massive objects and cosmic structures. 
Gravity theories beyond general relativity (GR) modify the GW propagation by altering the background on which GW propagate and introducing mixing among different polarizations. A theory for GW propagation should unify known propagation effects on FRW backgrounds with new interactions between different gravitational degrees of freedom around gravitational lenses, incorporating and generalizing the phenomenology of gravitational lensing. Formulating such a theoretical framework poses a significant challenge.

%sketch of approach
Here we have analyzed the propagation of gravitational radiation beyond GR in general space-times. We have first developed a model-independent framework and then applied it to Horndeski scalar-tensor theories, without assuming that GWs propagate at the speed of light. 
We addressed the mixing between different gravitational polarizations induced by lenses that locally break homogeneity and isotropy, working at leading order in derivatives. This approach allowed us to study the causal structure and thus the arrival time of different signals. It also provides a measure of the mixing between different radiative degrees of freedom, but not the corrections to their amplitudes.

%main conclusions
Our main conclusions can be summarized as follow:
\begin{itemize}
 \item Simplifications that allow the study of GWs in GR can not be generalized beyond. The traceless gauge can only be set as an initial condition. Non-radiative degrees of freedom, sourced by the GWs via constraint equations, become \emph{GW shadows}. 
 
 \item GW propagation is best described using propagation eigenstates $H_A$, which differs from the interaction basis $(h_{\mu\nu},\vp)$ in a space-dependent manner. Breaking the symmetry of the background is necessary for the $H_A$'s to mix the metric and scalar fields in Lorentz invariant theories.
 
 \item Propagation eigenstates can travel at different speeds, $c_1,c_2,c_3$. These depend on the theory and background solutions through the speeds for the metric, scalar \& mixing term ($c_h, c_s, c_m$), and the mixing amplitude $M_\phi$.

\item Gravitational lenses act like prisms, splitting the propagation eigenstates $H_I$ according to their speed. Differences in local speed and deflection angles contribute to lensing-induced time delays between $H_1, H_2, H_3$ and possible EM counterparts. 

\item The most promising novel observable is the \textit{birefringent time delay} between the two mostly-metric polarizations $H_1=h_\times$, $H_2\sim h_+$ (no EM counterpart needed).
A lensed GW signal can interfere with itself, causing a \emph{scrambling} of the wave-form, or be split into \emph{echoes} from the same event.
 
\item  GW birefringence provides novel tests of theories with \emph{screening mechanisms}. We present detailed predictions for a quartic Horndeski theory, showing how  GW lensing effects have the potential to probe regions of the parameter space beyond the stringent limits set by GW170817. 

\end{itemize}

%future direction
This work is a first step in developing a theory for GW propagation beyond GR including additional polarizations, and exploring the phenomenology of polarization mixing. Future work needs to include the evolution of the amplitude (perhaps at lower WKB orders) to derive complete predictions that can be tested against GW data.

%Other future directions to mention/discuss
 %Framework: 
The first obvious step at the theoretical level is extending the computation to next-to-leading order in the short-wave expansion and beyond. The full geometric optics framework is needed to reliably compute the amplitude and explore new effects that persist in the high-frequency limit. 
Additional, post-geometric optics corrections are frequency-dependent and could be very constraining, even if they're suppressed by inverse powers of the frequency. As argued in the text, birefringence may persist in the wave optics limit (at least for $f\gg 1/r_s, 1/r_V$): a complete treatment will allow new lenses to be used to test beyond GR theories, including stellar mass objects as lenses for LIGO/Virgo sources.

Another future direction is to link the general framework developed in section \ref{sec:lensing_beyond_gr} to other theories of gravity and place constraints on them. The example quartic Horndeski theory we have considered in section \ref{sec:screening} is already very constrained by GW170817, so further constraints require extreme lenses. However, in theories with multiple fields or Lorentz violation the cosmological/homogeneous deviation in the GW speed could be suppressed, allowing GW birefringence effects to place stringent constraints. As we have discussed in section \ref{sec:quintic_theories}, constraints may be derived also for theories with scalar hair, like scalar Gauss-Bonnet.

Future analyses should also test these novel beyond GR lensing effects against GW data. Under the assumptions outlined in section \ref{sec:observational_effects} birefringence predicts a very simple modification of the waveform, depending only on two parameters per lens. In the scrambling regime, the predictions can be tested against available GW data, including degeneracies with source parameters. Tests in the ``echoe'' regime, which splits the signal in two, are more subtle and rely on either on pairing events with related properties (similarly to searches for strongly lensed signals) or on an excess of edge on sources (if signals are lost). A robust statistical framework is needed to carry those tests, as well as to use them to further constrain theories of gravity. 

%Maps, new opportunities. 
The nature of birefringence beyond GR allows new opportunities with respect to ``traditional'' lensing studies. For instance, lenses near either the source or the observer (with very small Einstein radii) have a decent chance to produce birefringence through the Shapiro time delay. Correlating signals with maps of known nearby lenses may allow to refine constraints substantially (e.g. signals coming through the galaxy plane). 
Moreover, if a fraction of the population of binary black holes merge in the disk of active galactic nuclei, as it is suggested by the possible EM counterpart of GW190521 \cite{Abbott:2020tfl} discussed in \cite{PhysRevLett.124.251102}, these systems would be ideal for these tests. 
Similarly, identified strongly lensed GW pairs would be valuable probes since in that case it is also guaranteed that the GW has traveled close to the lens. 

%inspiring final paragraph
Altogether, gravitational lensing of GWs has the potential to become a fruitful laboratory in which to test gravity. It benefits from the precision of tests of the GW propagation, while avoids the necessity of identifying EM counterparts which limits the reach of tests of the cosmological GW propagation. Future GW observing runs will provide enough events for these and other novel lensing effects to be probed. 
This works represents a first step towards understanding the rich phenomenology of GW lensing beyond GR.

%-- Acknowledgments
\begin{acknowledgments}
We are grateful to Dario Bettoni and Kurt Hinterbichler for insightful conversations that helped seeding this project.
J.M.E. is supported by NASA through the NASA Hubble Fellowship grant HST-HF2-51435.001-A awarded by the Space Telescope Science Institute, which is operated by the Association of Universities for Research in Astronomy, Inc., for NASA, under contract NAS5-26555. He is also supported by the Kavli Institute for Cosmological Physics through an endowment from the Kavli Foundation and its founder Fred Kavli. 
%MZ is supported by...
\end{acknowledgments}

%--
%APPENDIX
%--
\appendix

%-------------
%MORE ON THE GAUGE
%-------------
\section{Alternative transverse gauges} \label{app:gauge_alternative}

It is possible to define \textit{alternative transverse gauges} (ATGs), relative to a (generic) metric $\tilde g_{\mu\nu}$ 
\begin{equation}\label{eq:alt_transverse_gauge}
 \tilde g^{\mu\alpha}\tilde\nabla_\alpha \tilde h_{\mu\nu} = 0\,.
\end{equation}
Here the trace-reversed metric is defined using the same tilde metric
\begin{equation}
 \tilde h_{\mu\nu} = h_{\mu\nu}-\frac{1}{2}\tilde g_{\mu\nu}\tilde h \,,\quad \tilde h = \tilde g^{\alpha\beta}h_{\alpha\beta}\,,
\end{equation}
This choice allows to express the residual gauge condition (preserving the ATG) as a wave equation at leading order in derivatives:
\begin{equation}\label{eq:alt_residual_gauge}
 \tilde \Box \xi_\mu + \cdots  
 = 0\,,
\end{equation}
where $\tilde \Box =  \tilde g^{\alpha\beta}\tilde\nabla_\alpha\tilde\nabla_\beta$ and $\cdots$ include both curvature and non-metricity terms that appear from re-arranging covariant derivatives and are lower order in derivatives. Explicitly
\begin{equation}
 \delta(\tilde\nabla^\alpha\tilde h_{\alpha\nu}) 
 =
 \tilde\Box\xi_\nu + \tilde R\ud{\nu}{\lambda}\xi_\lambda 
+ 2\tilde\nabla^\alpha(\mathcal{K}^\lambda_{\alpha\nu}\xi_\lambda)-\tilde\nabla_\nu(\mathcal{K}^\lambda\xi_\lambda)
 \,,
\end{equation}
where indices are lowered/raised with $\tilde g_{\mu\nu}$ and $\mathcal{K}^\lambda_{\alpha\beta}\equiv \bar\Gamma^\lambda_{\alpha\beta}-\tilde\Gamma^\lambda_{\alpha\beta}$, $\mathcal{K}^\lambda \equiv \tilde g^{\alpha\beta}\tilde\Gamma^\lambda_{\alpha\beta}$ encompass the difference between the connections (which are tensors, see Ref. \cite{Zumalacarregui:2013pma} for explicit expressions).%
\footnote{The ATG (\ref{eq:alt_transverse_gauge}) can be defined with a different covariant derivative. For instance, using $\tilde g^{\mu\alpha}\bar\nabla_\alpha \tilde h_{\mu\nu}$ (compatible with the background metric, as it emerges from the gauge transformation of $\bar h_{\mu\nu}$) yields
 \begin{eqnarray}
 \delta(\tilde g^{\alpha\beta}\bar\nabla_\beta\tilde h_{\alpha\nu})&=&
 \tilde g^{\alpha\beta}\bar\nabla_\alpha\bar\nabla_\beta\xi_\nu + \tilde g^{\alpha\beta}\bar R_{\alpha\phantom{\lambda}\beta\nu}^{\phantom{\alpha}\lambda}\xi_\lambda \\ && \nonumber 
 + \tilde g_{\nu\alpha}\mathcal{Q}^{\beta\alpha}_\beta(\tilde g^{\lambda\sigma}\bar\nabla_\lambda\xi_\sigma) 
 - \mathcal{Q}^{\lambda\sigma}_\nu(\bar\nabla_\lambda\xi_\sigma)\,,
\end{eqnarray}
where the non-metricity is defined as $\mathcal{Q}^{\alpha\beta}_\mu=\bar\nabla_\mu \tilde g^{\alpha\beta}$.
}

The next step is trying to fix other components (e.g. a trace) using the residual gauge. The residual ATG transformations can be written schematically as
\begin{equation}\label{eq:ATG_residual_schematic}
 \tilde\Box \xi_\nu + (\mathcal{K}\nabla\xi)_\nu + (\mathcal{M}\xi)_\nu = 0\,,
\end{equation}
where terms are arranged by number of derivatives in $\xi$ ($\mathcal{M}\supset \nabla\mathcal{K},R$) and the contracted indices have been omitted for conceptual simplicity.
To get a sense of the effect of this terms let's work on locally homogeneous space and define an even simpler version of the residual ATG equation
\begin{equation}
 \tilde g^{\mu\nu}\partial_\mu\partial_\nu \chi + k^\mu\partial_\nu\chi  + m^2\chi = 0 \,.
\end{equation}
This equation can be solved in Fourier space in the limit of high $|\vec k|$ as 
\begin{equation}\label{eq:ATG_residual_schematic_dispersion}
 \omega \approx \tilde c |\vec k| + \frac{i}{2}\Gamma + \frac{m^2}{2\tilde c|\vec k|}\,,
\end{equation}
where $\tilde c$ is the speed of sound of $\tilde g_{\mu\nu}$ (which may depend on $\hat k$) and $\Gamma \equiv \gamma^0 + \vec\gamma \hat k/\tilde c$.

The residual gauge allows us to fix the initial conditions of Eq. (\ref{eq:ATG_residual_schematic}) (real and imaginary parts of 4 $\xi_\nu$ components), which can be used to set 4 metric components to zero at some initial time
\begin{equation}\label{eq:ATG_residual_gauge_initial}
 h_X(t_0,\vec x) = 0\,,
\end{equation}
where $h_X$ can be a combination of metric perturbations (e.g. the trace $h$ or another trace such as $\tilde h$).
This condition will hold at later times only if $h_X$ obeys the same equation as the residual gauge. 
Let us assume that the solutions to 
$h_X \propto \int d^3k \tilde h_X(\vec k)e^{i (\omega_X t - \vec k \vec x)}$ 
follow a dispersion relation similar to Eq. (\ref{eq:ATG_residual_schematic_dispersion})
\begin{equation}\label{eq:ATG_metric_schematic_dispersion}
 \omega_X \approx c_X |\vec k| + \frac{i}{2}\Gamma_X + \frac{m^2_X}{2\tilde c|\vec k|}\,.
\end{equation}
The difference between the residual gauge and the physical mode solutions, Eqs. (\ref{eq:ATG_residual_schematic_dispersion},\ref{eq:ATG_metric_schematic_dispersion}) determine how far the residual gauge (\ref{eq:ATG_residual_gauge_initial}) can be extended beyond $t=t_0$:
\begin{enumerate}
 \item if $\tilde c \neq c_X$ the residual gauge can be fixed for $|\Delta x|\ll \Delta c/|\vec k|$, usually less than a wavelenght! In general, the freedom in choosing the ATG via $\tilde g_{\mu\nu}$ ensures that $\tilde c = c_X$ can be imposed 
 
\item if the friction differs $\Gamma \neq \Gamma_X$, fixing $X$ is a good approximation only in a region $|\Delta x^\mu|\ll \Delta\Gamma^{-1}$, which is determined by the non-metricity $\mathcal{K}$, but independent of the physical frequency.

\item if the mass term differs $m^2\neq m^2_X$ the fixing is good in a region $|\Delta x|\cdot \Delta m^2/|\vec k| \ll 1$, which becomes arbitrarily large at higher frequencies.

\end{enumerate}
Note that these conditions do not take into account the failure of the constant background assumption, which is independent of the GW frequency. Matter sources will also make it impossible to set $h_X = 0$ (just as in GR).

While fixing $\tilde c = c_X$ can be done in general (this is reason for defining an ATG), doing so introduces friction and and curvature terms that limit the validity of $h_X\approx 0$ (cases 2,3). While the mass condition (case 3) might be unimportant for sufficiently large frequencies, the friction condition (case 2) imposes a frequency-independent limit to the condition $h_X\approx 0$. Depending on the difference between $\bar g_{\mu\nu},\tilde g_{\mu\nu}$, this region may or not be large enough for the residual ATG to afford a valuable simplification.

\begin{widetext}
%-----------
%PROPAGATION EQUATIONS
%-----------
\section{Details on the propagation equations}
\label{app:equations}

In this appendix we provide further details on the equations of motion that we have used in the main text to compute the propagation eigenstates and mixing between the different polarizations. 
We will make use of the following perturbations of the Riemman tensor:
\begin{align}
&\delta R_{\mu\alpha\nu\beta}=-\frac{1}{2}\nabla_\nu\nabla_\mu h_{\beta\alpha}-\frac{1}{2}\nabla_\beta\nabla_\alpha h_{\nu\mu}+\nabla_\nu\nabla_{(\alpha}h_{\beta)\mu}+\nabla_\beta\nabla_{[\mu}h_{\nu]\alpha}+R^{\lambda}_{~\alpha\nu\beta}h_{\lambda\mu}\,, \\
&\delta R_{\mu\nu}=-\frac{1}{2}\Box\hmn+\nabla_{(\mu}\nabla^\alpha h_{\alpha\nu)}-\frac{1}{2}\nabla_\mu\nabla_\nu h+R^{\alpha}_{~(\mu}h_{\nu)\alpha}-R_{\mu\alpha\nu\beta}h^{\alpha\beta}\,, \\
&\delta R=-\Box h+\nabla^\alpha\nabla^\beta h_{\alpha\beta}-R_{\alpha\beta}h^{\alpha\beta}\,.
\end{align}
These identities can also be written in term of the trace-reversed perturbation $\hbmn$
\begin{align}
&\delta \bar R_{\mu\alpha\nu\beta}=-\frac{1}{2}\nabla_\nu\nabla_\mu \hb_{\beta\alpha}-\frac{1}{2}\nabla_\beta\nabla_\alpha \hb_{\nu\mu}+\nabla_\nu\nabla_{(\alpha}\hb_{\beta)\mu}+\nabla_\beta\nabla_{[\mu}\hb_{\nu]\alpha}+R^{\lambda}_{~\alpha\nu\beta}\hb_{\lambda\mu}-\frac{1}{2}R_{\mu\alpha\nu\beta}\hb \\
&~~~~~~~~~~~~~~~+\frac{1}{4}g_{\beta\alpha}\nabla_\nu\nabla_\mu \hb+\frac{1}{4}g_{\nu\mu}\nabla_\beta\nabla_\alpha \hb-\frac{1}{2}\nabla_\nu\nabla_{(\alpha}g_{\beta)\mu}\hb-\frac{1}{2}\nabla_\beta\nabla_{[\mu}g_{\nu]\alpha}\hb \\
&\delta \bar R_{\mu\nu}=-\frac{1}{2}\Box\hbmn+\nabla_{(\mu}\nabla^\alpha\hb_{\alpha\nu)}+\frac{1}{4}\gmn\Box\hb+R^{\alpha}_{~(\mu}\hb_{\nu)\alpha}-R_{\mu\alpha\nu\beta}\hb^{\alpha\beta} \\
&\delta \bar R=\nabla^\alpha\nabla^\beta\hb_{\alpha\beta}+\frac{1}{2}\Box\hb-R_{\alpha\beta}\hb^{\alpha\beta}+\frac{1}{2}R\hb
\end{align}

\subsection{Generalized Brans-Dicke}

We begin by considering the generalized Brans-Dicke theory presented in Eq. (\ref{eq:brans_dicke}). 
The metric EoM are given by
\be
G_4\Gmn+\gmn\lp G_{4\phi}\Box\phi-2XG_{4\phi\phi}\rp-G_{4\phi}\phi_{\mu\nu}-G_{4\phi\phi}\phi_\mu\phi_\nu-\frac{1}{2}\gmn G_2-\frac{1}{2}G_{2X}\phi_\mu\phi_\nu=0\,,
\ee
while the scalar EoM reads
\be
G_{4\phi}R+G_{2X}\Box\phi-G_{2XX}\pP=0\,.
\ee
By computing the perturbation of these equations and focusing in the leading derivative part, we can rewrite the EoM as
\be
\lb\bpm G_4\bD_{\alpha\beta}^{~~~\mu\nu} & 0 & G_{4,\phi}(g_{\alpha\beta}\Box-\nabla_\alpha\nabla_\beta) \\  -G_4\nabla^\nu\nabla^\mu & -G_4\Box/2 & 3G_{4,\phi}\Box \\ G_{4,\phi}\nabla^\nu\nabla^\mu & G_{4,\phi}\Box/2 & G_{2X}\Box-G_{2XX}\phi^\alpha\phi^\beta\nabla_\alpha\nabla_\beta\epm+
\cdots\rb\bpm \hbmn \\ \hb \\ \vp\epm =0\,\,,
\ee
where we have introduced the trace-reversed perturbation (\ref{eq:trace-reversed}) and the differential operator $\bD$ defined in Eq. (\ref{eq:diff_op}). Here we have introduced a matrix notation to highlight the diagonalization process. One should note that the second row is nothing but the trace of the tensor equation. 
From this equation it is then direct to see that one can reabsorb the scalar perturbation terms in the metric equation by introducing a new perturbation as given by (\ref{eq:redefinition_bd}). Then applying the transverse condition to the new perturbation one completely diagonalizes the problem. 

\subsection{Kinetic Gravity Braiding}

We can follow a similar approach for Kinetic Gravity Braiding, the cubic Horndeski theory defined in Eq. (\ref{eq:kgb}). 
The metric EoM of this theory are given by
\be
G_4\,\Gmn+\frac{1}{2}G_{3,X}\Box\phi\,\phi_\mu\phi_\nu-G_{3,X}\phi^\alpha\phi_{\alpha(\mu}\phi_{\nu)}+\frac{1}{2}g_{\mu\nu}G_{3,X}\pP=0\,
\ee
and the scalar EoM follows
\be
G_{3,X}\phi^\mu\phi^\nu R_{\mu\nu}+G_{3,X}\lp(\Box\phi)^2-[\phi^2]\rp+G_{3,XX}\lp\pPP-\pP\Box\phi\rp=0\,.
\ee
Again, the leading derivative part of the EoM for the perturbations can be written in matrix form as
\be
\lb\bpm 2G_4\bD_{\alpha\beta}^{~~~\mu\nu} & 0 & G_{3X}\lp\phi_\alpha\phi_\beta\Box -2\phi^\mu\phi_{(\alpha}\nabla_{\beta)}\nabla_\mu+\gab\phi^\mu\phi^\nu\nabla_\mu\nabla_\nu\rp \\  -2G_4\nabla^\nu\nabla^\mu & -G_4\Box & G_{3X}\lp-2X\Box +2\phi^\mu\phi^\nu\nabla_\mu\nabla_\nu\rp \\ G_{3X}\lp-\phi^\mu\phi^\nu\Box+2\phi^\alpha\phi^\beta\nabla_{(\alpha}\nabla^\mu\delta_{\beta)}^{\nu}\rp & -XG_{3X}\Box & \dpp \epm+
\cdots\rb\bpm \hbmn \\ \hb \\ \vp\epm =0\,,
\ee
where we have introduced the scalar term
\be
\dpp=G_{3,X}\lp2\Box\phi\Box-2\phi^{\mu\nu}\nabla_\mu\nabla_\nu\rp+G_{3,XX}\lp\lp2\phi_\gamma\phi^{\gamma(\alpha}\phi^{\beta)}-\Box\phi\phi^\alpha\phi^\beta\rp\nabla_\alpha\nabla_\beta-\pP\Box\rp\,.
\ee
From here we se again that the scalar perturbation terms in the metric equation can be reabsorbed in a field redefinition as given by (\ref{eq:redefG3}). Then applying the transverse condition one fully diagonalizes the leading derivative interactions.

\subsection{Shift-symmetric quartic Horndeski}

Moving to a non-luminal theory, we consider now a shift-symmetric quartic Horndeski theory with generalized kinetic term for the scalar as given by equation \ref{eq:quartic_horndeski}. 
At leading order in derivatives for the linear perturbations, the EoM are given by
\begin{align}
&G_4\delta G_{\mu\nu}+G_{4X}\delta\mathcal{R}_{\mu\alpha\nu\beta}\phi^\alpha\phi^\beta+(G_{4X}\mathcal{C}_{\mu\nu}^{~~\alpha\beta}+G_{4XX}\mathcal{E}_{\mu\nu}^{~~\alpha\beta})\nabla_\alpha\nabla_\beta\vp=0\,, \\
&\mathcal{G}_s^{\alpha\beta}\nabla_\alpha\nabla_\beta\vp+2G_{4X}\phi^{\mu\nu}\delta G_{\mu\nu}-2G_{4XX}\phi^{\mu\nu}\delta\mathcal{R}_{\mu\alpha\nu\beta}\phi^\alpha\phi^\beta=0\,,
\end{align}
where we have defined two tensors of the metric perturbations $\hmn$
\begin{align}
\delta G_{\mu\nu}&\equiv \delta R_{\mu\nu}-\frac{1}{2}g_{\mu\nu}\delta R\,, \\
\delta\mathcal{R}_{\mu\alpha\nu\beta}\phi^\alpha\phi^\beta&\equiv2\phi_{(\mu}\delta R_{\nu)\lambda}\phi^\lambda+\delta R_{\mu\alpha\nu\beta}\phi^\alpha\phi^\beta-\delta R_{\alpha\beta}\phi^{\alpha}\phi^\beta g_{\mu\nu}-\frac{1}{2}\delta R \phi_\mu\phi_\nu
\end{align}
and two tensors contracted with the scalar perturbations $\vp$
\begin{align}
\mathcal{C}_{\mu\nu}^{~~\alpha\beta}\vp_{\alpha\beta}\equiv&(\Box\phi\Box\varphi-\phi^{\alpha\beta}\varphi_{\alpha\beta})g_{\mu\nu}-(\varphi_{\mu\nu}\phi^{\alpha\beta}+\phi_{\mu\nu}\varphi^{\alpha\beta})g_{\alpha\beta}+2\phi^{\alpha}_{(\mu}\varphi_{\nu)}^{\beta}g_{\alpha\beta})\,,
\end{align}
\begin{equation}
\begin{split}
\mathcal{E}_{\mu\nu}^{~~\alpha\beta}\vp_{\alpha\beta}\equiv&(\phi^{\alpha\beta}\varphi_{\alpha\beta}-\Box\phi\Box\varphi)\phi_\mu\phi_\nu+(\varphi_{\mu\nu}\phi^{\alpha\beta}+\phi_{\mu\nu}\varphi^{\alpha\beta})\phi_\alpha\phi_\beta \\
+&2\phi^{\alpha}_{(\mu}\varphi_{\nu)}^{\beta}(-\phi_\alpha\phi_\beta)-(\Box\varphi\phi_{\alpha\beta}\phi^\alpha\phi^\beta+\Box\phi\varphi_{\alpha\beta}\phi^\alpha\phi^\beta-2\phi^\alpha\varphi_{\alpha\beta}\phi^{\beta\gamma}\phi_\gamma)g_{\mu\nu} \\
+&2\phi^\gamma(\Box\phi\varphi_{\gamma(\mu}\phi_{\nu)}+\Box\varphi\phi_{\gamma(\mu}\phi_{\nu)}-\varphi_{\gamma\sigma}\phi^{\sigma}_{(\mu}\phi_{\nu)}-\phi_{\gamma\sigma}\varphi^{\sigma}_{(\mu}\phi_{\nu)})\,,
\end{split}
\end{equation}
and the scalar effective metric
\begin{equation} \label{eq:full_scalar_metric}
    \begin{split}
\mathcal{G}_s^{\mu\nu}=& 2G_{4X}G^{\mu\nu}+G_{4XX}(-4\phi_\lambda R^{\lambda(\mu}\phi^{\nu)}+\phi^{\mu}\phi^{\nu} R+2\phi^\alpha\phi^\beta R_{\alpha\beta}g^{\mu\nu}-2\phi_{\alpha}\phi_{\beta} R^{\mu\alpha\nu\beta}) \\
-&G_{4XX}(3(\Box\phi^{2}-\nabla\nabla\phi^{2})g^{\mu\nu}-6\Box\phi\cdot\phi^{\mu\nu}+6\phi_\lambda^{~\mu}\phi^{\nu\lambda}) \\
+&G_{4XXX}((\Box\phi^{2}-\nabla\nabla\phi^{2})\phi^\mu\phi^\nu+2\phi^\alpha\phi^\beta\phi_{\alpha\beta}(\Box\phi g^{\mu\nu}-\phi^{\mu\nu}) \\
-&2(2\Box\phi\phi^{(\mu}\phi^{\nu)\lambda}\phi_{\lambda}-\phi_\lambda\phi^{\lambda\mu}\phi^{\nu\gamma}\phi_\gamma-2\phi^{(\mu}\phi^{\nu)\lambda}\phi_{\lambda\gamma}\phi^\gamma+\phi^\alpha\phi_{\alpha\beta}\phi^{\beta\gamma}\phi_\gamma g^{\mu\nu}))\,.
\end{split}
\end{equation}

If we restrict to a theory where $G_{4XX}=G_{4XXX}=0$, then we do not have to consider $\mathcal{E}_{\mu\nu}^{~~\alpha\beta}$ and the effective scalar metric simplifies to
\begin{align}
\mathcal{G}_s^{\alpha\beta}\vp_{\alpha\beta}=& \lp G_{2X}g^{\alpha\beta}-G_{2XX}\phi^\alpha\phi^\beta+2G_{4X}G^{\alpha\beta}\rp\nabla_\alpha\nabla_\beta\vp\,.
\end{align}
It is useful to rewrite the tensors of the metric perturbations in terms of the trace-reversed metric (\ref{eq:trace-reversed}), 
so that we obtain
\begin{align}
\delta G_{\mu\nu}&=\bD_{\mu\nu}^{\alpha\beta}\hb_{\alpha\beta}=-\frac{1}{2}\Box\hbmn+\nabla_{(\mu}\nabla^\alpha \hb_{\alpha\nu)}-\frac{1}{2}\nabla^\alpha\nabla^\beta\hb_{\alpha\beta}\,, \\
\delta\mathcal{R}_{\mu\alpha\nu\beta}\phi^\alpha\phi^\beta&=
2\phi_{(\mu}\lp-\frac{1}{2}\Box \hb_{\nu)\lambda}+\nabla_{(\nu)}\nabla^\alpha \hb_{\alpha\lambda)}+\frac{1}{4}g_{\nu)\lambda}\Box\hb\rp\phi^\lambda \\
+&\Huge(-\frac{1}{2}\nabla_\nu\nabla_\mu \hb_{\beta\alpha}-\frac{1}{2}\nabla_\beta\nabla_\alpha \hb_{\nu\mu}+\nabla_\nu\nabla_{(\alpha}\hb_{\beta)\mu}+\nabla_\beta\nabla_{[\mu}\hb_{\nu]\alpha} \nonumber\\
+&\frac{1}{4}g_{\alpha\beta}\nabla_\nu\nabla_\mu\hb+\frac{1}{4}\gmn\nabla_\alpha\nabla_\beta\hb-\frac{1}{2}\nabla_\nu\nabla_{(\alpha}g_{\beta)\mu}\hb-\frac{1}{2}\nabla_\beta\nabla_{[\mu}g_{\nu]\alpha}\hb\Huge)\phi^\alpha\phi^\beta \nonumber\\
+&\lp\frac{1}{2}\Box \hb_{\alpha\beta}-\nabla_{(\alpha}\nabla^\rho \hb_{\rho\beta)}-\frac{1}{4}g_{\alpha\beta}\Box\hb\rp\phi^{\alpha}\phi^\beta g_{\mu\nu} \nonumber\\
-&\frac{1}{4}\lp\Box \hb+\nabla^\alpha\nabla^\beta \hb_{\alpha\beta}\rp \phi_\mu\phi_\nu \nonumber
\end{align}
In the transverse gauge for the trace reversed perturbations, $\nabla^\mu\hbmn=0$, the EoM of the $G_{4XX}=G_{4XXX}=0$ theory simplify to 
\begin{align}
G_4\Box\hbmn&+G_{4X}\Huge(\phi^\alpha\phi^\beta\nabla_\beta\nabla_\alpha \hbmn+2\phi_{(\mu}\Box \hb_{\nu)\lambda}\phi^\lambda-\phi_{(\mu}g_{\nu)\lambda}\Box\hb\phi^\lambda -\frac{1}{2}\gmn\phi^\alpha\phi^\beta\nabla_\alpha\nabla_\beta\hb \\
-&g_{\mu\nu}\lp\phi^{\alpha}\phi^\beta\Box\hb_{\alpha\beta}+X\Box\hb\rp +\frac{1}{2}\phi_\mu\phi_\nu\Box\hb\Huge)-2G_{4X}\mathcal{C}_{\mu\nu}^{~\alpha\beta}\vp_{\alpha\beta}=0\,, \nonumber\\
\mathcal{G}_s^{\alpha\beta}\varphi_{\alpha\beta}&-G_{4X}\phi^{\mu\nu}\Box\hbmn=0\,.
\end{align}

\end{widetext}

%-------------------
%APP: DEGREES OF FREEDOM
%-------------------
\section{Local diagonalization of the propagating degrees of freedom}
\label{app:local_diagonalization}

As discussed in section \ref{sec:radiative_dof}, in order to solve the local propagation we have to diagonalize a system of $11\times11$ equations of motion for $\Phi$, $w_i$, $s_{ij}$, $\Psi$ and $\vp$. This calculation will make use of the following perturbations of the Riemann tensor
\begin{align}
\delta R_{0j0l}&=\partial_j\partial_l\Phi+\partial_0\partial_{(j}w_{l)}-\frac{1}{2}\partial_0\partial_0 h_{jl}\,, \\
\delta R_{0jkl}&=\partial_{j}\partial_{[k}w_{l]}-\partial_{0}\partial_{[k}h_{l]j}\,, \\
\delta R_{ijkl}&=\partial_{j}\partial_{[k}h_{l]i}-\partial_{i}\partial_{[k}h_{l]j}\,,
\end{align}
the Ricci tensor
\begin{align}
\delta R_{00}&=\nabla^{2}\Phi+\partial_0\partial_{k}w^{k}+3\partial_0^2 \Psi\,, \\
\delta R_{0j}&=-\frac{1}{2}\nabla^{2}w_{j}+\frac{1}{2}\partial_j\partial_k w^k+2\partial_{0}\partial_{j}\Psi+\partial_{0}\partial_{k}s^{k}_{j}\,, \\
\delta R_{ij}&=-\partial_{i}\partial_{j}(\Phi-\Psi)-\partial_{0}\partial_{(i}w_{j)}+\Box\Psi\delta_{ij}\\
&\quad\quad\quad\quad\quad\quad\quad-\Box s_{ij}+2\partial_{k}\partial_{(i}s_{j)}^{k}\,,\nonumber
\end{align}
and of the Ricci scalar
\begin{equation}
\delta R=-2\nabla^2\Phi-2\partial_0\partial_kw^k-6\partial_0^2\Psi+4\nabla^2\Psi+2\partial^k\partial^j s_{kj}\,.
\end{equation}
We have denoted $\partial_0^2=\partial_0\partial_0$ and $\nabla^2=\partial_i\partial^i$.

The first thing to notice is that the above equations do not contain second order time derivatives of $w_i$ or $\Phi$. This means that for theories with EoM that are linear in the perturbed Riemann tensor, these modes will not propagate. They can be written in terms of the other propagating DoF.
Fortunately, this is the case of Horndeski theory and we do not need to worry about these modes. The only caution to take is that, although not propagating, they can be sourced by the scalar background for instance. Thus, if we want to keep the analysis fully general, we cannot set them to zero. 

In the following we provide further details on the diagonalization of a quartic Horndeski theory discussed in section \ref{sec:quartic}. In particular, we will detail the equations needed to solve the propagation in the \emph{absence} of scalar perturbations. The main operator that we need to compute are $\delta G_{\mu\nu}$ and $\delta\mathcal{R}_{\mu\alpha\nu\beta}\phi^\alpha\phi^\beta$. 
Let's begin with the 00-equation. The relevant terms are
\be
\begin{split}
\delta G_{00} & = \delta R_{00} - \frac{1}{2}\eta_{00}\delta R =2\nabla^2\Psi+\partial^k\partial^j s_{kj} %+ \cdots
\end{split}
\ee
and
\be
\begin{split}
\delta&\mathcal{R}_{0\alpha0\beta}\phi^\alpha\phi^\beta  = \delta R_{0i0j}\phi^i\phi^j - \eta_{00}\delta R_{ij}\phi^i\phi^j \\
&=\phi_i\phi^i\nabla^2\Psi+\phi^i\phi^j \lp\partial_i\partial_j\Psi-\nabla^2s_{ij}+2\partial_k\partial_{i} s_{j}^{\ k}\rp %+ \cdots
\end{split}
\ee
Therefore, as in GR, the 00-equation tell us that the spatial trace $\Psi$ follows a Poisson-like equation where the source are the components $s_{ij}$ modulated by the background.  
This implies that for this theory only $s_{ij}$ contains propagating DoF. Note that in the case of having a scalar perturbation present this conclusion would not change. 

In the transverse gauge, for quartic Horndeski in vacuum and $\vp=0$, we can write the solution of the $00$-equation, presented in Eq. (\ref{eq:G4X00}), as
\be
\Psi\sim \frac{G_{4X}\phi^i\phi^j s_{ij}}{2G_4 + G_{4X}\lp\phi_\pe^2+2\phi_\pa^2\rp}\,.
\ee

We apply a similar strategy to the other equations. To simplify let us fix $\partial^iw_i=0$. For the the $0j$-equations, the relevant terms are
\be
\delta G_{0j}=\delta R_{0j}-\frac{1}{2}\eta_{0j}\delta R = -\frac{1}{2}\nabla^2w_j+2\partial_0\partial_j\Psi
\ee
and
\be
\begin{split}
\delta&\mathcal{R}_{0\alpha j\beta}\phi^\alpha\phi^\beta = \phi_j\delta R_{0k}\phi^k + \delta R_{0kjl}\phi^k\phi^l \\
&=\phi_j\phi^k\lp\partial_0\partial_k\Psi-\frac{1}{2}\nabla^2w_k\rp +\phi_k\phi^k\partial_0\partial_j\Psi \\
&+\phi^k\phi^l\lp\partial_k\partial_{[j}w_{l]}-2\partial_0\partial_{[j}s_{l]k}\rp\,.
\end{split}
\ee
Therefore we obtain the constraint equation for $w_j$ presented in equation \ref{eq:G4X0j}. 

Next we move to the $ij$-equations. The two parts are 
\be \label{eq:ij_1}
\begin{split} 
\delta G_{ij} &= \delta R_{ij}-\frac{1}{2}\eta_{ij}\delta R=-\partial_i\partial_j(\Phi-\Psi)-\partial_0\partial_{(i}w_{j)}\\
+&\Box\Psi\delta_{ij}-\Box s_{ij}+\lp\nabla^2\Phi+3\partial_0^2\Psi-2\nabla^2\Psi\rp\delta_{ij}\,,
\end{split}
\ee
and
\begin{widetext}
\be \label{eq:ij_2}
\begin{split} 
\delta&\mathcal{R}_{i\alpha j \beta}\phi^\alpha\phi^\beta = 2\phi_{(i}\delta R_{j)k}\phi^k+\delta R_{ikjl}\phi^l\phi^k -\delta R_{kl}\phi^k\phi^l\delta_{ij}-\frac{1}{2}\delta R \phi_i\phi_j \\
&=-2\phi_{(i}\partial_{j)}\partial_k\Phi\phi^k-\phi_{(i}\partial_0\partial_{j)}w_k\phi^k-\phi_{(i}\partial_0\partial_kw_{j)}\phi^k -2\phi_{(i}\Box s_{j)k}\phi^k + 2\phi^l\phi^k\Huge(\partial_k\partial_{[j}s_{l]i} \\
&-\partial_i\partial_{[j}s_{l]k}+\frac{1}{2}\partial_i\partial_{j}\delta_{lk}\Psi\Huge)-\phi^k\phi^l\lp\Box\Psi\delta_{kl}-\Box s_{kl}-\partial_k\partial_l\Phi-\partial_0\partial_{(k}w_{l)}\rp\delta_{ij} +\phi_i\phi_j \lp\nabla^2\Phi+\partial_0^2\Psi\rp\,.
\end{split}
\ee
\end{widetext}
With all these calculations we can compute the trace of the $ij$-equations which determine the evolution of $\Phi$ given in equation \ref{eq:G4Xtrace}.

%------------

\bibliography{gw_refs}

\end{document}